\documentclass[fleqn,usenatbib]{mnras}

\usepackage{newtxtext,newtxmath,hyperref}
\usepackage[T1]{fontenc}
\usepackage{graphicx}	
\usepackage{amsmath}	
\usepackage{amssymb}	
\usepackage[english]{babel}
\usepackage{gensymb}

\title{The atomic Hydrogen content of the post-reionization Universe}

\author[Spinelli et al.]{Marta Spinelli$^{1,3,4}$\thanks{E-mail: marta.spinelli@inaf.it}, Anna Zoldan$^{1}$, Gabriella De Lucia$^{1,4}$, Lizhi Xie$^{5}$ and Matteo Viel$^{2,1,3,4}$ 
\\
$^{1}$INAF-Osservatorio Astronomico di Trieste, Via G.B. Tiepolo 11, 34143 Trieste, Italy\\
$^{2}$SISSA, International School for Advanced Studies, Via Bonomea 265, 34136 Trieste TS, Italy\\
$^{3}$INFN, Sezione di Trieste, Via Valerio 2, 34127 Trieste TS, Italy\\
$^{4}$IFPU - Institute for Fundamental Physics of the Universe, Via Beirut 2, 34014 Trieste, Italy\\
$^{5}$Tianjin Astrophysics Center, Tianjin Normal University, Binshuixidao 393, 300384, Tianjin, China\\
}
\date{\today}
\pubyear{2019}
\begin{document}
\label{firstpage}
\pagerange{\pageref{firstpage}--\pageref{lastpage}}
\maketitle

\begin{abstract}
We present a comprehensive analysis of atomic hydrogen (HI) properties using a semi-analytical model of galaxy formation and N-body simulations covering a large cosmological volume at high resolution. 
We examine the HI mass function and the HI density, characterizing both their redshift evolution and their dependence on hosting halo mass. 
We analyze the HI content of dark matter haloes in the local Universe and up to redshift $z=5$, discussing the contribution of different galaxy properties. We find that different assembly history plays a crucial role in the scatter of this relation. We propose new fitting functions useful for constructing mock HI maps with HOD techniques.
We investigate the HI clustering properties relevant for future $21$~cm Intensity Mapping (IM) experiments, including the HI bias and the shot noise level.
The HI bias increases with redshift and it is roughly flat on the largest scales probed. The scale dependency is found at progressively larger scales with increasing redshift, apart from a dip feature at $z=0$. The shot-noise values are consistent with the ones inferred by independent studies, confirming that shot-noise will not be a limiting factor for IM experiments.  
We detail the contribution from various galaxy properties on the HI power spectrum and their relation to the halo bias. We find that HI poor satellite galaxies play an important role at the scales of the 1-halo term. Finally, we present the $21$~cm signal in redshift space, a fundamental prediction to be tested against data from future radio telescopes such as SKA. 
\end{abstract}



\section{Introduction}\label{sec:intro}
In the attempt to unravel the mysteries of the dark Universe, mapping its large scale structure on progressively larger volumes is crucially important. The ubiquitous cosmic neutral hydrogen (HI), seen through the $21$~cm line, is one of the best candidate to illuminate the underlying dark matter density distribution. In the post reionization Universe, most neutral hydrogen is expected to be in dense systems inside galaxies, like Damped Lyman-$\alpha$ absorbers, where the HI has been shielded from ionizing Ultra Violet (UV) photons, thus tracing well the galaxy distribution \citep[e.g.,][]{Furlanetto2006,Pritchard2010}. The $21$~cm redshifted signal (from the spin flip transition between the hyperfine states of the neutral hydrogen) therefore provides the possibility of performing tomographic studies of cosmic structures. Using radio telescopes at frequencies between $1420$~MHz to $250$~MHz, one can probe the 21cm line up to the end of the reionization epoch. 

To observe the required large volumes within reasonable amounts of telescope time, one solution is to integrate the signal in large angular pixels of the sky, without resolving individual HI galaxies, whose signal is already faint ($\sim\mu$Jy) at $z\sim 1.5$. This relatively new technique, called Intensity Mapping (IM), can be used to probe the large scale structure of the Universe, and constrain cosmology with competitive precision \citep[e.g.,][]{Bharadwaj2001,Battye2004,Loeb2008,Chang2008}.

Although the measurement of the auto-power spectrum of HI IM survey is currently challenged by the contamination from foreground residuals \citep{Switzer2013}, statistical detection of the  signal has been achieved in cross-correlation:
first measurements of the $21$~cm IM signal have been obtained using the Giant Meterwave Radio Telescope (GMRT) in cross-correlation with the DEEP2 optical redshift survey \citep{Chang2010} and with the emission line galaxy redshift survey WiggleZ \citep{Masui2013}, and using the Parkes telescope in cross-correlation with the 2dF survey \citep{Anderson2018}. Several other surveys are already ongoing or have been proposed, such as the Canadian Hydrogen Intensity Mapping Experiment \citep[CHIME,][]{Bandura2014}, the Tianlai cylinder array \citep{Xu2015}, the Hydrogen Intensity and Real-time Analysis eXperiment \citep[HYRAX,][]{Newburgh2016}, and surveys using MeerKAT \citep{Santos2017}. Ultimately, the Square Kilometre Array (SKA), combining a SKA1-MID WideBand survey ($20,000$~deg$^2$ and $10,000$~hrs integration time in single dish mode) with a SKA1-LOW deep-like survey ($100$~deg$^2$ sky coverage and $5,000$~hrs integration time), will provide a unique picture of HI on cosmological scales over a wide redshift range ($0<z<6$). 
While ultradeep radio continuum surveys planned on SKA will be crucial 
to characterize the statistical properties of the SFG population and thus the coevolution between galaxies and supermassive BHs \citep[e.g.][]{Mancuso2017}, IM surveys will have a great impact on cosmology. Indeed, the IM signal will constrain the expansion history of the Universe and the growth of structures, allowing us to measure neutrino masses \citep{VN2015}, to test dark matter nature at small scales \citep{Carucci2015}, to look for signatures of inflation in the power spectrum \citep{Xu2016,Ballardini2018}, and to probe the geometry of the Universe with Baryon Acoustic Oscillations \citep[BAOs,][]{Bull2015,VN2017bao}.

The primary tool adopted to describe the distribution of atomic hydrogen in the Universe is the 21~cm power spectrum $P_{21\mathrm{cm}}(k,z)$, that is a function of scale $k$ and of redshift. In redshift space, matter peculiar velocities result in an apparent enhancement of clustering on large scales \citep{Kaiser1987}, and the signal can be modeled as
\begin{equation}
    P_{21\mathrm{cm}}(z,k)=T_{\mathrm b}^2\left[\left(b_{\mathrm{HI}(k)}^2+f\mu_k^2\right)^2P_{\mathrm m}(z,k) +P_{\mathrm{SN}}\right]\, , \label{eq:P_21tot}
\end{equation}
where $T_{\rm b}$ is the mean brightness temperature that depends on the cosmic density of neutral hydrogen $\rho_{\mathrm{HI}}(z)$, $b_{\mathrm{HI}}(k)$ is the HI bias, $f$ is the linear growth rate, $\mu_k=\hat{k}\cdot \hat{z}$, $P_{\rm m}(z,k)$ is the linear matter power spectrum, and $P_{\mathrm{SN}}$ is the HI shot noise. In real space, there is a degeneracy between $\rho_{\mathrm{HI}}(z)$ and the HI bias that can be broken using Redshift Space Distortion \citep[RSD,][]{Masui2013}. The IM signal can be completely specified knowing the HI density, the bias, and the shot noise. An accurate modeling of these quantities is thus crucial in preparation of future survey data exploitation.

A convenient and fast approach to model the HI distribution on large scales is to use Halo Occupation Distribution (HOD) techniques. In these methods, the HI content of dark matter haloes depends only on halo mass - this defines the HI halo mass function $M_{\rm HI}(M_h)$ - and all the quantities described above can be related to this quantity \citep[e.g.][]{Santos2015, VN2018}. Halo catalogues obtained with approximate methods such as Lagrangian Perturbation Theory \citep[e.g][]{Monaco2002} can be used to populate the dark matter haloes with HI, in combination with parametrizations for the HI halo mass function. This allows the creation of a large number of HI mock catalogues, with relatively small computational resources. Standard HOD approaches, however, do not model the spatial distribution of HI within dark matter haloes and generally neglect possible environmental dependencies and assembly bias \citep[e.g.][]{Gao2007}. This can have important consequences when using this approach to carry out precision cosmology experiments. 

An alternative approach to model the HI distribution in a cosmological context is based on hydro-dynamical simulations \citep[e.g.][]{Duffy2012,Zavala2016,Dave2013,VN2018}. Physical processes like star formation, feedback from stellar winds, supernovae and Active Galactic Nuclei (AGN), black hole accretion, etc. are included using `sub-grid' models, while the approach allows an explicit and self-consistent treatment of the gas-dynamics. From the computational point of view, large high-resolution simulations require large investment of resources. In addition, a self-consistent modeling of the HI content is not typically accounted for as it would require, in principle, a consistent treatment of the formation of molecular hydrogen on dust grains, and of the transition from molecular to atomic hydrogen including a proper treatment for photo-dissociation and self-shielding. Therefore, HI is usually computed in post-processing \citep[e.g.][]{Duffy2012,Lagos2015}.
An independent and more efficient approach is provided by semi-analytic models of galaxy formation (SAMs), coupled to merger trees extracted from high-resolution N-body simulations. In this case, the advantage is a significantly reduced computational cost and the access to a large dynamic range in mass and spatial resolution. The physical processes driving the evolution of the baryonic components of dark matter haloes are included using simple yet physically and/or observational motivated prescriptions, that are equivalent to the sub-grid modeling used in hydro-dynamical simulations (in fact, often some prescriptions are constructed using controlled numerical experiments). The most recent renditions of several independent SAMs have included prescriptions to partition the cold gas in its atomic and molecular components, based either on empirical relations or on results from sophisticated numerical simulations  \citep[e.g.][]{Fu2010,Lagos2011,Kim2011,Somerville2015,Stevens2017,Xie2017,Zoldan2017,Cora2018}. These models can provide reliable mock $21$~cm maps, thereby helping to understand the relevant processes determining the observed $21$~cm signal.

In this paper, we take advantage of the state-of-the-art semi-analytic model {\sc GAEA}, whose relevant features/details are described in Section~\ref{sec:sim}. In section~\ref{sec:HI_func}, we analyze the HI mass function both as a function of halo mass (section~\ref{sec:HI_cond}), and as function of redshift (section~\ref{sec:HI_z}). In section~\ref{sec:HI_rho} we discuss the HI density, while we investigate the HI content of dark matter haloes in section~\ref{sec:HI_halo} focusing in particular on the halo HI mass function  ($M_{\rm HI}(M_h)$, section~\ref{sec:MHI}) and on the shape of the HI profile (section~\ref{sec:HIprof}). What learned from this analysis helps understanding the clustering signal of neutral hydrogen (section~\ref{sec:clustering}). We discuss both shot noise and the HI bias (in sections~\ref{sec:SN} and ~\ref{sec:bias}, respectively) and how the clustering signal varies as a function of halo mass (section~\ref{sec:Pk_Mh}), galaxy type (section~\ref{sec:Pk_censat}), HI mass (section~\ref{sec:Pk_MHI}), and color (section~\ref{sec:Pk_redblue}). In section~\ref{sec:RSD}, we discuss the effect of the redshift space distortion on the HI power spectrum, and give our predictions for the $21$~cm signal. We draw conclusions in section~\ref{sec:conlusion}.

\section{Simulations}\label{sec:sim}
This work is based on outputs from the GAlaxy Evolution and Assembly ({\sc GAEA}) semi-analytic model. The model and all its upgrades are described in details in several papers \citep[e.g.][]{Delucia2014,Hirschmann2016,Xie2017,Zoldan2017}. We summarize here the main features that are relevant for the analysis presented in this paper. 

{\sc GAEA} has been run on the merger trees of two large scale dark matter cosmological simulations: the Millennium (MI) simulation \citep{Springel2005} and the Millennium II (MII) simulation \citep{Boylan-kolchin2009}. While the latter has a better mass resolution, the former covers a larger volume. The main parameters of the simulations are listed in table~\ref{tab:M}. Both simulations are based on a WMAP1 cosmological model \citep{Spergel2003} with $\Omega_{\rm m}=0.25$, $\Omega_{\rm b}=0.045$, $h=0.73$ and $\sigma_8=0.9$. Although latest Planck measurements \citep{Planck2018,PlanckXIII} point towards a lower value for $\sigma_8$, and a higher value of $\Omega_{\rm m}$, these differences are not expected to have a major impact on the predictions from our semi-analytic models \citep{Wang2008,Guo2013}.

\begin{table}\caption{Main parameters of the simulations used in this study: number of particles $N_p$, box size $\ell_{box}$, and minimum mass of dark matter substructures  min($M_h$). In the last column, we give the minimum stellar mass min($M_{s}$) that we consider in this analysis.}\label{tab:M}\begin{tabular}{c|cccc}
Simulation  & $N_p$ & $\ell_{box}$ & min($M_h$) & min($M_{s}$)  \\
 & & $[h^{-1}\mathrm{Mpc}]$ & $[h^{-1}{\rm M}_\odot]$ & $[h^{-1}{\rm M}_\odot]$ \\\hline
MI & $2160^3$ & $500$ & $1.7\times 10^{10}$  &  $10^8$  \\
MII & $2160^3$ & $100$ & $1.4\times 10^8$ & $10^6$  \\ \hline
\end{tabular}\end{table}

The SUBFIND algorithm \citep{Springel2001} has been used to identify bound substructures (sub-haloes) within standard Friend-of-Friend (FoF) dark matter haloes. The bound part of the FOF group represents what is referred to as the `main subhalo', and hosts the central galaxy, while satellite galaxies are associated to all other bound subhaloes. Merger trees are constructed identifying a unique descendant for each subhalo, by tracing the majority of the most bound particles. When a halo is accreted onto a more massive system (i.e. it becomes a substructure), it suffers significant stripping by tidal interaction. In particular, it can be stripped below the resolution of the dark matter simulation, when still at a significant distance from the parent halo center, i.e. when the merger between the galaxy it hosts and the central galaxy is incomplete. The semi-analytic model accounts for this by introducing `orphan' galaxies (Type II in our jargon, versus Type I that are associated to distinct dark matter substructures). The evolution of these galaxies is traced by following the most bound particle of the parent substructure before it disappeared, and the galaxy is assigned a residual merger time that is given by a variation of the dynamical friction formula \citep{Delucia2007,Delucia2010}.

The GAEA semi-analytic model follows the evolution of the different baryonic components of model galaxies. Different reservoirs are considered: the stars in the bulge and in the disk, the cold gas in the galactic disk, a diffuse hot gas component and an ejected gas component both associated to the parent dark matter haloes. The model has prescriptions for the transfer of mass, metals and energy between these different components, including gas cooling, star formation and stellar feedback. When a halo is identified for the first time, it is assigned a hot gas reservoir that is proportional to its dark matter mass (via the universal baryon fraction). This gas can then cool onto the galactic disk of the central galaxy either via a "rapid" or a "slow" accretion mode. The cooling is "rapid" if the characteristic cooling radius, that depends mainly on the temperature and the metallicity of the gas, is larger than the halo virial radius, and "slow" otherwise. Generally, gas cooling is rapid for small haloes at early epochs, and becomes slower for more massive haloes and at lower redshift. 

The collapse of the cold gas leads to the formation of stars. Observations have shown that there is a strong correlation between the surface density of the molecular gas and that of star formation, against no significant correlation between HI content and star formation \citep[e.g][]{Wong2002,Kennicutt2007,Leroy2008}. This is accounted for in the GAEA model through an explicit dependence of star formation on the molecular gas content \citep{Xie2017}. At each time-step, the galactic disk is divided into concentric rings and, in the model version we are using, the cold gas is partitioned into atomic (HI) and molecular (H$_2$) gas using an empirical power law relation: ${\rm H}_2/\mathrm{HI}=(P_{ext}/P_0)^{\alpha}$ between the molecular to atomic ratio (H$_2$/HI) and the hydrostatic mid-plane pressure of the disk \citep[$P_{ext}$,][]{Blitz2006}. The index $\alpha$ and the pivot scale $P_0$ are tuned to observational data in the local Universe. In each ring, the star formation rate density is assumed to be proportional to the molecular hydrogen surface density, and the total star formation rate is obtained summing the contributions from all rings. A star formation efficiency parameter regulates how much H$_2$ is converted into stars, and its value is tuned to reproduce the observed HI mass function in the local Universe (see section~\ref{sec:HI_func}). Our H$_2$ based star formation law predicts an increasing molecular fraction with increasing redshift, in qualitative agreement with what inferred from observational data \citep[see also discussion in][]{Popping2015}. With respect to the case of fixed molecular fraction, this lowers the number density for HI-rich galaxies at high redshift. Moreover, the molecular fraction depends on stellar mass, which also disproves the idea of a fixed molecular fraction.

Our model includes a treatment for specific angular momentum exchanges between different components \citep{Xie2017}. In particular, the hot gas halo is assumed to have the same specific angular momentum of the parent halo. This specific angular momentum is transferred to the cold gas disk proportionally to the cooled mass, and then to the stellar disk proportionally to the mass of formed stars. The specific angular momentum of a disk is used to consistently estimate its scale radius, that is proportional to the angular momentum divided by the circular velocity of the halo (at the time of accretion for satellite galaxies). 

At the end of their life-cycle, stars eject mass, energy, and metals through winds or supernovae explosions. These stellar feedback events are assumed to reheat the cold gas of the disk and to eject part of the gas outside the galaxy parent halo. Reheated and ejected gas are both assigned to the reservoir of the central galaxy of each halo, even if coming from satellites. The gas can be re-accreted onto the central galaxy on time scales inversely proportional to the virial mass of the halo \citep{Henriques2013,Hirschmann2016}. The GAEA model also includes a complex treatment of chemical enrichment, accounting for the non-instantaneous recycling of gas, metals and energy from Asymptotic Giant Branch stars (AGBs), supernovae type Ia (SnIa) and supernovae type II (SNII). The model also includes a treatment for black holes and AGN feedback. At high redshift, black hole mass seeds are placed at the center of haloes with virial temperature above $10^4$~K. The seed mass scales with that of the parent haloes and ranges from $10^3$ to $10^5$ ${\rm M}_\odot$ in the MI, and from $10$ to $10^4$ ${\rm M}_\odot$ in the MII. Note that the different values for the seeds in the MI and MII are due to their different resolution.
Black holes then grow through mergers (quasar mode) and hot gas accretion (radio mode). The latter is implemented in GAEA following \citet{Croton2006}, thus relating the growth of a back hole to its mass, the virial velocity of the parent halo, and hot gas fraction. Accretion of gas from the hot gas atmosphere associated with dark matter haloes generates a mechanical energy, that reduces the cooling rate. This effect is important to suppress the HI content of massive haloes. Finally, in case of a merger event, the cold gas of the secondary galaxy (that is always a Type II) is added to the total cold gas disk of the primary (that can be a Type I or a central) and a star-burst is triggered, with the new stars remaining in the disk of the remnant \citep{Somerville2001,Cox2008}. Depending on the masses of the galaxies involved, the merging event can be considered a major or a minor event. In case of a major event, all stars (including the new ones) are moved to a spheroidal bulge. 

\section{The HI mass function}\label{sec:HI_func}

In this section, we discuss the HI mass function (HIMF), i.e. the number density of galaxies with different HI mass, in the MI and MII simulations, focusing in particular on the role of central and satellite galaxies. In section~\ref{sec:HI_cond}, we analyze the HI conditional mass function and its dependence on hosting halo mass, while in section~\ref{sec:HI_z} we study the HI mass evolution with redshift. As discussed in section~\ref{sec:sim}, the star formation efficiency is tuned to reproduce the local Universe HI mass function measured by \citet{Zwaan2005} and \citet{Martin2010}, using the blind HI surveys HIPASS \citep{Meyer2004} and ALFALFA \citep{Giovannelli2005}, These are limited to redshifts $z < 0.04$ and $z < 0.06$, respectively. We do not retain all model galaxies, but apply a cut on their stellar mass. This cut is evaluated, for the MII, on the basis of the good agreement between the predicted and observed stellar mass function \citep{Baldry2012, Moustakas2013}, down to a certain stellar mass threshold. In this work we adopt a mass threshold of $M_s > 10^6 {\rm M}_\odot$, although the stellar mass function of the model overpredicts the number of galaxies with $M_s \lesssim 10^{7.5} {\rm M}_\odot$ of $\sim 0.5$ dex.
For the MI, the stellar mass cut is based on the convergence with the higher resolution simulation: we find good agreement between MI and MII down to $M_s \sim 10^8 {\rm M}_\odot$.
Previous work based on the GAEA model \citep[e.g.][]{Xie2017,Zoldan2017} adopted more conservative stellar mass limits ($\sim 10^9 {\rm M }_\odot$ for the MI, and $\sim 10^8 {\rm M}_\odot$ for the MII).
In this work, we lower the stellar mass thresholds (see table~\ref{tab:M}) in order to access lower HI masses, as the minimum HI available is connected to the stellar mass selection.  
\citet{Kim2017} show that low HI mass galaxies play a key role for the clustering signal, and therefore for the intensity mapping measurements we are interested in. 
As the measured HI mass function is well reproduced by our model, we do not expect the overestimation of the stellar mass function in the low mass end to influence our results. 

\begin{figure}
\includegraphics[width=\columnwidth]{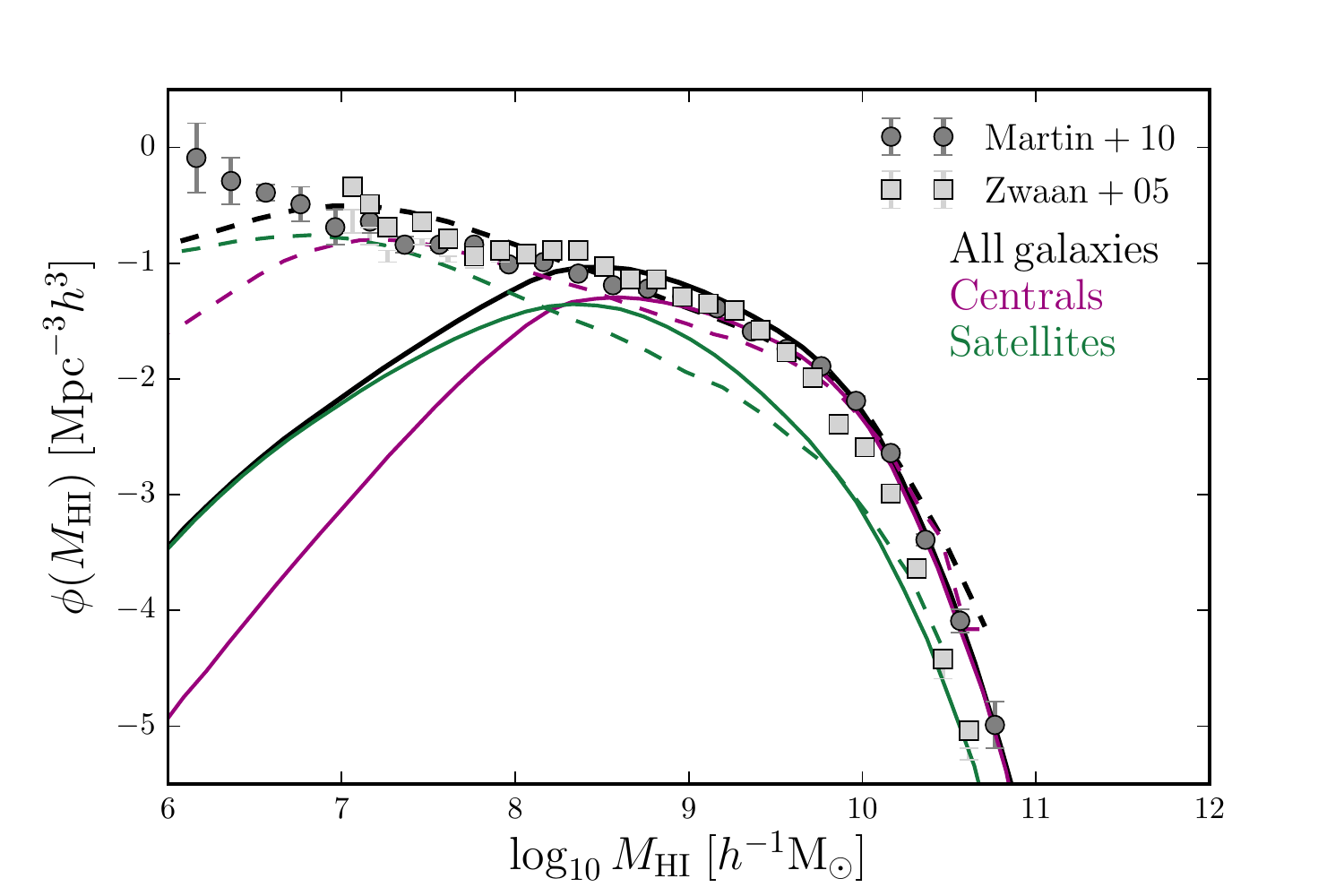}
\caption{The HI mass function, i.e. the number density of galaxies with different HI
mass, in the Millennium I (solid lines) and Millennium II (dashed lines) simulation at redshift zero. Magenta and green lines are used for centrals and satellites respectively, while black lines are for all galaxies. Squares and circles with error bars show the data measured by \citet{Zwaan2005} and \citet{Martin2010} using the blind HI surveys HIPASS \citep[][limited to $z < 0.04$]{Meyer2004} and ALFALFA \citep[][limited to $z < 0.06$]{Giovannelli2005}. In this work, we use $M_s>10^8 {\rm M}_\odot$ for the MI and $M_s>10^6 {\rm M}_\odot$ for the MII, as a compromise between resolution requirements and the inclusion of low HI mass galaxies in our sample.} \label{fig:HI_mass_func_cen_sat_z0}
\end{figure}

In figure~\ref{fig:HI_mass_func_cen_sat_z0}, we show the HI mass function for the MI and MII simulations at redshift zero. As discussed above, the model has been tuned to reproduce this quantity. Below $\sim 10^{8} h^{-1}{\rm M}_\odot$ the MI starts deviating from the MII, underestimating the number density with respect to the data. 
Predictions from the MII follow rather well the observational measurements down to HI masses $\sim 10^{7} h^{-1}{\rm M}_\odot$. While the agreement between our model predictions and observational data is not entirely surprising, we note that other models are characterized by an excess of galaxies at low HI masses. For example, the GALFORM model over-predicts by a factor 3 the HI number density around $\sim 10^{8} h^{-1}{\rm M}_\odot$ \citep[e.g][]{Baugh2019}. The excess is ascribed to halo mass resolution and to the implementation adopted for photo-ionization feedback \citep{Kim2015}. In the model presented by \citet{Popping2015}, there is a similar excess due to low mass galaxies ($M_s < 10^7 {\rm M}_\odot$) residing in low mass haloes ($M_h < 10^{10} {\rm M}_\odot$). 
This excess (or the lack of the excess in our model) has important consequences on the model predictions that we will discuss in the following sections.

In figure~\ref{fig:HI_mass_func_cen_sat_z0}, we further separate the contribution from central and satellite galaxies. The low mass end of the HIMF is dominated by satellite galaxies, starting from $M_{\rm HI} \sim 10^{8} h^{-1}{\rm M}_\odot$ for the MI,  and from  $M_{\rm HI} \sim 10^{7} h^{-1}{\rm M}_\odot$ for the MII. \citet{Lagos2011} argue that, in their model, central galaxies do not contribute much in this mass range because of the reionization scheme adopted: hot gas in small haloes is not allowed to cool, thus there is less HI in centrals. In contrast, satellites have formed before reionization, and therefore before this effect becomes efficient. 
For intermediate and high HI masses, the HIMF is  dominated by central galaxies. This behavior (i.e central galaxies being more important at medium at high masses and satellites at small masses) is quite general and found also in other models \citep[e.g. GALFORM,][]{Kim2017}.

\subsection{The HI conditional mass function}\label{sec:HI_cond}
\begin{figure*}
\includegraphics[width=\columnwidth]{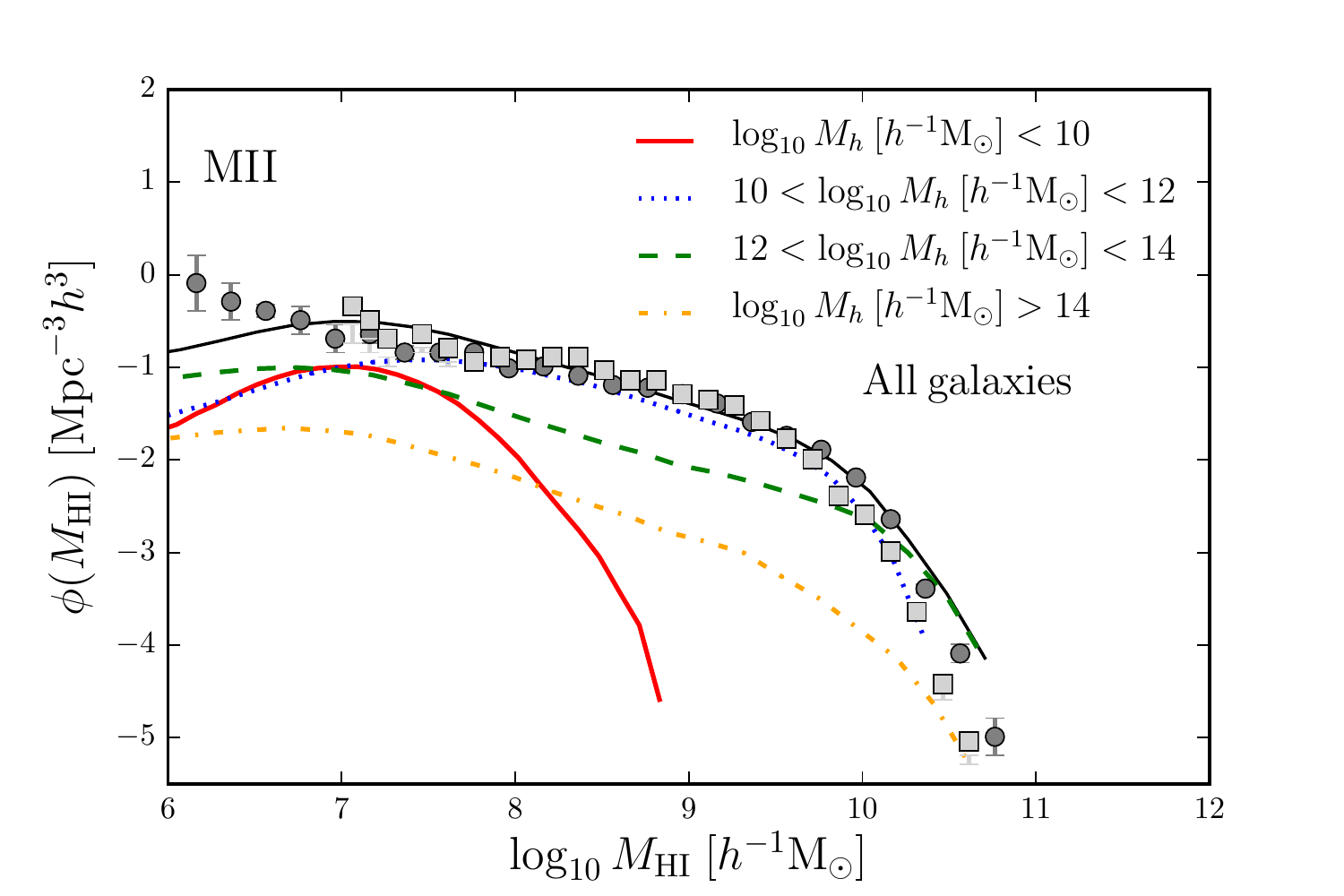}
\includegraphics[width=\columnwidth]{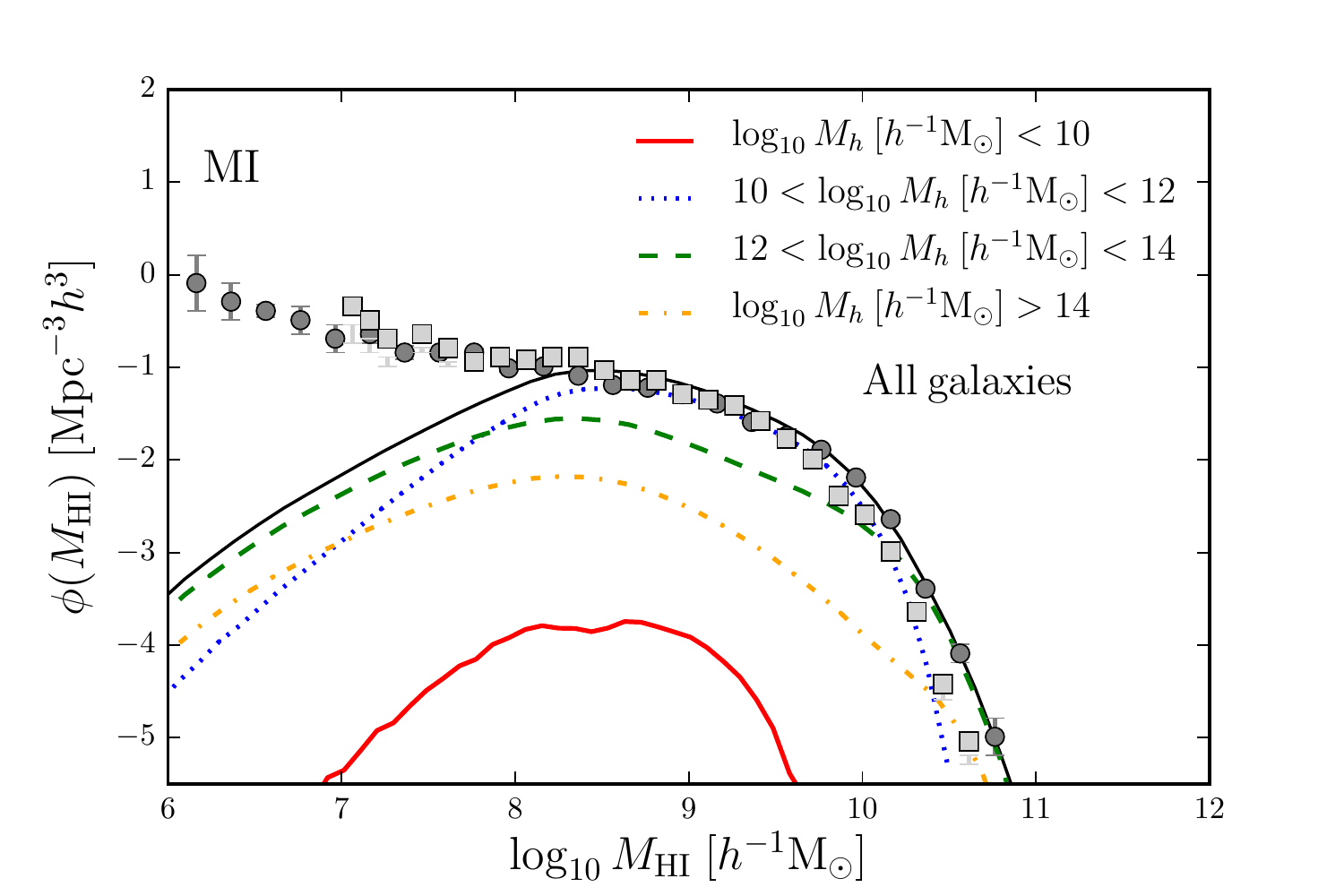}
\includegraphics[width=\columnwidth]{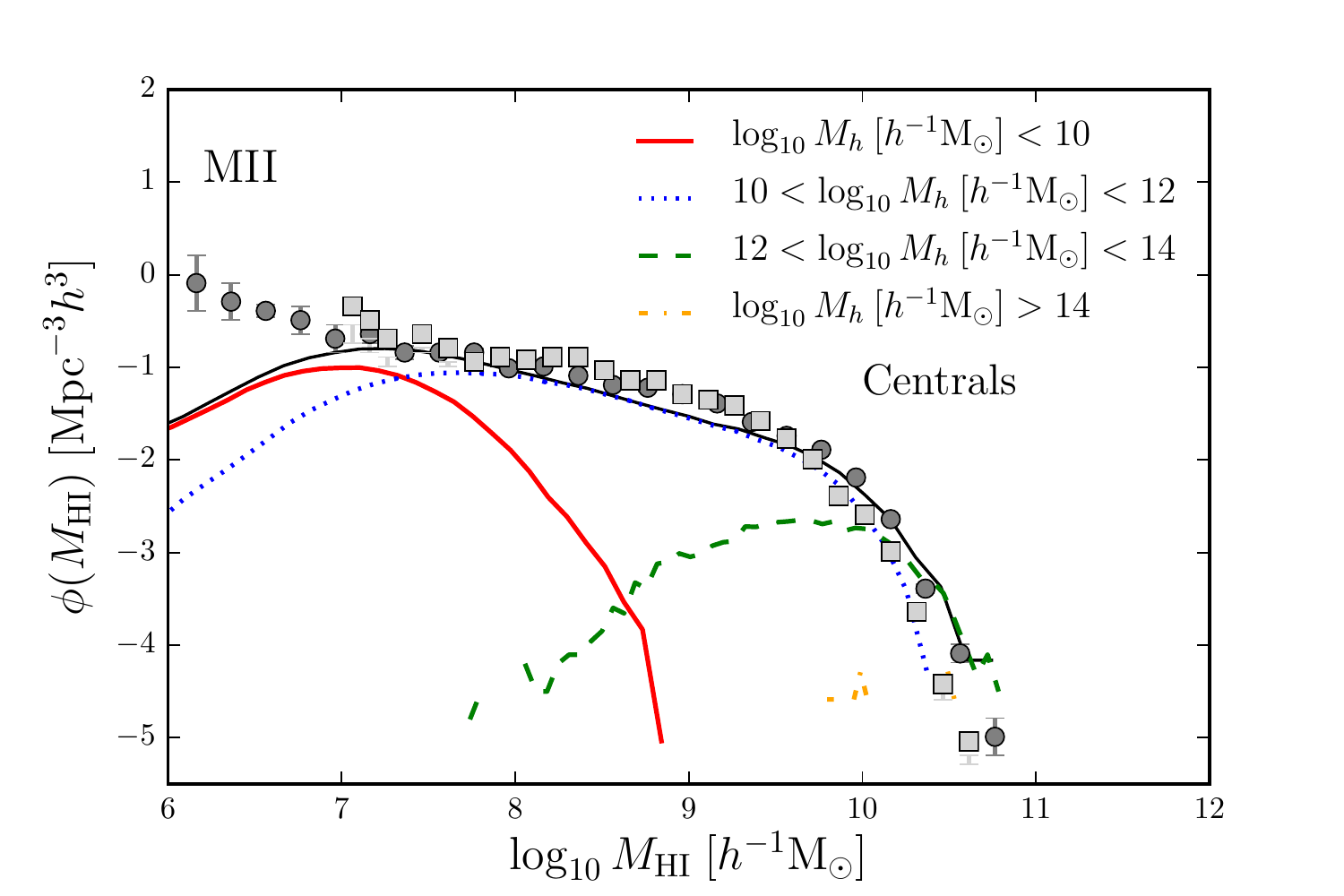}
\includegraphics[width=\columnwidth]{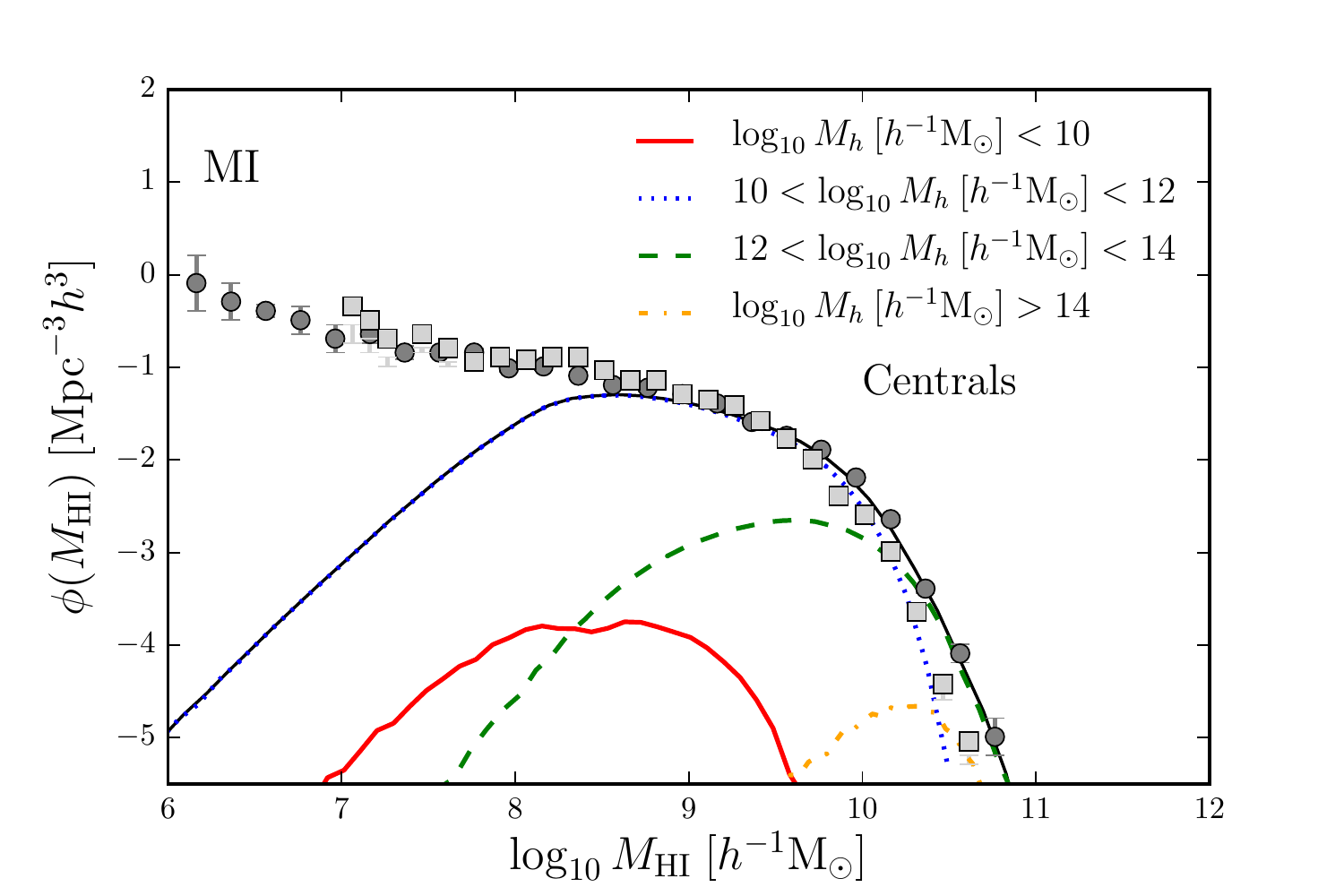}
\includegraphics[width=\columnwidth]{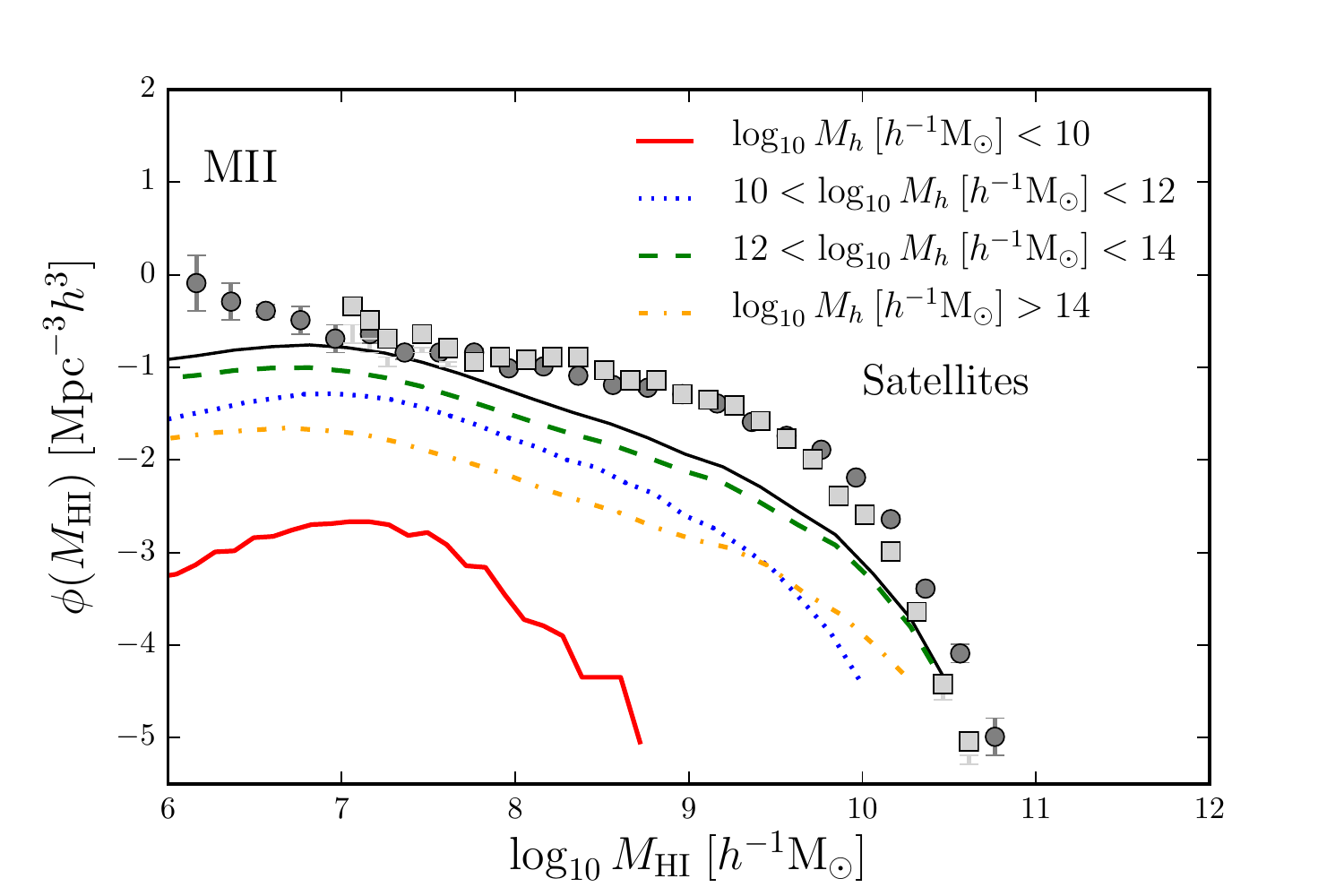}
\includegraphics[width=\columnwidth]{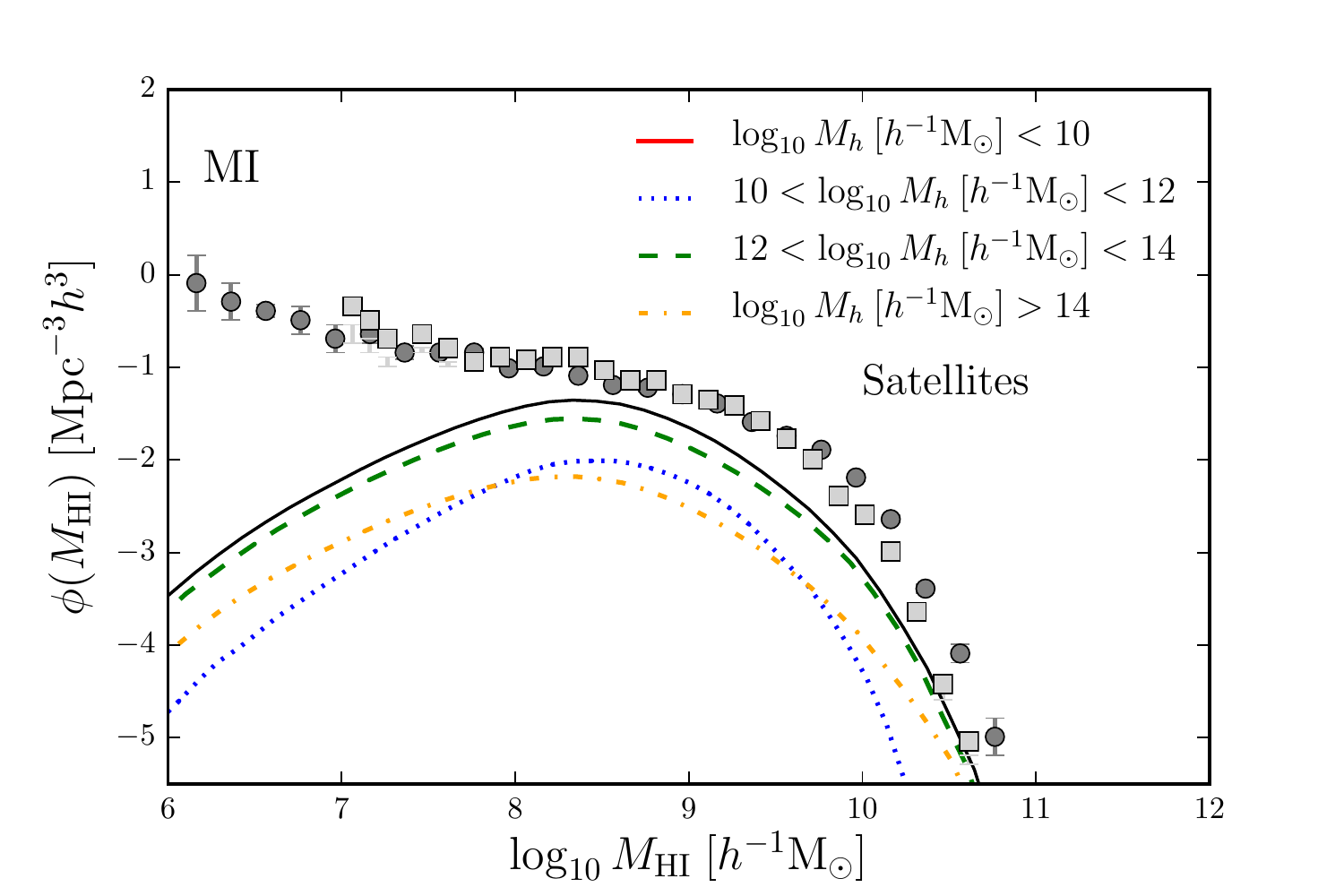}
\caption{The predicted HI conditional mass function for the MII (left column) and the MI (right column). In the middle and bottom panels we separate the contribution from central and satellite galaxies, respectively. The solid black line in each panel is the sum of the contributions from different host dark matter haloes. Squares and circles with error bars show observational measurements by \citet{Zwaan2005} and \citet{Martin2010} in the local Universe.}\label{fig:HI_mass_cond}
\end{figure*}

Figure~\ref{fig:HI_mass_cond} shows the contribution to the HI mass function of galaxies hosted by haloes of different mass at $z=0$. We show the same figure at $z=4$ in appendix~\ref{app:z4}. As above, we further separate the contribution from central and satellite galaxies. 

The conditional HI mass function emerges from a non trivial convolution of the halo number density, baryon fraction, and gas cooling. The high mass end of the HI mass function is dominated by HI in central galaxies hosted by massive haloes ($10^{12} < M_{h} [h^{-1}{\rm M}_\odot]< 10^{14}$), in both the MI and MII. Just below the knee, in the mass range $10^{8} < M_{\rm HI}(h^{-1}{\rm M}_\odot) < 10^{10}$, the main contributors are central galaxies in lower mass haloes ($10^{10} < M_{h} [h^{-1}{\rm M}_\odot]< 10^{12})$, with satellite galaxies in haloes with mass $10^{12} < M_{h} [h^{-1}{\rm M}_\odot]< 10^{14}$  also giving a non negligible contribution. The lower mass end of the HI mass function is dominated by these satellite galaxies, with contributions from satellites in less (more) massive haloes for the MI (MII). Centrals from the lowest mass haloes are also important at these low masses, as also found by \citet{Popping2015}. In the MII, the dominant contribution comes from the smallest haloes  ($M_{h} < 10^{10} h^{-1}{\rm M}_\odot$); in the MI case, these haloes are not resolved and the main contribution comes from the haloes with larger mass ( $10^{10} < M_{h} [h^{-1}{\rm M}_\odot] < 10^{12} $).

\subsection{Redshift evolution of the HI mass function}
\label{sec:HI_z}

\begin{figure}
\includegraphics[width=\columnwidth]{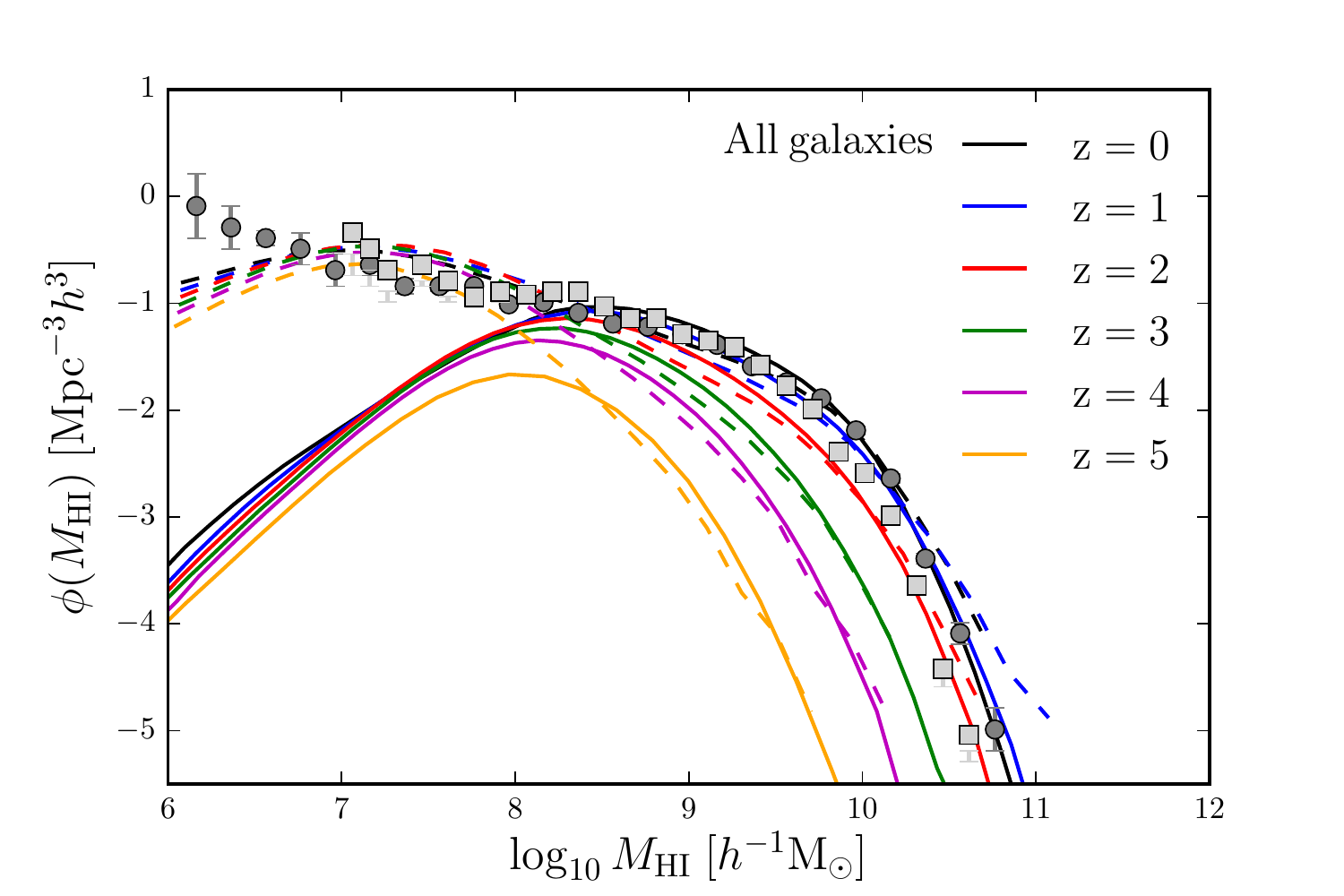}
\includegraphics[width=\columnwidth]{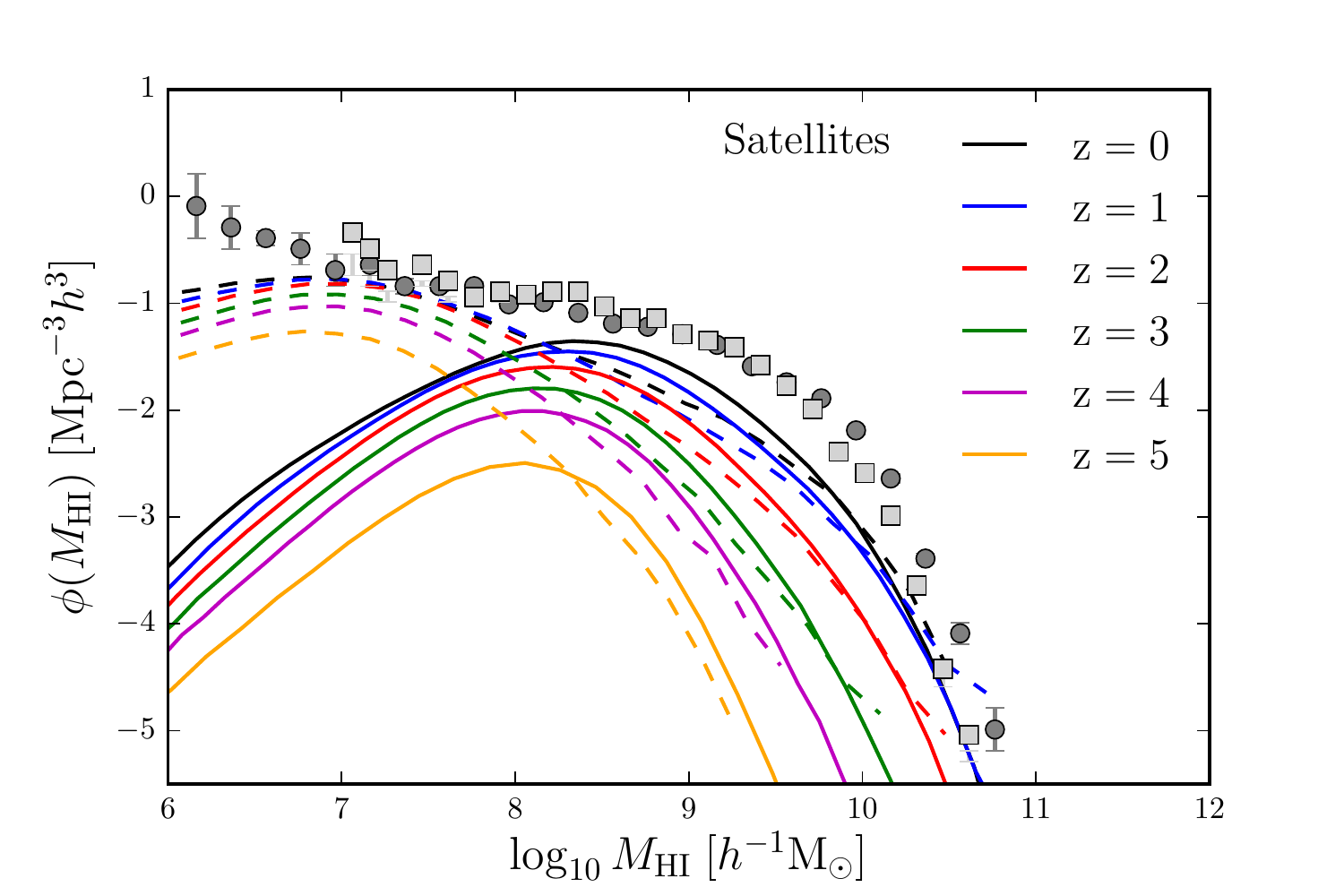}
\caption{{\it Top panel}: The predicted HI mass function as a function of redshift for the MI (solid lines) and the MII simulation (dashed lines). {\it Bottom panel}: Contribution from satellite galaxies. For reference, we show with squares and circles the observational measurements of the HI mass function in the local Universe by \citet{Zwaan2005} and \citet{Martin2010}, as in figure~\ref{fig:HI_mass_func_cen_sat_z0}.}\label{fig:HI_mass_z}
\end{figure}

In figure~\ref{fig:HI_mass_z}, we show the evolution of the HI mass function in the redshift range $0< z< 5$. The high mass end grows with decreasing redshift until $z\sim 1$, tracing the formation of progressively more massive haloes. At $z=0$, the suppression of gas cooling due to AGN feedback (see section~\ref{sec:sim}) reverses slightly this trend. In the HI mass range $10^{8} < M_{\rm HI} [h^{-1}{\rm M}_\odot] < 10^{10} $, where the HI mass function is dominated by central galaxies in much lower mass haloes (where AGN feedback is not efficient), the number density of galaxies is largest at $z=0$.

As for the $z=0$ case discussed earlier, the resolution does not affect significantly the number density of galaxies with HI mass above $\sim 10^{8} h^{-1}{\rm M}_\odot$ in the MI, although this limit deteriorates slightly with increasing redshift, reaching $\sim 10^{8.5} h^{-1}{\rm M}_\odot$ at $z=5$. Below  $\sim 10^{8} h^{-1}{\rm M}_\odot$, our model predicts a mild dependence on redshift for the global population and a stronger evolution for satellite galaxies, whose number is expected to increase with decreasing redshift. As discussed above, other models tend to over-predict the number density of galaxies with HI masses in the range $10^{7} \lesssim M_{\rm HI} [h^{-1}{\rm M}_\odot] \lesssim 10^{9} $ at $z=0$. This excess becomes more prominent and shifts towards lower HI mass values with increasing redshift \citep[e.g.][]{Lagos2011,Dave2017,Baugh2019}. It is difficult to understand the origin of the different behavior between GAEA and independent models, as it likely originates from the complex and non-linear interaction between different physical processes implemented (in particular, photo-ionization feedback, star formation, and stellar feedback). Future observational programs that will allow the HI mass function to be measured beyond the local Universe, will provide important constraints on the model predictions discussed here. 


\section{HI density}
\label{sec:HI_rho}

\begin{figure}
\includegraphics[width=\columnwidth]{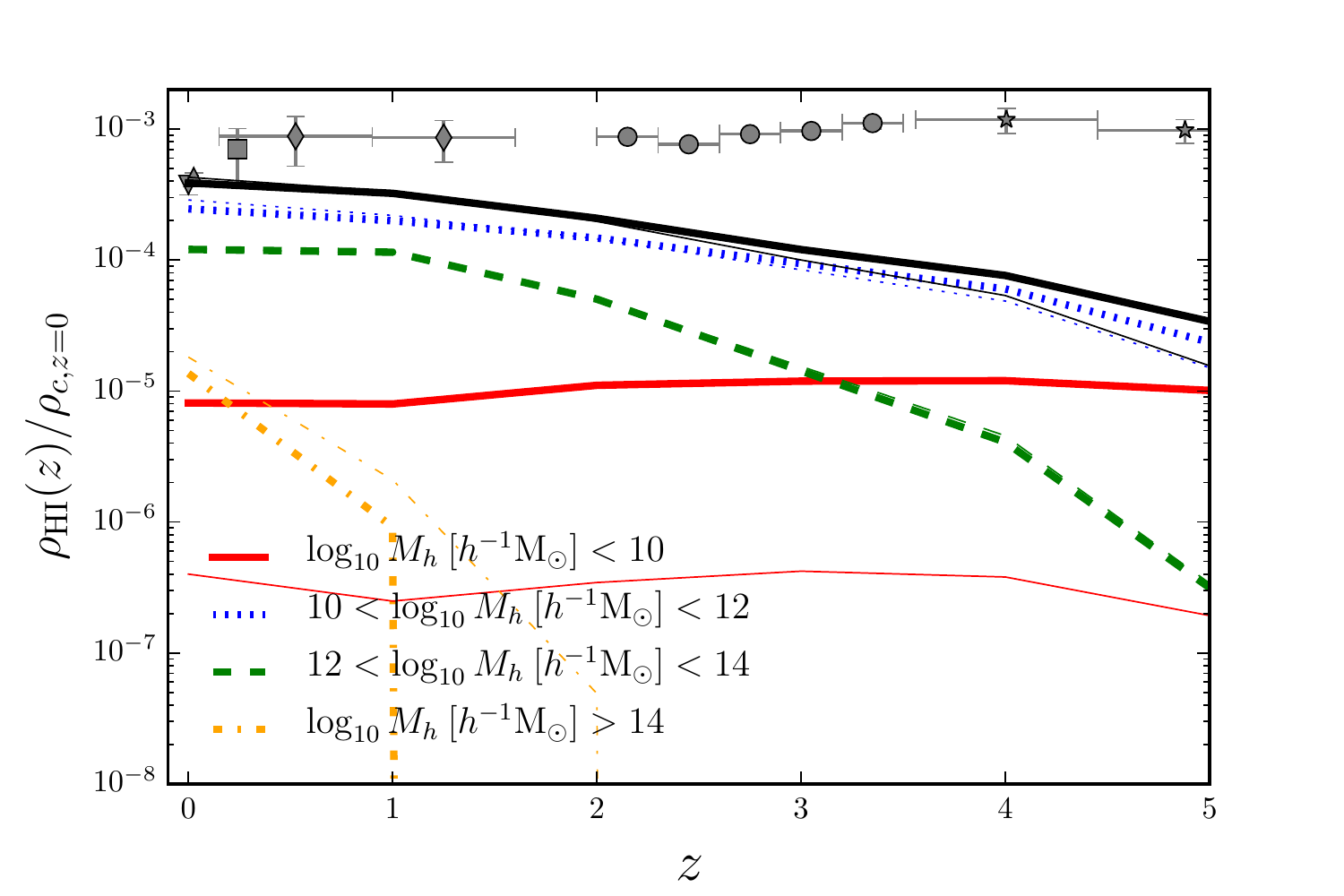}
\caption{Evolution with redshift of the HI density $\rho_{\mathrm{HI}}/\rho_{c,z=0}$ for the MI (thin lines) and MII (thick lines), divided into the contributions for different host dark matter halo masses. Differences between the two simulations can be appreciated only for very high and very low mass haloes.  The GAEA model prediction are compared with observational data (\citealt{Crighton2015} and references therein).}\label{fig:rho_HI_z}
\end{figure}

In figure~\ref{fig:rho_HI_z}, we show the abundance of neutral hydrogen $\rho_{\rm HI}(z)/\rho_{c,z=0}$ for the MI and MII simulations, as a function of redshift. Our results are tuned to agree with ALFALFA and HIPASS measurements at $z\sim 0$ (see also figure~\ref{fig:HI_mass_func_cen_sat_z0}). At higher redshift, our model predicts a clear decrease of $\rho_{\rm HI}(z)$ with respect to the observational data. This behavior is shared  by independent semi-analytic models \citep[e.g.][]{Lagos2014} that, however, typically start from a larger value of the HI density at $z=0$. In contrast, hydro-dynamical simulations generally predict  an opposite trend or significantly weaker evolution \citep[e.g.][]{Dave2017,VN2018}. At high redshift, the observational measurements are based on Damped Lyman-$\alpha$ systems (DLAs) identified in the spectra of bright quasars. Although the physical origin of these systems is still debated \citep[e.g.][]{Pontzen2008,Tescari2009,Rahmati2013,Berry2014}, the measurements based on DLAs are robust as they are relatively easy to identify in quasar spectra thanks to their prominent damping wings.

As discussed in section~\ref{sec:HI_z}, the HI mass function in our model does not evolve significantly with redshift for $M_{\rm HI} < 10^{8} h^{-1}{\rm M}_\odot$, while independent models tend to predict larger number densities for low HI masses at higher redshift (see for example figure 7 and 12 of \citealt{Popping2014}, figure 4 in \citealt{Baugh2019}, or figure 7 and 9 in \citealt{Dave2017}). These enhanced number densities clearly play a key role in the final HI content as a function of redshift, explaining in part the origin of the disagreement between our model predictions and observational data. In addition, the cosmic HI density depends significantly on the resolution of the simulations used. To emphasize this, in figure~\ref{fig:rho_HI_z}, we show the contribution to the total HI density from haloes of different mass. As expected from hierarchical structure formation, larger haloes collapse later so that above $z\sim 5$ haloes smaller than $\sim 10^{10} h^{-1}{\rm M}_\odot$ already represent an important contribution to the cosmic number density of HI. These are, however, below the resolution of the MI simulation. 

We could partially fix the decreasing trend of HI with redshift, by assuming that the missing HI is contained in small, unresolved haloes.
However, this solution would not be physically motivated and rather unrealistic, as these haloes should contain excessively large quantities of HI. 
In summary, although resolution plays a role in the HI density determination, it seem unlikely that it is the main driver of the discrepancy with observational data. 
Note that, however, a decreasing HI content is partially expected in our model since, closer to the epoch of reionization, the neutral hydrogen should be progressively found outside haloes as diffuse gas and filaments in the IGM. At $z=5$, for example, the total HI inside galaxies should account for only the $80\% $ of the total HI \citep{VN2018}.
Further uncertainties can be associated to the specific post-processing framework adopted for the HI evaluation in simulations \citep{Diemer2018}.
On the other end, one should bare in mind that the measurements of HI density are based on two different observational strategies: at redshift $z \gtrsim 2$ they are based on DLAs (in absorption) while at lower redshift are based on the local HI mass function (in emission).  
In the next future, thanks to next generation radio telescopes like SKA, we will be able to measure the HI quantity using both emission/absorption measurements on the same region. 
At the moment, the two estimates are independent, and we cannot definitely asses that they are measuring exactly the same quantity, both in terms of sources and intensity. 

\section{HI content in dark matter haloes}\label{sec:HI_halo}
A detailed characterization of the HI content of dark matter haloes is a fundamental ingredient to make predictions for the $21$~cm signal. This is particularly relevant for intensity mapping, especially in the framework of the halo model \citep{Castorina2017,VN2018}. Hydro-dynamical simulations and SAMs are not only a valuable tool to model the relation between HI and halo mass, but also offer the possibility of investigating the physical origin of this relation. In this section, we analyze the HI halo mass function $M_{HI}(M_h)$ (section~\ref{sec:MHI}) and its redshift evolution in the post reionization Universe. We provide a  fitting formula for it, discuss the different contributions from centrals and satellites, and the dependence on the assembly history of dark matter haloes. In section~\ref{sec:HIprof}, we focus on the HI spatial distribution within dark matter haloes, analyzing the contribution from different satellite types. 

\subsection{The HI halo mass function}\label{sec:MHI} 
\begin{figure}
\includegraphics[width=\columnwidth]{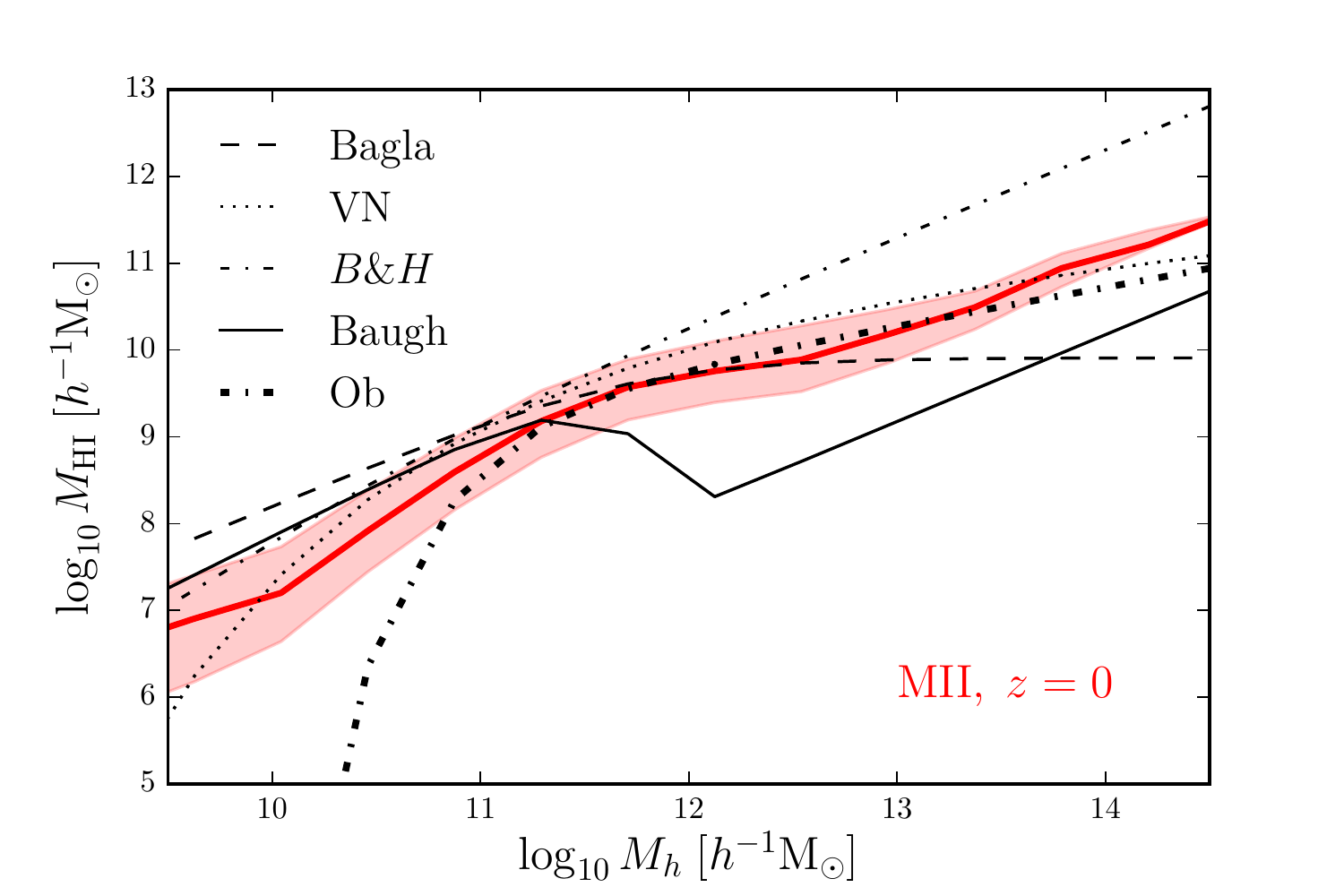}
\caption{The median HI content of dark matter haloes in the MII simulation as a function of halo mass (red), at $z=0$. The relation is compared to different parametrizations that have been introduced in the recent literature: \citet[][dashed]{Bagla2010}; \citet[][dotted]{VN2018}; \citet[][thin dashed dotted]{Barnes2010}; \citet[][solid]{Baugh2019}; and \citet[][thick dashed dot]{Obuljen2019}. The red shaded area shows the 16th and 84th percentiles of the model distribution.}\label{fig:MHI_models}
\end{figure}

\begin{figure}
\includegraphics[width=\columnwidth]{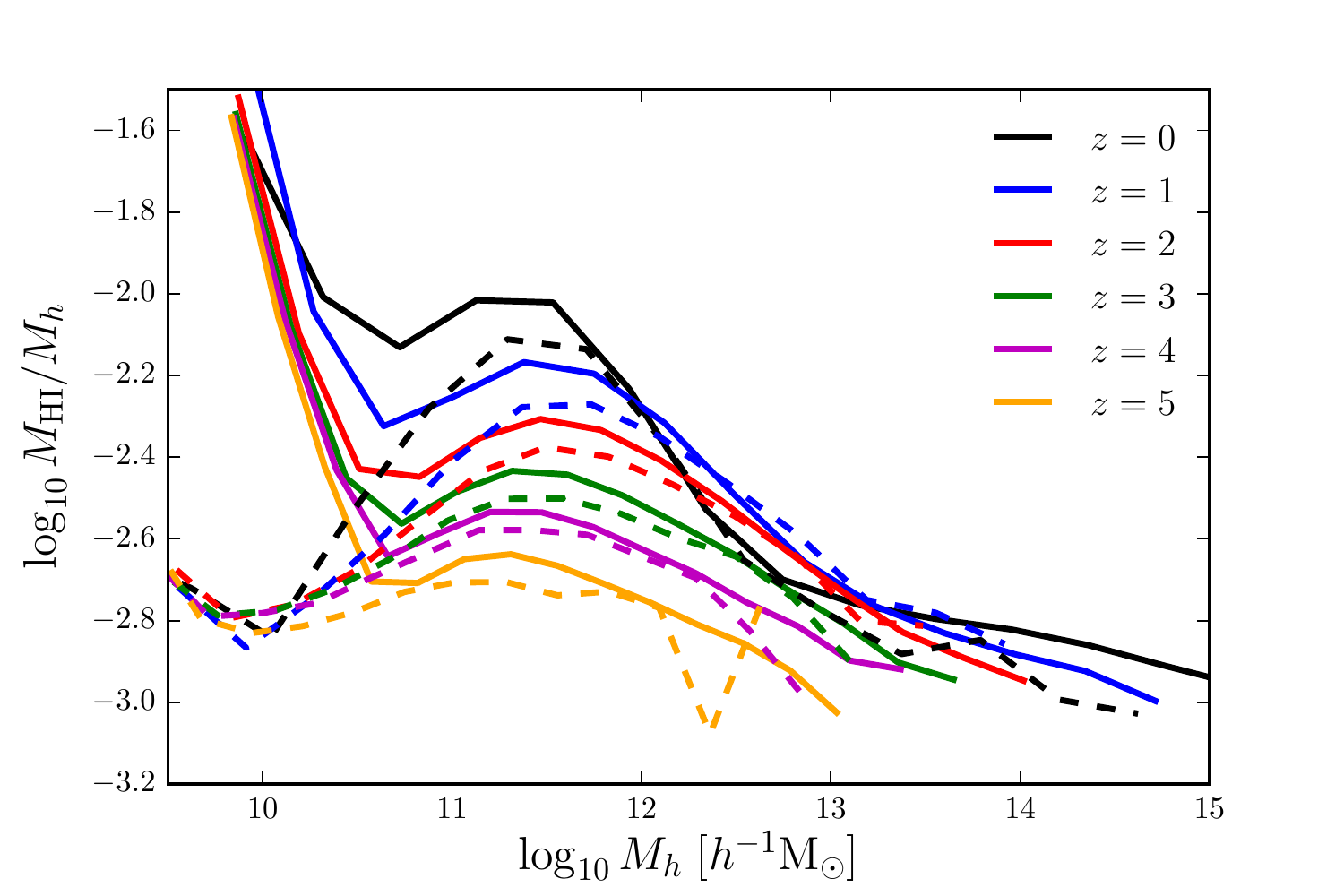}
\caption{The fraction $M_{\rm HI}/M_h$ as a function of $M_h$, measured from the MI (solid lines) and the MII simulations (dashed lines), as a function of redshift (different colors).}\label{fig:frac_HI_Mhalo_z}
\end{figure}

In this section, we analyze the average HI mass $M_{\rm HI}$ hosted by a halo of mass $M_h$ at redshift $z$; this represents the so called HI halo mass function, $M_{\rm HI}(M_h, z)$. Empirical models have been proposed for this relation by many authors \citep[e.g.][]{Bagla2010,Santos2015, Barnes2010, Padmanabhan2017,VN2018,Obuljen2019,Baugh2019}. The expectation is a linear relation between halo mass and HI mass, at least above a certain mass threshold. Due to different astrophysical processes like tidal stripping and photo-ionization, low mass haloes are not expected to host large amounts of HI. Therefore, most parametrizations account for a cutoff at low masses. In figure~\ref{fig:MHI_models}, we show the $M_{\rm HI}(M_h)$ at $z=0$, estimated from our higher resolution simulation (i.e. MII), compared to other models that have been proposed and used in the recent literature. 
For halo masses $M_h > 10^{12.5} h^{-1}{\rm M}_\odot$, the slope of the predicted relation becomes somewhat flatter than for lower mass haloes. This coincides with the significant `dip' that has been reported by \citet{Baugh2019}, and corresponds to the halo mass where AGN feedback becomes efficient.
This behavior is not described by a simple power law model at high halo masses, that is instead typical of most parametrizations (see figure~\ref{fig:MHI_models}).  
For halo masses in the range $10^{10} < M_h < 10^{11.5} h^{-1}{\rm M}_\odot$, the HI content drops. 
This physically motivated trend is shared by most parametrizations although the rate is different: some models predict almost no HI in small haloes \citep[see][in the figure]{Obuljen2019}, while other models have a smoother decline  \citep[see][in the figure]{Barnes2010}. 
In this low mass range, \citet{Bagla2010} have a cutoff at an even smaller mass (note however that the value in the figure is that at $z\sim 3$, but the authors speculate a shift to higher values at lower redshift), while \citet{Baugh2019} do not parametrize any cutoff. 
To fit our model predictions, we extend the parametrization by \citet{Baugh2019} to include a cutoff at low masses: 
\begin{equation}
    M_{\mathrm{HI}}(M_h)=M_h \left[a_1 \left(\frac{M_h}{10^{10}}\right)^{\beta} e^{-\left(\frac{M_h}{M_{\mathrm{break}}}\right)^\alpha}+a_2\right] e^{-\left(\frac{M_{\mathrm{min}}}{M_h}\right)^{\gamma}},\label{eq:M_HI}
\end{equation}
where $a_1$, $\beta$, $\alpha$, $M_{\rm break}$, $a_2$, and $M_{\rm min}$ are free parameters. The value of $\gamma$ is kept fixed  to $0.5$, which we find provides the best description of our model predictions, and is half way between $1$ (the most common choice in literature) and the $0.35$ adopted recently by \citet{VN2018}. 
The parametrization proposed assumes that the HI mass is proportional to the halo mass $M_h$ at the high mass end, and proportional to $M_h^{1+\beta}$ for intermediate masses. We list the best fit values of the free parameters that we find for our model predictions in table~\ref{tab:M_HI}. At $z=0$, the value of $M_{\mathrm{break}}$ is $\sim 10^{12} h^{-1}{\rm M}_\odot$, slightly larger than that reported in \citet{Baugh2019}, in agreement with what we see in figure~\ref{fig:MHI_models}. The value of $\beta$ is negative, which reflects the fact that the HI mass increases more slowly for intermediate mass haloes than for the most massive ones. $M_{\mathrm{min}}$ is $10^{11.4} h^{-1}{\rm M}_\odot$, compatible with the values reported in \citet{VN2018}. In figure~\ref{fig:frac_HI_Mhalo_z}, we show $M_{\mathrm{HI}}(M_h)$ for both the MI and MII simulations, as a function of redshift. We plot the fraction $M_{\mathrm{HI}}/M_h$, to better differentiate results at different redshifts. As discussed in section~\ref{sec:HI_func}, we can clearly see the hierarchical growth of structures at the high mass end, and the AGN feedback inverting the trend at $z=0$. This is in agreement with the results listed in table~\ref{tab:M_HI}: the value of $\beta$ increases from $z=5$ to $z=1$ before dropping at $z=0$, and the value of $M_{\mathrm{break}}$ decreases slowly from $\sim 10^{12} h^{-1}{\rm M}_\odot$ and rises again at $z=0$. The presence of the cutoff is more evident at $z=0$, while a simpler parametrization (without a cutoff) could be adopted at higher redshift. 
Note that low halo mass end rise of the $M_{\mathrm{HI}}/M_h$ fraction for the MI, reflects its poorer resolution that causes a flattening in the $M_{\mathrm{HI}}-M_h$ relation near the resolution limit.

To quantify the scatter of the $M_{\mathrm{HI}}(M_h)$, we show in figure~\ref{fig:M_HI_scatter} in the appendix~\ref{app:M_HI_halo} the density distribution of the HI mass hosted by haloes of different mass, at different redshifts. 


\begin{table}
\caption{Best fit values for the parameters of the HI halo mass function $M_{\mathrm{HI}}(M_h)$ parametrized in  equation~\ref{eq:M_HI}, at different redshifts.}\label{tab:M_HI}
\begin{tabular}{c|cccccc}
z & $a_1$ & $a_2$ & $\alpha$ & $\beta$ & $\log_{10}(M_{\mathrm{break}})$  & $\log_{10}( M_{\mathrm{min}})$  \\
& & & & & $(h^{-1}{\rm M}_\odot)$  &  $(h^{-1}{\rm M}_\odot)$ \\ \hline
0 & 0.42 & 8.7e-4 & -3.7e-05 & -0.70 & 12.1 & 11.4 \\
1 & 3.8e-3 & 1.6e-3 & 0.24 & 1.70 & 8.30 & -1.3 \\
2 & 5.8e-4 & 1.5e-3 & 0.52 & 0.63 & 11.66 & -3.11 \\
3 & 1.7e-3 & 4.4e-4 & 0.47 & 0.23 & 12.30 & -2.23 \\
4 & 1.7e-3  & 3.4e-4 & 0.55 & 0.19 &  12.26 & -2.75 \\
5 & 5.2e-3 &  -5.5e-4 &  0.050 & 0.04 & 12.20 & -3.71 \\ \hline
\end{tabular}\end{table}

Equation~\ref{eq:M_HI} can be used, as has been done for alternative parametrizations, to create $21$~cm mocks from dark matter catalogues. One can use for this a classic halo occupation distribution (HOD) approach where each dark matter halo is assigned a total HI content that depends on the halo mass as described by equation~\ref{eq:M_HI}. One can also add the information about the scatter (figure~\ref{fig:M_HI_scatter}). More complex halo occupation models can be constructed by adding additional information that can be extracted from our model. 

\begin{figure}
\includegraphics[width=\columnwidth]{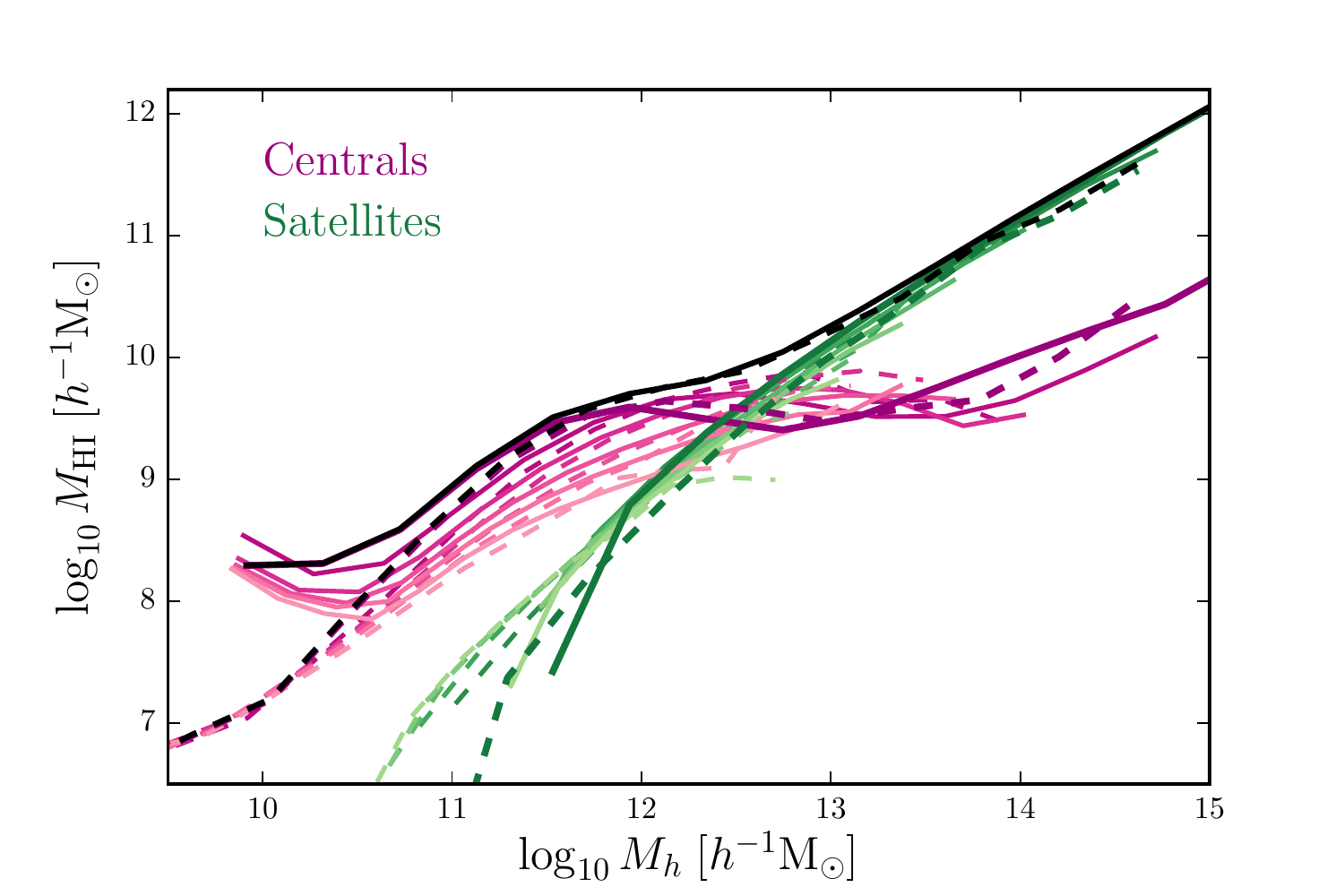}
\caption{The median HI content of dark matter haloes for the MI (solid lines) and for the MII simulations (dashed lines) as a function of halo mass, for central galaxies (violet) and satellites (green). We report the results for redshift 0,1,2,3,4 and 5, with darker colors corresponding to lower redshifts. We show in black also the $M_{\rm HI}(M_h)$ obtained considering all galaxies at redshift $z=0$. At all redshifts, HI is found predominantly in satellite galaxies with halo masses $M_h > 10^{12.5}~h^{-1}{\rm M}_\odot$. For haloes of low and intermediate masses, instead, the dominant contribution comes from central galaxies.}\label{fig:median_halo_HI_cen_sat}
\end{figure}

In figure~\ref{fig:median_halo_HI_cen_sat}, we show the median HI content of dark matter haloes as a function of halo mass for centrals and satellites. At all redshifts, the relation is dominated by central galaxies for small haloes, while for haloes more massive than $M_h\sim 10^{12.5} h^{-1}{\rm M}_\odot$ the satellites give the dominant contribution. In appendix~\ref{app:M_HI_halo}, we propose a parametric model considering separately centrals and satellites. We also give in the appendix the best fit values of the parameters for the redshift range $0<z<5$, as done in table~\ref{tab:M_HI} for the relation obtained considering all galaxies.

\begin{figure}
\includegraphics[width=\columnwidth]{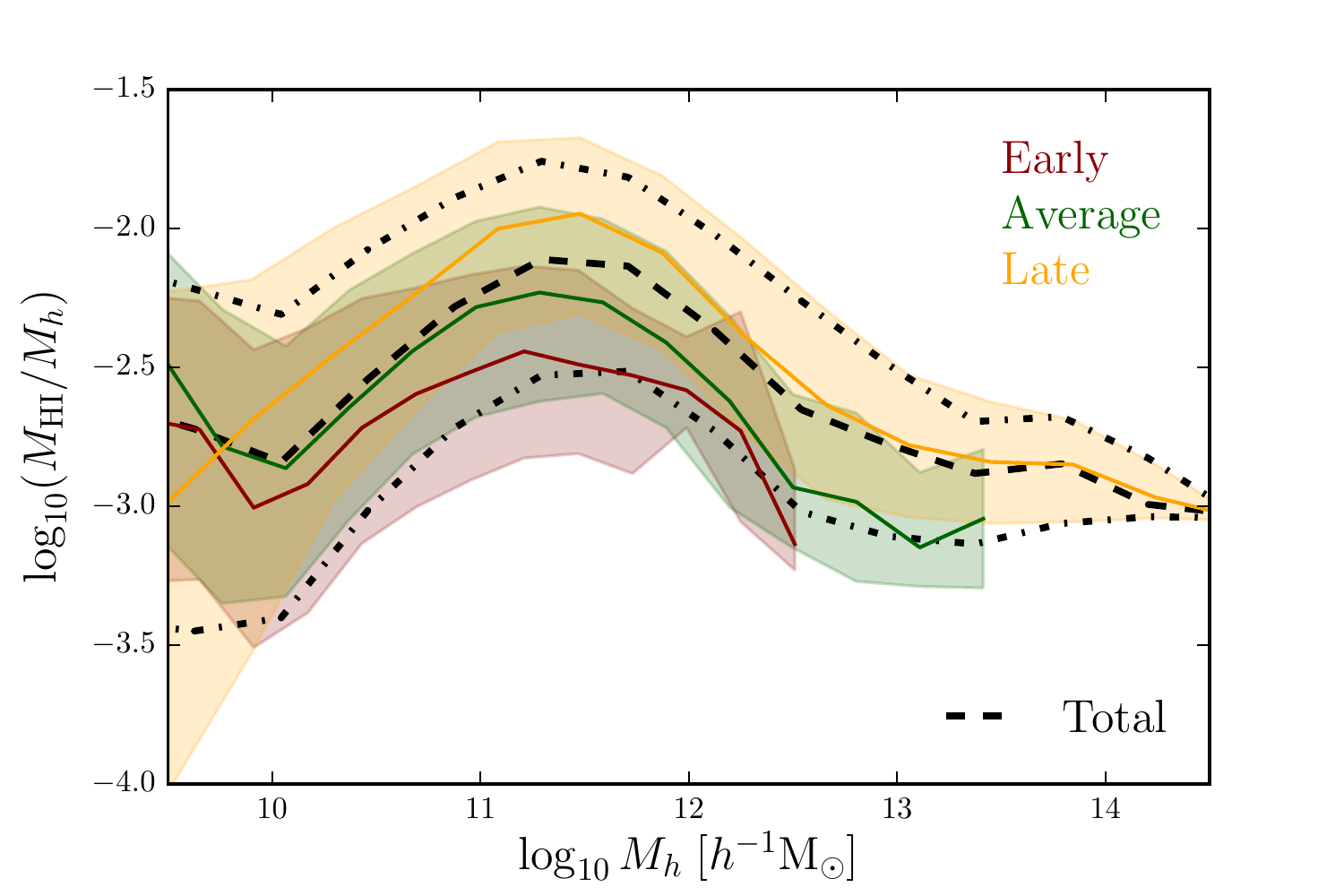}
\caption{The ratio between the median HI halo mass function $M_{\rm HI}(M_h)$ and $M_h$ for the MII simulation, at redshift $z=0$ (black dashed), and its $16$th and $84$th percentiles (black dot-dashed lines). Red, green, and orange lines correspond to the same quantities computed for early, average, and late assembly haloes (see text for definitions).}\label{fig:assembly}
\end{figure}

Another interesting property that can be studied with SAMs is the dependence of HI halo mass function on halo formation time. We use as a proxy for assembly history the redshift at which a halo has acquired half of its final mass, i.e. $z_{50}$. We then consider three different samples: (1) old/early assembled haloes as those whose $z_{50}$ falls within the $33$th percentile of the distribution; (2) average age haloes as those whose $z_{50}$ is within the $33$th and the $66$th percentiles; (3) young/late assembly haloes as those whose $z_{50}$ ranges above the $66$th percentile. In figure~\ref{fig:assembly}, we show the ratio between the median HI halo mass function and $M_h$ for the MII simulation at redshift $z=0$, compared to the same relation obtained for haloes with an early, average, and late assembly. 
In appendix~\ref{app:M_HI_halo}, we report the fitting formulas for the relations obtained for the three cases. 

The normalization of the relation increases with formation time, explaining in part the scatter of the relation obtained when considering all haloes. 
For low mass haloes the total relation coincides with that of haloes with an average assembly history. 
As halo mass increases the total relation is closer to that obtained for haloes with late assembly times. 
Comparing figures~\ref{fig:median_halo_HI_cen_sat} and \ref{fig:assembly}, we note that central galaxies dominate the haloes with average formation time at the low mass end, while at increasing halo mass, the relation is driven by satellites, that are the main contributors to the HI content of late formed haloes. 
At fixed halo mass, we can interpret the different normalizations of early, average and late formed haloes, as a proxy for the depletion of HI in their satellites. 
In fact,  in our model, cooling is possible only on central galaxies, implying that the HI content of satellites can only be depleted. 
As a result, satellites that reside in same mass haloes that formed later, have had less time to loose HI, dominating the HI content.


\subsection{The HI density profile}\label{sec:HIprof}
\begin{figure*}
\includegraphics[width=\textwidth]{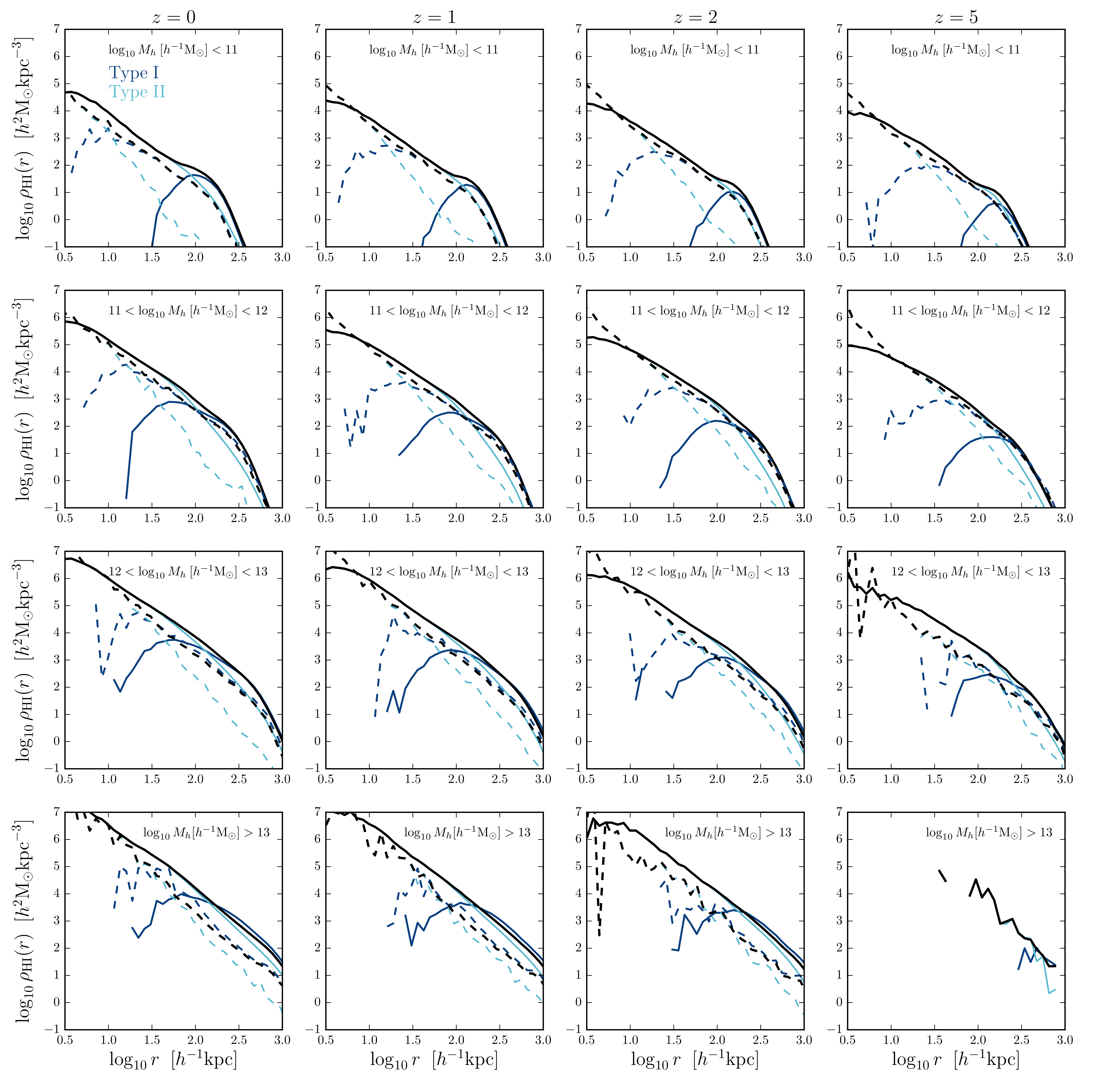}
\caption{HI profiles of the FoF haloes in the MI (solid lines) and the MII (dashed lines) simulations. FoF haloes are separated in different mass bins at different redshifts, and $\rho_{\rm HI}(r)$ is computed averaging the HI content in spherical shells as a function of the distance from the central galaxy. We also show the HI radial profile obtained considering only Type I (dark blue) and Type II (light blue) satellites. Type I satellites shape the HI profile far from the center of the haloes, while Type II get much closer. The transition occurs at $r \gtrsim 100h^{-1}$kpc for the MI simulation, and at few tens of $h^{-1}$kpc for the MII.
}\label{fig:HI_profile}
\end{figure*}

As seen in section~\ref{sec:MHI}, an accurate description of the $21$~cm signal down to small scales requires the knowledge of the the spatial distribution of HI inside dark matter haloes (the so called "1-halo term").
We compute the HI density profile by summing up the HI mass of satellites in thin spherical shells around the central galaxy of each halo, up to few virial radii.
We then divide these profiles according to the total halo mass at different redshifts, and show the average results in figure~\ref{fig:HI_profile}. We do not show error bars because the scatter is large and would make the figure too busy. One reason for the large scatter is the non universality of the HI profiles with respect to dark matter ones (as found also by \citealt{VN2018}), and the large binning we have used for halo mass. 
The HI density increases towards the halo center. This is true down to the very central regions for the MII, while there is some flattening in the innermost regions ($r<1 h^{-1}\mathrm{kpc}$) for the MI, that is likely due to resolution. Indeed, in the MII, the subhaloes are traced down to smaller masses and, consequently, identified down to smaller distances from the halo center. 
Generally, we notice that the MII profiles are more concentrated towards the innermost regions of haloes, while the MI profiles tend to be more rounded. Nevertheless, the HI profiles obtained from the MI and MII are roughly in agreement, particularly for haloes of intermediate mass, at all redshifts. 

Radial profiles of low mass haloes are characterized by a knee-like shape far from the center followed by a pronounced ankle-like feature. This behavior is less  pronounced (in particular for the MI) at higher halo mass. As expected, the profiles are more extended for more massive haloes. We do not find a strong redshift dependence. 

As discussed in section~\ref{sec:sim}, satellite galaxies are divided in two different types: galaxies residing in distinct bound substructures are called Type I, while galaxies whose parent dark matter subhalo has been stripped below the resolution of the simulation are called Type II. Since there is a strong correlation between the time of subhalo accretion and their cluster-centric distance \citep{Gao2004b}, Type II satellites are expected to be more concentrated in the inner regions of dark matter haloes \citep{Gao2004a}. Due to the different resolution limits of the MI and MII, this effect is expected to appear at different radii for these simulations. 

To emphasize this behavior, we consider in figure~\ref{fig:HI_profile} also the contribution from only Type I or Type II satellites. In the outskirts of haloes of all masses, HI is found mainly in Type I satellites. Closer to the center of haloes, HI is present only in the Type II satellites (the only types of satellites present in the inner regions). The transition occurs at radii $\gtrsim 100h^{-1}$kpc for the MI simulation, and at few tens of $h^{-1}$kpc for the MII. At lower halo masses, the transition is more drastic giving rise to the ankle-like shape in the HI profile.


\section{Clustering}
\label{sec:clustering}

The main statistical quantity that can be used as cosmological tool within future $21$~cm experiments is the power spectrum of neutral hydrogen. Higher order statistics can also be used, and can provide important constraints on e.g. non-gaussianity \citep[e.g.][]{Cooray2006,Pillepich2007,Majumdar2018}. 

As discussed when introducing equation~\ref{eq:P_21tot}, several ingredients contribute to the HI power spectrum. In this section, we analyze in detail each of them. In particular, we start giving an estimate of the shot noise in section~\ref{sec:SN}, and we study the bias in section~\ref{sec:bias}.

The first step is the calculation of the HI clustering signal. We use the cloud-in-cell (CIC) interpolation scheme on a $512^3$ grid, and assign to each cell a weight that is equal to the total HI mass hosted by all galaxies in the cell under consideration. We then compute the power spectrum via a Fast Fourier Transform (FFT) of the values in the grid. Since the MII has a higher mass resolution, the power spectrum extends to smaller scales than for the MI. On the other hand, the larger volume of the MI allows larger scales to be sampled (table~\ref{tab:M}). Using our semi-analytic model, we can analyze in detail thee contribution to $P_{\rm HI}(k)$ of sub-samples selected according to different physical properties. In the following, we will focus, in particular, on the clustering signal due to haloes of different mass (section~\ref{sec:Pk_Mh}), on the relative contribution of central and satellite galaxies (section~\ref{sec:Pk_censat}), and on the dependence as a function of HI mass (section~\ref{sec:Pk_MHI}) and color (section~\ref{sec:Pk_redblue}). We will see how the results discussed in the previous sections drive the trends that we will discuss in the following. We will often compare $z=0$ and $z=4$ to convey a sense of the redshift evolution of some of the properties that we analyze. If not otherwise specified, we use galaxy positions in real space.  In section~\ref{sec:RSD}, we will relax this assumption and study the clustering in redshift space. This will lead us to the computation of the observable $21$~cm power spectrum.

\subsection{Shot Noise}
\label{sec:SN}

\begin{figure}
\includegraphics[width=\columnwidth]{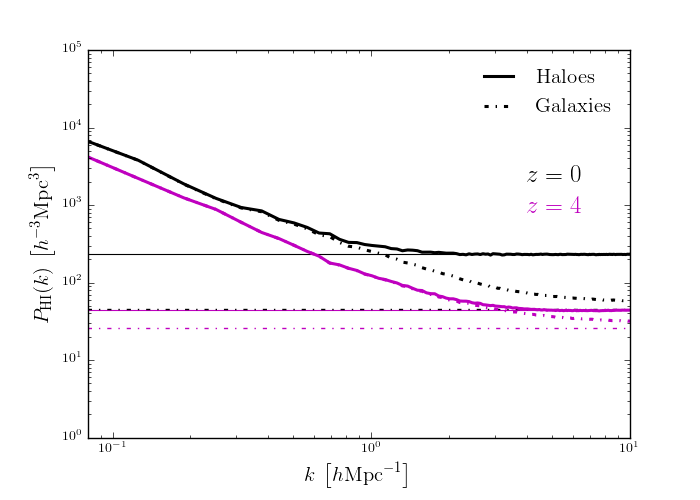}
\caption{The real space power spectrum of galaxies (dot-dashed lines) for the MII simulation, weighted by their HI mass. Black lines and magenta lines are for $z=0$ and $z=4$, respectively. At high $k$, this can be used to infer the level of shot noise and agrees, as expected, with the estimate that can be obtained using equation~\ref{eq:SN} (thin dot-dashed lines). We compare it with the "halo" power spectrum at z=0 (solid black) and z=4 (solid magenta). We define "halo" power spectrum the one computed considering all the HI content of galaxies at the center of their hosting halo. The information on the spatial distribution of galaxies inside the halo is thus collapsed onto the central galaxy and the halo power spectrum flattens at larger scales. The shot noise associated with "haloes" can be estimated from the power spectrum at small scales or using  equation~\ref{eq:SN} (thin solid lines).}\label{fig:SN}
\end{figure}

An important quantity for intensity mapping is the shot noise contribution, which is linked to the discrete nature of the measurement.  In figure~\ref{fig:SN}, we show the power spectrum $P_{\rm HI}$ for the MII simulation at $z=0$ and $z=4$. Since we are dealing with individual galaxies, the shot noise can be measured from the small scale value of the power spectrum. This value is in agreement with the theoretical expectation, i.e.: 
\begin{equation}\label{eq:SN}
    P_{\mathrm{SN}}=\ell_{\mathrm{box}}^3\frac{\sum{M_{\mathrm{HI},i}^2}}{(\sum{M_{\mathrm{HI},i}})^2},
\end{equation}
where the index $i$ runs over all the HI selected galaxies. The shot noise values obtained using equation~\ref{eq:SN} are listed in table~\ref{tab:SN}. We do not provide the values for the MI simulation since the power spectrum does not flatten significantly at small scales and there is also the contribution from aliasing; indeed the values obtained with equation~\ref{eq:SN} are around a factor of two smaller than the small scales value of $P_{\rm HI}$ measured for the MI. 

In the halo model, the shot noise is defined as the limit, for $k \rightarrow 0$, of the 1-halo term of the HI power spectrum \citep[e.g][]{Castorina2017}
\begin{equation}\label{eq:SN_1h}
    P_{\mathrm{SN}}(z)=\lim_{k \rightarrow 0}P^{1h}_{\mathrm{HI}}(k,z)=\frac{\int n(M_h,z)M_{\rm HI}^2(M_h,z)dM_h}{[\int n(M_h,z)M_{\rm HI}(M_h,z)dM_h]^2}
\end{equation}
Following \citet{VN2018} and \citet{Baugh2019}, we re-compute the HI power spectrum, concentrating all the HI of satellite galaxies in the central galaxy of the corresponding parent halo, thus effectively removing the 1-halo contribution. The results, listed in table~\ref{tab:SN} and shown in figure~\ref{fig:SN}, are in agreement with what expected from equation~\ref{eq:SN} (in this case, the index $i$ runs over all haloes, and $M_{\mathrm{HI},i}$ represents the total HI mass in the $i$-th halo). A good agreement is obtained also for the MI, as expected given the fact that we are now considering the shot noise on larger scales. 

\begin{figure}
\includegraphics[width=\columnwidth]{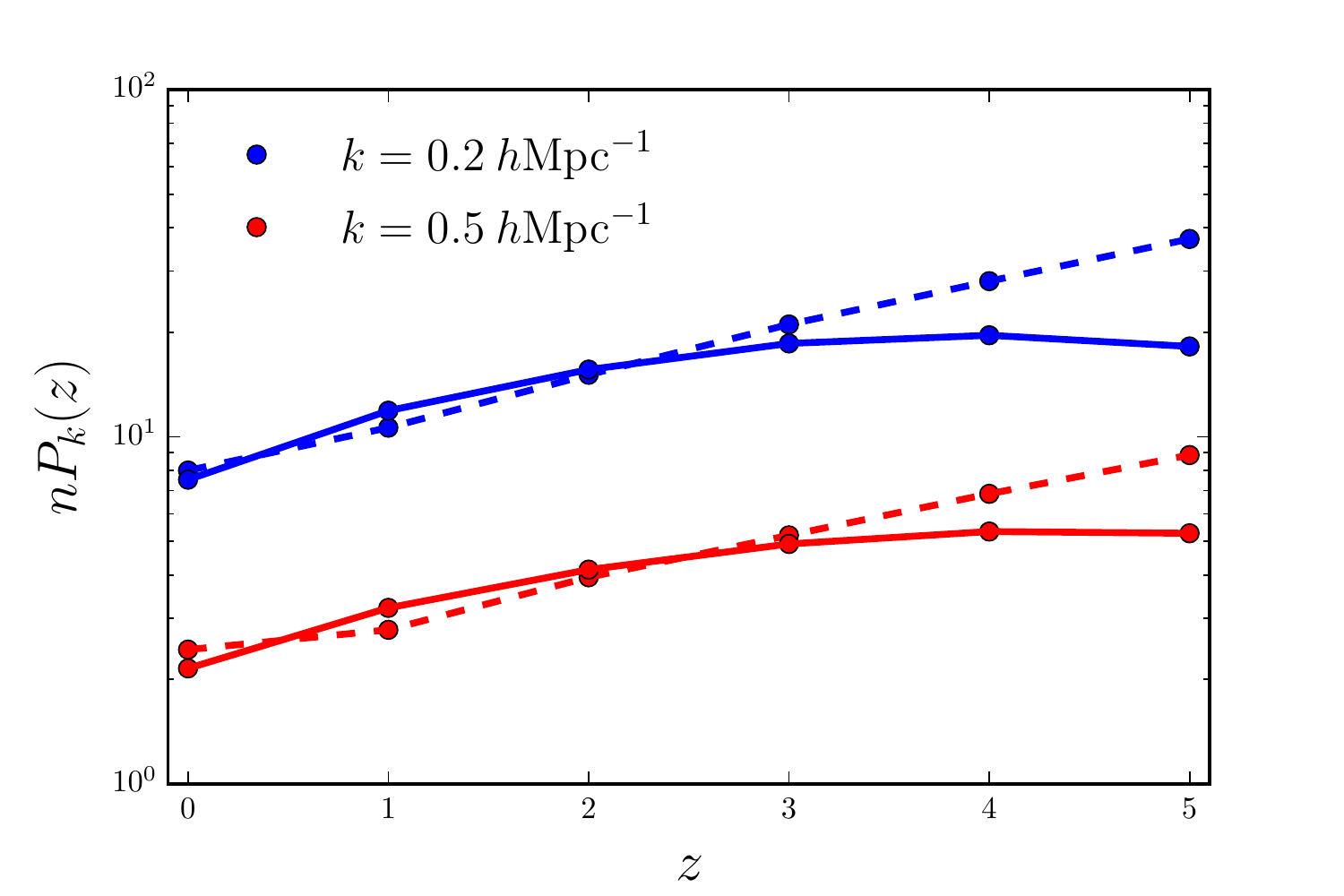}
\caption{The value of $nP_{0.2}$ (defined in equation~\ref{eq:nP}) as a function of redshift, for the MI (blue solid line) and the MII (blue dashed). As done in \citet{VN2018}, we also give the values of $nP_{0.5}$, i.e. $nP_k(z)$ computed at the smaller scale of $k=0.5 h \mathrm{Mpc}^{-1}$ (red lines). These high values show that shot noise should not be a limitation for $21$~cm intensity mapping at these scales.}\label{fig:nP}
\end{figure}

On large scales, the HI power spectrum obtained for "galaxies" is identical to that obtained when considering "haloes", by construction. On smaller scales, where the contribution from satellites is important, the HI power spectrum measured for "haloes" is flatter than that measured for "galaxies".
At $z=0$, this difference is much more pronounced because satellites play an important role. At high redshift, and especially for the MI, the differences become less important. For completeness, we also compute $P_{\mathrm{SN}}(z)$ employing equation~\ref{eq:SN_1h}. We use the Sheth and Tormen \citep{Sheth1999} halo mass function for $n(M_h,z)$, while for $M_{\rm HI}$ we use the fitting formula given in equation~\ref{eq:M_HI} and the values in table~\ref{tab:M_HI}. The results obtained are in very good agreement with the ones computed directly from the simulations, demonstrating that of our fitting formula provides a good description of the predicted HI halo mass function. When considering galaxies in the MII simulation,  the shot noise increases monotonically going to lower redshift up to $z=1$, while decreasing at $z=0$. When considering haloes, it decreases steadily in the MII, while it starts rising again at $z=3$ for the MI. These complex trends are deeply linked to the evolution of the HI mass function, of the HI density (section~\ref{sec:HI_func}), and of the HI halo mass function (section~\ref{sec:MHI}). 
One of the main goal for present and future radio telescopes is the detection of the BAO feature using intensity mapping technique. BAOs can be used to constrain the Hubble rate, the angular diameter distance and  the growth rate from RSDs \citep{Bull2015,Bacon2018}. 
The feasibility of this measurement relies on the strength of the HI power spectrum with respect to the shot noise level.
Following \citet{Castorina2017} and \citet{VN2018}, we compute an approximate quantity that is a proxy for the signal-to-noise: 
\begin{equation}\label{eq:nP}
    nP_{0.2}(z)=P_{\rm HI}(k=0.2 h \mathrm{Mpc}^{-1},z)/P_{\rm SN}(z).
\end{equation}
We show how this quantity evolves as a function of redshift in figure~\ref{fig:nP}, for both the MI and the MII. The values are consistent, although slightly lower than those reported in \citet{VN2018}, showing that the shot-noise contamination should not be a problem at the BAO scale. The same quantity computed at $k=0.5 h \mathrm{Mpc}^{-1}$ shows that also smaller scales should be available for $21$~cm intensity mapping.

\begin{table}
\caption{Shot noise $P_{\mathrm{SN}}$ $[h^{-3} \mathrm{Mpc}^3]$ as a function of redshift, computed both considering the HI mass in galaxies or collapsing the total HI mass in a halo in the central galaxy. For the MI case, we only list values corresponding to the latter case.}\label{tab:SN}
\begin{tabular}{c|cccccc}
z & 0 & 1 & 2 & 3 & 4 & 5\\\hline
MII (gals) & 46 & 61 &  46 & 32 & 26 &  22 \\
MII (haloes) & 232 &  156 &  94 & 61 & 44 &  34\\\hline
MI (haloes) & 292 &  144 &  114 &  104 &   107 & 134 \\\hline
\end{tabular}\end{table}

\subsection{HI Bias}
\label{sec:bias}

\begin{table}
\caption{The approximate value of the HI bias, $b_{\rm HI}$, computed as in equation~\ref{eq:bias} using predictions from the linear theory prediction for the matter power spectrum. At large scales, the bias is about constant, and there is good agreement between the MI and the MII.}
\label{tab:bias}
\begin{tabular}{c|cccccc}
z & 0 & 1 & 2 & 3 & 4 & 5\\\hline
MI  & 0.87 & 1.22 & 1.76  & 2.36 & 2.98  & 3.73 \\
MII  & 0.89 & 1.31 &  1.74 & 2.18 &  2.65  & 3.18 \\\hline
\end{tabular}
\end{table}

\begin{figure}
\includegraphics[width=\columnwidth]{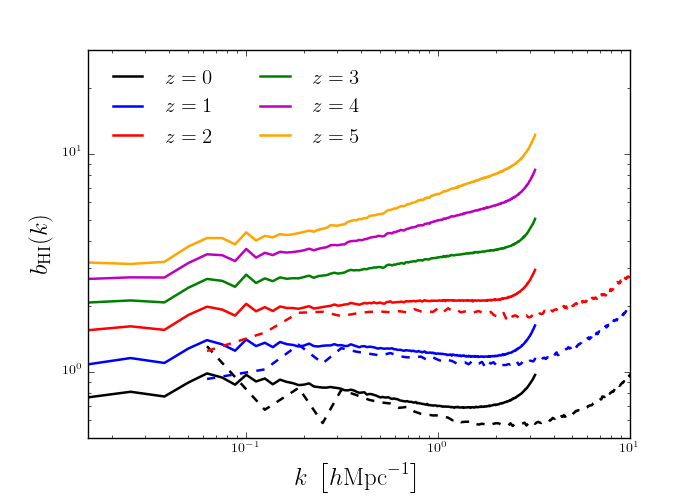}
\caption{The HI bias defined as the square root of the ratio between the HI power spectrum and the dark mater power spectrum, for the MI (solid lines) and the MII (dashed lines), at different redshifts (different colors). 
}\label{fig:bias}
\end{figure}

Although an intensity mapping HI survey will collect information on the neutral hydrogen content of the Universe, the information about the cosmological parameters is ultimately carried by the dark matter distribution. To understand the relation between the HI clustering and the underlying dark matter distribution, it is thus crucial to study the amplitude and shape of the HI bias, $b_{\rm HI}$, defined in equation~\ref{eq:P_21tot}. One can write
\begin{equation}
\label{eq:bias}
   b_{\rm HI}(k)=\sqrt{(P_{\rm HI}(k)-P_{\mathrm{SN}})/P_m(k)},
\end{equation}
where $P_{\mathrm{SN}}$ is the shot noise contribution described in section~\ref{sec:SN}. The bias can be measured on simulations simply computing the HI power spectrum (shot noise subtracted) and the matter power spectrum $P_m(k)$. On large scales, $b_{\rm HI}(k)$ is expected to be roughly constant, while at small scales it has a more complex behavior \citep[e.g][]{GuhaSarkar2012,Camera2013,Penin2018}. This is shown in figure~\ref{fig:bias}, where the large scale measurements are noisy due to sample variance. 
Indeed, if we compute the bias using the theoretical $P_m(k)$ instead of that measured directly from simulations\footnote{For the MII, measured $P_m(k)$ are publicly available only for some of the redshifts used in this study.}, the large scale value of the bias becomes constant and agrees quite well between the MI and MII simulation. We list the value of the bias as a function of redshift in table~\ref{tab:bias}. Our results are in quite good agreement with those found in \citet{VN2018}. Similarly to what found in other studies \citep[e.g][]{Marin2010,VN2018,Baugh2019,Ando2019}, the bias grows with redshift, showing that HI is more clustered than dark matter, and its $z=0$ value at large scales is roughly in agreement with the standard value of $0.85$ \citep[e.g][]{Marin2010}. The significant increase of $b_{\rm HI}$ with redshift is important for IM experiments, since it will make the $21$~cm signal stronger. At $z=0$, the bias predicted from our model shows a scale dependence starting from $k\sim0.1 \;h\mathrm{Mpc}^{-1}$, with a dip around $k\sim1 \;h\mathrm{Mpc}^{-1}$. This has been found also in preliminary observational measurements \citep{Anderson2018}, and in independent studies based on hydro-dynamical simulations \citep{VN2018}. We will discuss this further in section~\ref{sec:Pk_redblue}. Up to $z=2$, the bias is roughly constant down to $k\sim2 \;h\mathrm{Mpc}^{-1}$, while at higher redshift the scale dependence is already noticeable at $k\sim0.3 \;h\mathrm{Mpc}^{-1}$. 

\subsection{Clustering and halo mass}
\label{sec:Pk_Mh}
In this section, we analyze the HI mass weighted galaxy power spectrum $P_{\rm HI}(k)$, focusing on its dependence on the halo mass. These results can be understood from the conditional HI mass fucntion (section~\ref{sec:HI_cond}) and the HI halo mass function (section~\ref{sec:MHI}).

\begin{figure}
\includegraphics[width=\columnwidth]{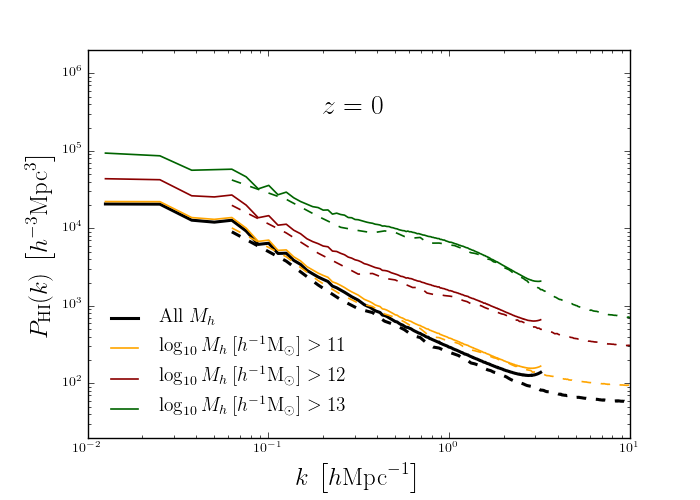}
\includegraphics[width=\columnwidth]{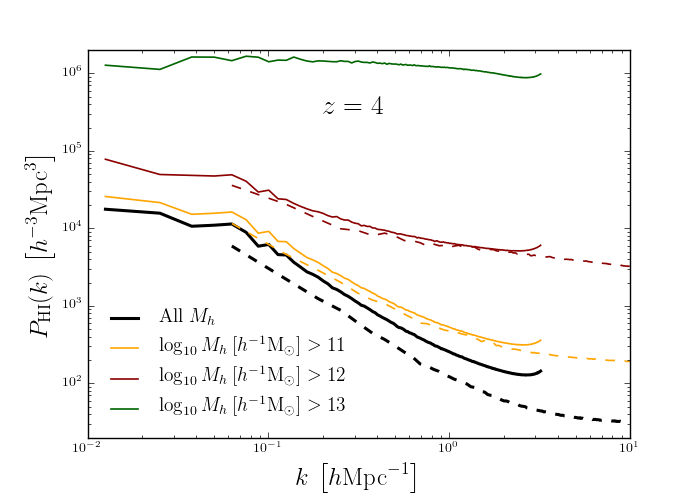}
\caption{{\it Top panel}: The HI power spectrum considering all galaxies (black lines) in the MI (solid) and MII (dashed lines) simulations, at $z=0$. We further consider the contribution due to haloes of increasing mass (different colors, as in the legend). {\it Bottom panel}: Same as for the top panel, but at $z=4$. }\label{fig:Pk_Mhgt}
\end{figure}

In figure~\ref{fig:Pk_Mhgt}, we show the HI power spectrum $P_{\rm HI}(k)$ at $z=0$ (top) and $z=4$ (bottom panel), computed by considering HI in galaxies hosted by haloes of increasing minimum mass. 
These power spectra are compared to the total neutral hydrogen power spectrum, that is calculated using all the galaxies in the simulations (selected using the stellar mass cuts presented in table~\ref{tab:M}). 
At $z=0$, the total neutral hydrogen power spectra of the MI and MII are in good agreement for all the scales that they have in common, indicating that they are describing similar HI distributions\footnote{The deviation between the MI and the MII around $k\sim 3 h\mathrm{Mpc}^{-1}$ is due to aliasing from unresolved small-scale modes due to the finite FFT grid \citep{Sefusatti2016}, and should not be considered in the discussion.}.
Rising the halo mass threshold, the amplitude of the power spectrum increases revealing the halo bias, namely more massive haloes are expected to be more clustered.
Excluding low mass haloes ($M_h > 10^{11} h^{-1}{\rm M}_\odot$), the clustering increases only on small scales, indicating that the large scale structure is not driven by these small haloes in either the MI and MII simulation. 
As discussed in section~\ref{sec:MHI}, most of the HI in the most massive haloes ($M_h > 10^{13} h^{-1}{\rm M}_\odot$) is hosted in an increasing number of HI-poor satellite galaxies, i.e. this is sampling the 1-halo term.
We will discuss the role of satellites further in section~\ref{sec:Pk_censat}.

At higher redshift ($z=4$ in the bottom panel of figure~\ref{fig:Pk_Mhgt}), the $P_{\rm HI}(k)$ evaluated considering all haloes from the MI and MII do not agree. 
The reason for this can be found in the different HI mass functions for the  MI and the MII (see figure~\ref{fig:HI_mass_z}). 
At $z=0$, most of the galaxies of both simulations have HI masses in the convergent part of the HI mass functions. 
At high redshift, high HI mass galaxies are not yet formed, resulting in suppressed tails of the HI mass functions. 
This enhances the relative importance of the low mass end, where MI and MII diverge. This is in agreement with what expected from figure~\ref{fig:HI_mass_z}: the convergence limit between the MI and MII is higher at higher redshift.

Rising the minimum halo mass above $10^{11} h^{-1}{\rm M}_\odot$ will select indirectly galaxies with HI masses mostly $\gtrsim 10^{8} h^{-1}{\rm M}_\odot$ (see figure~\ref{fig:M_HI_scatter}), avoiding the divergent part. 
As a consequence, the halo mass selection erases the differences between MI and MII power spectra. 
The normalization increases again with halo mass threshold, as a result of halo bias. 
Note that there are only very few very massive haloes ($M_h > 10^{13} h^{-1}{\rm M}_\odot$) in MI at this redshift, and none in the volume of MII. 
For this reason, we can see only a highly biased shot noise in the case of MI.

\subsection{Clustering of central and satellite  galaxies}
\label{sec:Pk_censat}

\begin{figure}
\includegraphics[width=\columnwidth]{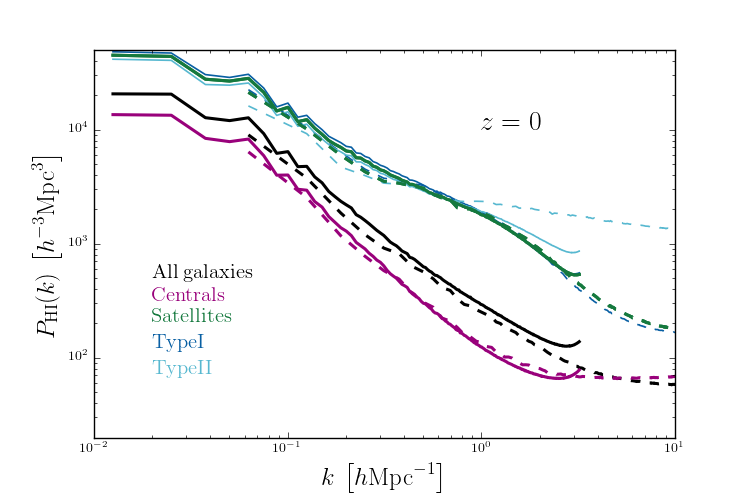}
\includegraphics[width=\columnwidth]{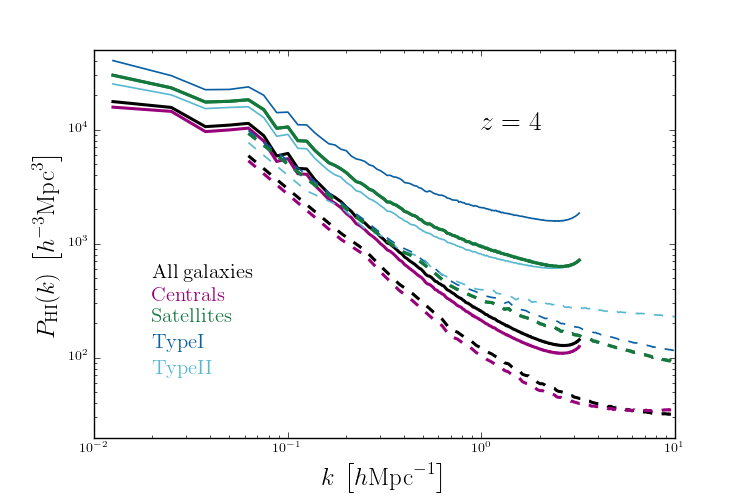}
\caption{{\it Top panel}: The power spectrum of HI selected galaxies at $z=0$ in the MI (solid) and MII (dashed lines) simulations, considering separately the contribution from centrals (magenta) and satellites (green). Satellite galaxies are further divided into Type I (dark blue) and Type II (light blue). The total HI power spectrum (black) is the same as the one in figure~\ref{fig:Pk_Mhgt}. {\it Bottom panel}: Same as in the top panel, but at $z=4$.  }\label{fig:Pk_censat}
\end{figure}

To build a clearer picture of the role of satellite and central galaxies in the distribution of neutral hydrogen, we show in figure~\ref{fig:Pk_censat} the HI power spectra computed considering these different types of galaxies, at $z=0$ and $z=4$ (top and bottom panels). 
The black lines in this figure represent the total HI power spectrum, as in figure~\ref{fig:Pk_Mhgt}. 
The HI power spectra  of satellite and central galaxies have, respectively, higher and lower amplitude than the total HI power spectrum, for both MI and MII, at all redshifts. 
As discussed in section~\ref{sec:MHI}, the neutral hydrogen is hosted mainly by central galaxies in small haloes, while large amounts of HI is hosted by satellite galaxies in the the most massive haloes (see figure~\ref{fig:median_halo_HI_cen_sat}). This implies that the $P_{\rm HI}(k)$ of satellites should be higher than that of centrals, because it is due a population with larger bias. 
The difference is more pronounced at $z=0$ than at $z=4$, because of the larger number densities of massive haloes. 
At $z=0$, we also expect the power spectrum of satellites to be similar to that obtained when considering only massive haloes (top panel of figure~\ref{fig:Pk_Mhgt}): this is indeed the case, and we recognise the shape of the 1-halo term.

At $z=4$, in the MII satellites and massive haloes are far less numerous than in the MI. 
As a consequence, there is no strong correlation between satellites and massive haloes in the MII, namely a selection based on galaxy type does not necessarily translate into a selection in halo mass. 
Indeed, in the bottom panel of figure~\ref{fig:Pk_censat}, we see a clear deviation between the HI power spectra of the satellites in MI and MII. 
This behavior is found also when comparing the HI mass functions of satellites in the MI and MII at $z=4$, as shown in figure~\ref{fig:HI_mass_cond_z4}. 
In the same figure, the differences between the HI mass functions of central galaxies in MI and MII can justify the different power spectra we obtain at $z=4$ in figure~\ref{fig:Pk_censat}, since the bulk of the centrals population in MII is peaked around a lower $M_{\rm HI}$ with respect to MI.

We then compute the $P_{\rm HI}(k)$ considering separately Type I and Type II satellites. Type I are accreted  more recently and are more numerous than Type II, especially at high redshift. As can be seen also in figure~\ref{fig:HI_profile}, Type I satellites hosting HI are more common in massive haloes. Since dark matter halo bias is stronger for more massive haloes, the neutral hydrogen power spectrum of Type I satellites is larger than that of Type II, at all redshifts and for both MI and MII, for $k<0.6\;h\mathrm{Mpc}^{-1}$.  In section~\ref{sec:HIprof}, we have seen that the inner part of HI profiles is dominated by Type II satellites that have had the time to drift to the inner regions of haloes, in particular for the MII. In terms of small scale power spectrum, this  effect translates into a relatively  high shot noise term, since we are effectively almost tracing the power spectrum of massive haloes. 

\subsection{Clustering as a function of HI mass}
\label{sec:Pk_MHI}

\begin{figure}
\includegraphics[width=\columnwidth]{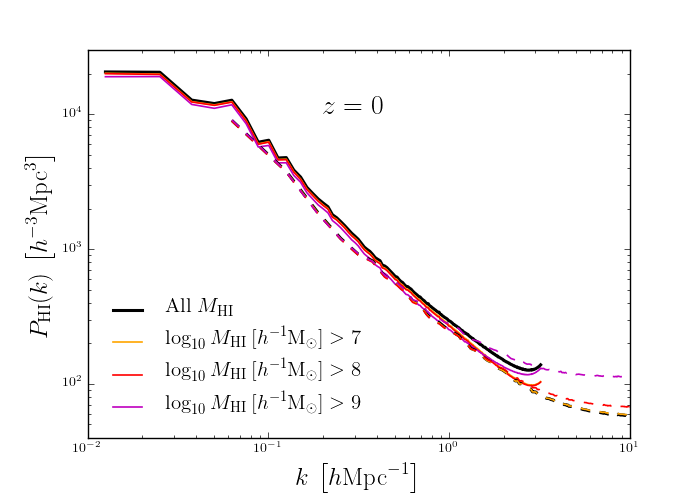}
\includegraphics[width=\columnwidth]{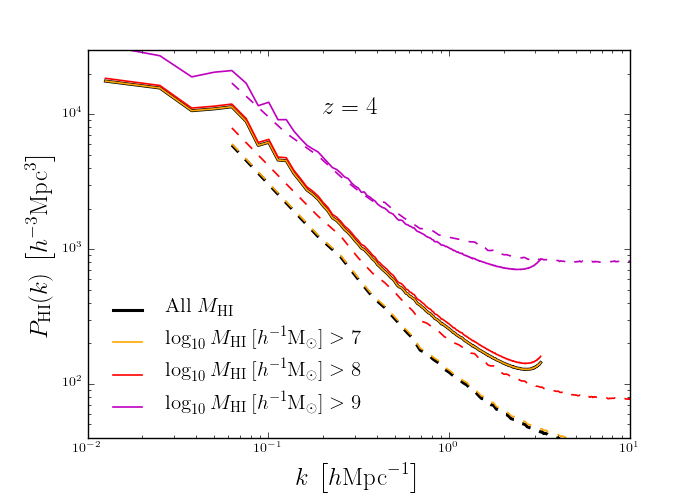}
\caption{{\it Top panel}: The power spectrum of HI selected galaxies at $z=0$  in the MI (solid) and MII (dashed lines) simulations, computed selecting galaxies with progressively larger HI mass. The total HI power spectrum (black) is the same as the one in figure~\ref{fig:Pk_Mhgt}. {\it Bottom panel}: Same as in the top panel, but at $z=4$. }\label{fig:Pk_HIgt}
\end{figure}

In this section, we study how the HI power spectrum depends on the HI threshold adopted, similarly to what has been done in e.g. \citet[][see also \citealt{Zoldan2017}]{Kim2015}. 
We show in figure~\ref{fig:Pk_HIgt} the HI power spectra calculated selecting galaxies using  three different HI mass thresholds and compare them to the total HI power spectrum (in black, as in figure~\ref{fig:Pk_Mhgt}).

At $z=0$, for both the MI and MII, the  $P_{\rm HI}(k)$ does not depend significantly on the HI mass threshold adopted for $k\lesssim 1\;h\mathrm{Mpc}^{-1}$. 
Indeed, as discussed in section~\ref{sec:HI_halo}, galaxies of different HI mass are distributed evenly across haloes of different mass (see figure~\ref{fig:HI_mass_cond}). 
Only the highest threshold ($M_{\rm HI}>10^{9} h^{-1}{\rm M}_\odot$) leads to a rise of the shot noise in the MII simulation on small scales. This happens because this cut removes completely the lowest mass haloes. 

At $z=4$, we get higher amplitudes of the HI power spectra for higher HI mass thresholds, in particular for $M_{\rm HI}\gtrsim 10^{8} h^{-1}{\rm M}_\odot$. These moderately HI-rich galaxies populate mainly massive haloes at this redshift, and their power spectra reflect the halo bias. 
This is valid for both MI and MII, and increasing the HI mass threshold, the amplitudes of the MI and MII power spectra get into good agreement, because the threshold progressively matches the convergence mass between the two simulations.

\begin{figure}
\includegraphics[width=\columnwidth]{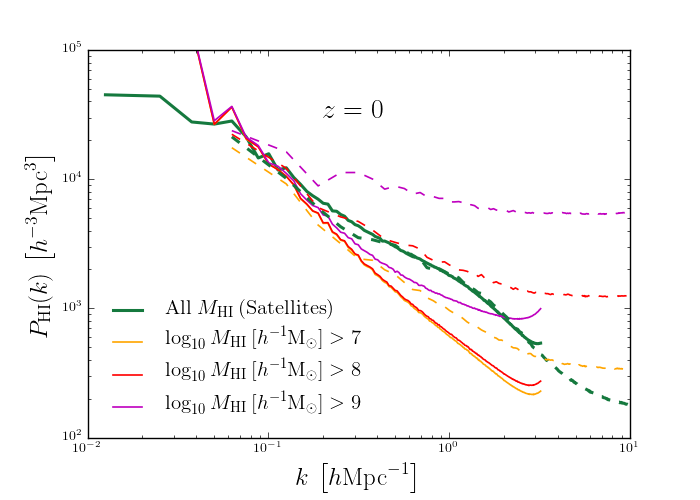}
\caption{The power spectrum of HI selected satellite galaxies at $z=0$  in the MI (solid) and MII (dashed lines) simulations, computed selecting satellites with progressively larger HI mass. The total satellite HI power spectrum is the same as the green one in figure~\ref{fig:Pk_censat}. }\label{fig:satPk_HIgt}
\end{figure}

In figure~\ref{fig:satPk_HIgt}, we analyze the same relations but this time considering only satellite galaxies at $z=0$. 
We show as a comparison the satellite total HI power spectrum (the green line as in top panel of figure~\ref{fig:Pk_censat}).
Increasing the HI mass threshold does not change significantly the $P_{\rm HI}(k)$ at intermediate scales, since satellites are distributed in haloes of all masses. 
The increasing threshold has an important effect on smaller scales: removing an increasing number of galaxies rises the shot noise contribution.
Indeed, HI poor galaxies are fundamental in describing the 1-halo term, as is evident from the comparison with the satellite total HI power spectrum.

\subsection{Red and Blue galaxies}
\label{sec:Pk_redblue}

As seen in section~\ref{sec:bias}, the $z=0$ bias has a spoon shape scale dependency that seems to be present in observational data (see fig.~\ref{fig:bias}).  Indeed, \citet{Anderson2018} have found a statistically significant decrement of the $21$~cm intensity$\,\times\,$galaxy cross power spectrum at  $k\sim 1.5\;h\mathrm{Mpc}^{-1}$, at $z\sim 0.08$. They ascribe this behavior to a combination of lack of HI clustering and lack of correlation between HI and optical galaxies. In particular, they find a scale and color dependent correlation coefficient between HI and galaxies. The spoon shape present in our results is indeed a sign of lack of HI clustering at those scales. In this section, we investigate further the decrease of the HI bias at $k\sim 1-2\;h\mathrm{Mpc}^{-1}$, and how it relates to the red or the blue galaxy population.

\begin{figure}
\includegraphics[width=\columnwidth]{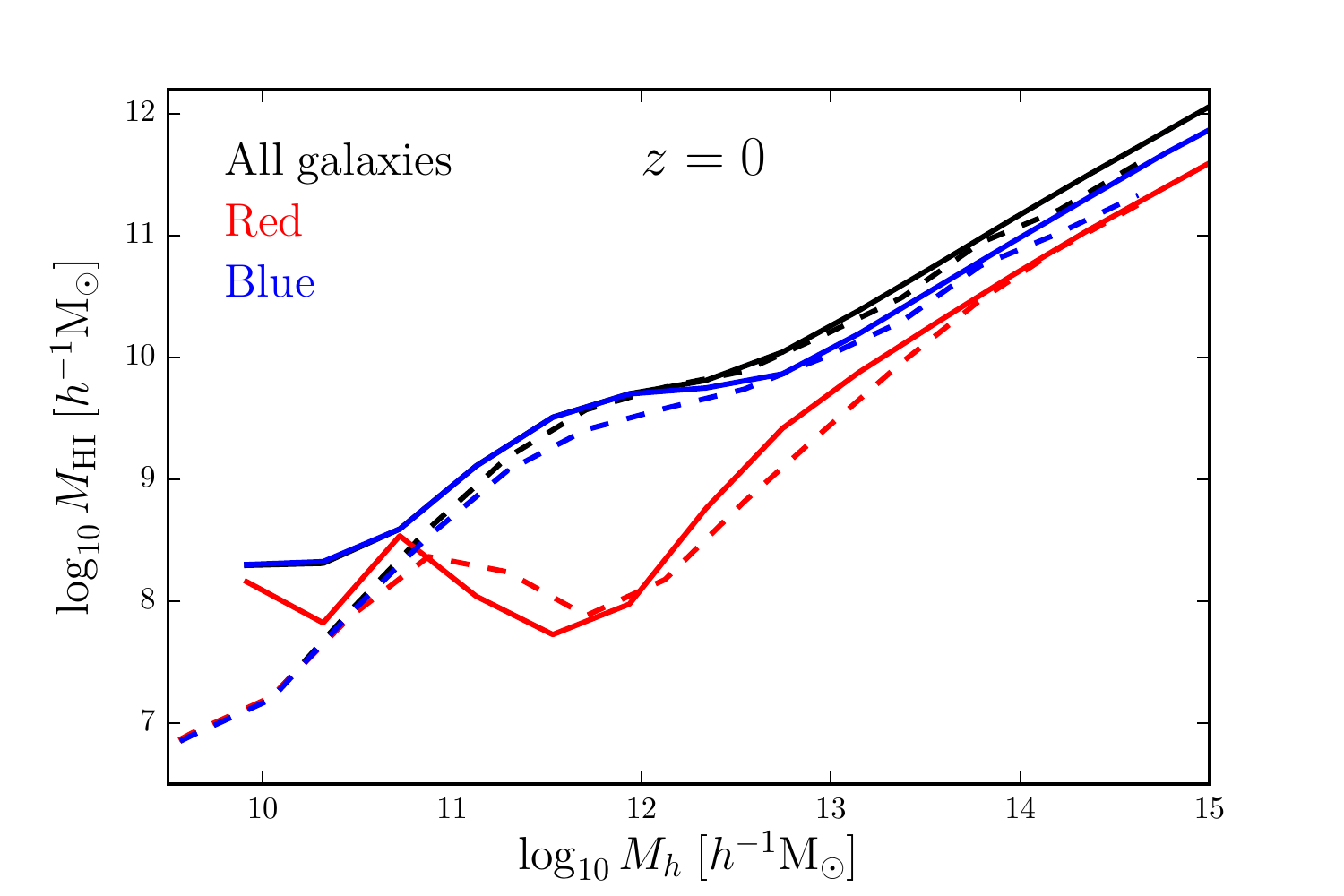}
\includegraphics[width=\columnwidth]{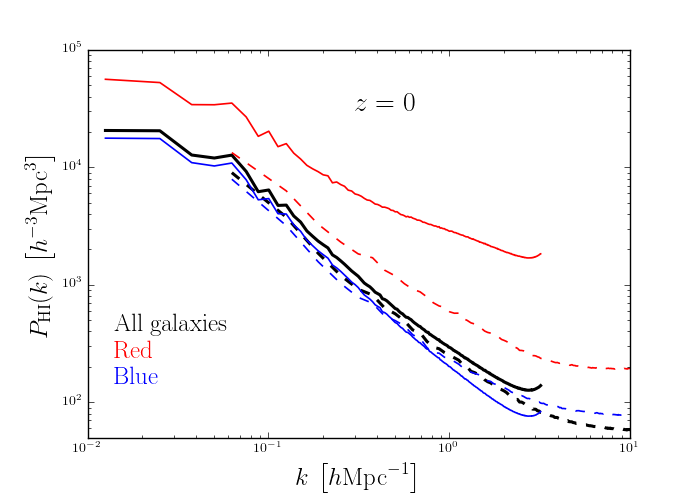}
\caption{
\textit{Top panel}: The HI content of dark matter haloes as a function of halo mass, separating the contribution from red and blue galaxies (lines of corresponding colors). We consider as blue galaxies those with sSFR$>0.3/t_H \;\mathrm{Gyr}^{-1}$. Blue galaxies dominate the HI content of intermediate mass haloes, while red galaxies are mostly found in very massive haloes.
\textit{Bottom panel}: The power spectrum of the red and blue populations (color coded accordingly) compared to the total one for the MI (solid lines) and the MII (dashed lines).}\label{fig:halo_HI_mass_red_blue}
\end{figure}

We focus our analysis at $z=0$, and select red and blue galaxies using a cut in specific star formation rate, i.e. defining a galaxy as blue if sSFR$>0.3/t_H\;\mathrm{Gyr}^{-1}$, where $t_H$ is the Hubble time \citep[see e.g.][]{Gonzalez2017}. In our model, the fraction of red passive galaxies is around $20\%$ for $M_s<10^{10.5} \; h^{-1}{\rm M_{\odot}}$, while steadily rising to more than $80\%$ for the most massive galaxies. Despite the presence of massive passive galaxies, red galaxies overall represent less than the $20\%$ of our sample at $z=0$. 
The majority of these red galaxies have low HI masses ($M_{\rm HI}<10^8 \;h^{-1}{\rm M_{\odot}}$).
In the top panel of figure~\ref{fig:halo_HI_mass_red_blue}, we show the HI halo mass function for red and blue galaxies. 
The red population is mostly found in massive haloes with high halo bias, as most satellites in these massive haloes are red galaxies. 
This can be seen in the bottom panel of figure~\ref{fig:halo_HI_mass_red_blue},
where we show the HI power spectra for the selected red and blue galaxies. 
The HI power spectrum of red galaxies has a larger amplitude than found for the total and blue HI power spectra. 
For the MI, the presence of a visible 1-halo term indicates a large population of old HI poor red galaxies that populate the inner regions of massive haloes.
The blue star forming population dominates the HI content of medium mass haloes, and follows very well the clustering obtained when considering the total HI power spectrum. 
The $P_{\rm HI}(k)$ is, in this case, lower than the total one at all scales because the red population gives an important contribution in the very massive haloes.
In addition, the blue population is less clustered than the overall population in the MI at small scales, while in the MII there is a rise of the shot noise.

\begin{figure}

\includegraphics[width=\columnwidth]{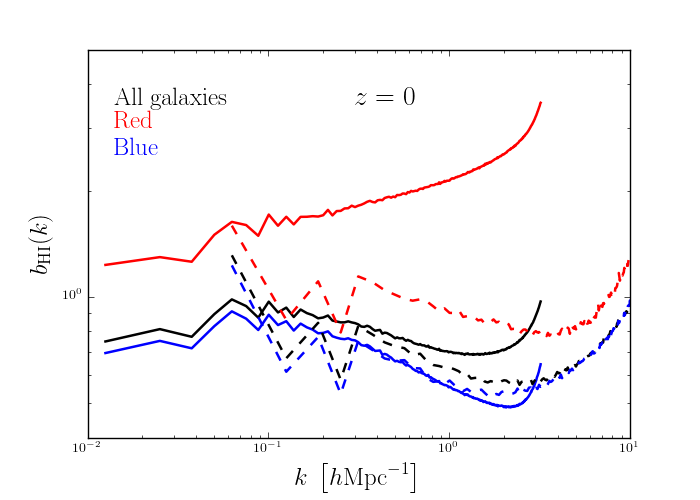}
\caption{The total HI bias $b_{\rm HI}$ (see equation~\ref{eq:bias}) compared to the one computed selecting the blue and red population, for both the MI (solid lines) and the MII (dashed lines). A galaxy is defined blue if its specific star formation rate is larger than sSFR$>0.3/t_H\;\mathrm{Gyr}^{-1}$, where $t_H$ is the Hubble time.}\label{fig:Pk_redblue}
\end{figure}

For completeness, we compute, using equation~\ref{eq:bias}, the different bias for the HI rich blue galaxies and HI poor red galaxies. The results are shown in figure~\ref{fig:Pk_redblue}.
We see that the lack of clustering at small scales for the blue population results in  an enhanced spoon shape at $k\sim 1-2\;h\mathrm{Mpc}^{-1}$. 
This is in agreement with what found by \citet{Anderson2018}, i.e HI is strongly correlated to blue star-forming galaxies.   
We see a different behavior for the red population, especially at the scales sampling the 1-halo term and the shot noise. This difference is more pronounced for the MI simulation, as expected from the important contribution of very massive haloes.

\subsection{HI signal in redshift space}
\label{sec:RSD}

\begin{figure}
\includegraphics[trim=0cm 1cm 0cm 1cm, clip, width=0.46\textwidth]{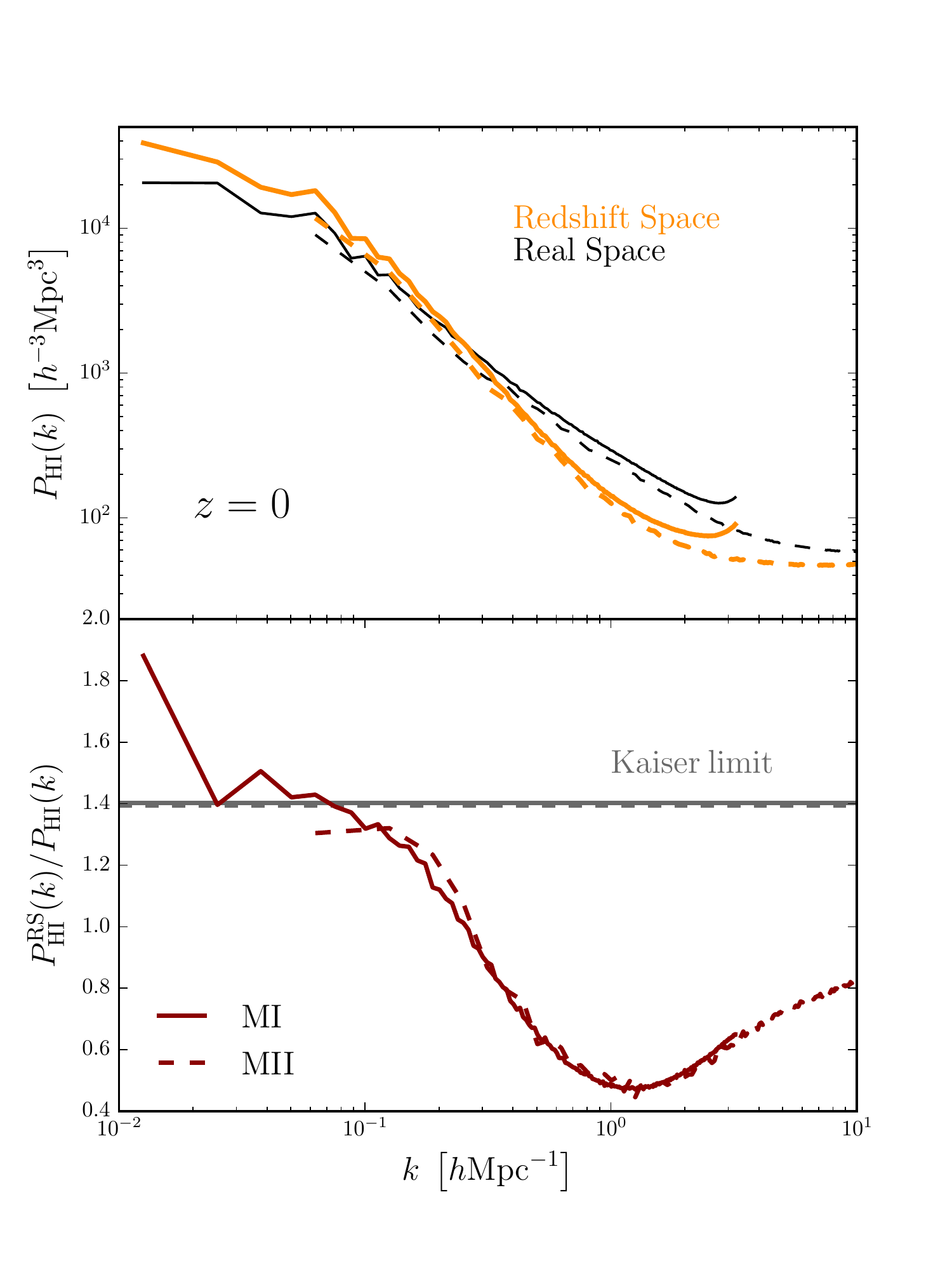}
\includegraphics[trim=0cm 1cm 0cm 1cm, clip, width=0.46\textwidth]{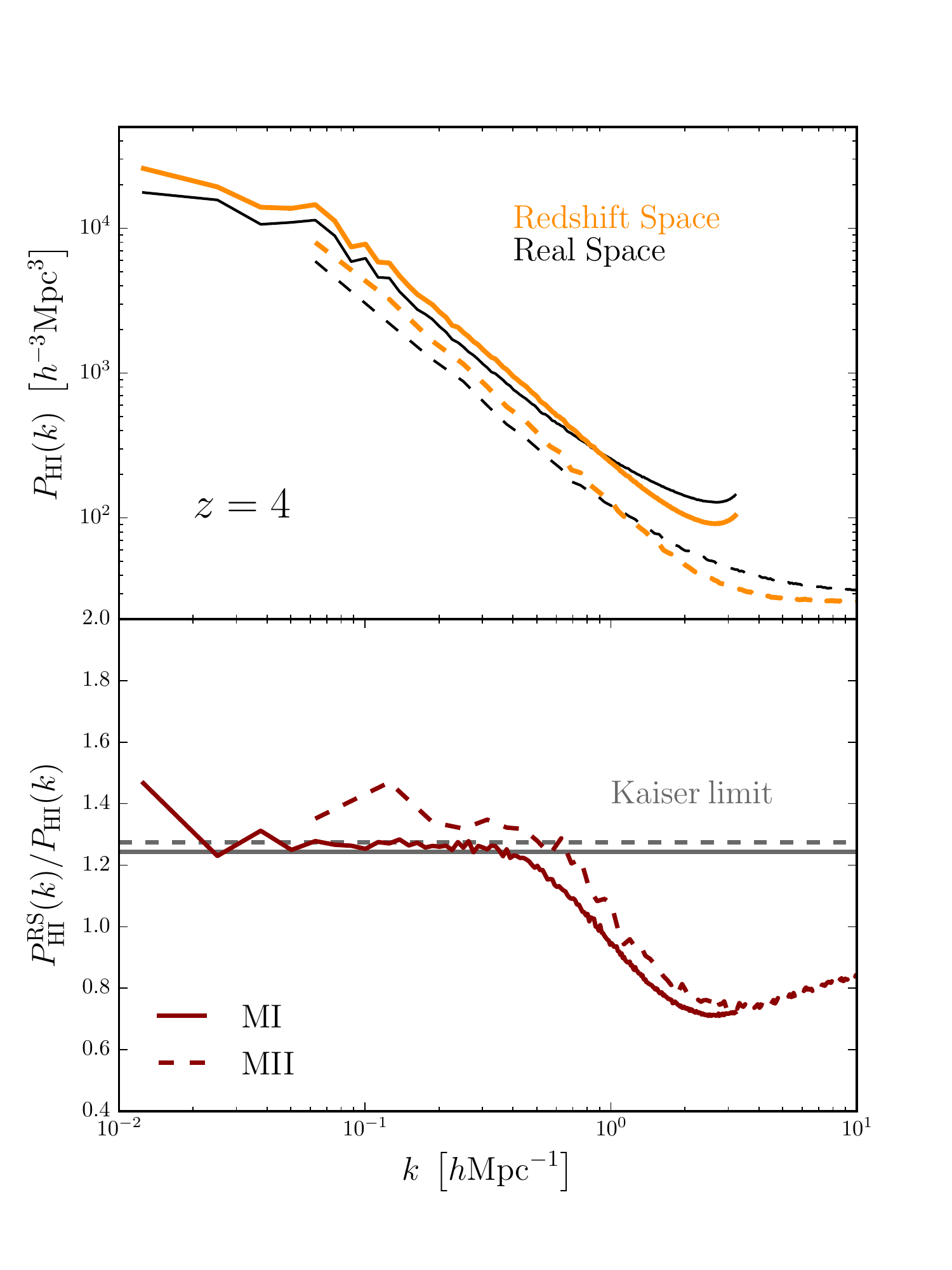}
\caption{Comparison between the HI power spectrum computed in real space (black) and in redshift-space (orange), for the MI (solid) and MII simulations (dashed lines), for $z=0$ (top) and $z=4$ (bottom panels). 
In the lower sub-panel of each panel, we show the Kaiser limit for both the MI (gray solid) and the MII (gray dashed), computed as in equation~\ref{eq:Kaiser}, and compared to what is measured in simulations. }\label{fig:Pk_RS}
\end{figure}

It is well known that peculiar velocities of galaxies cause a Doppler shift that distorts the shape of the observed power spectrum, enhancing the clustering on large scales \citep{Kaiser1987}, and suppressing it on the smaller scales (the Finger-of-God effect). In this section, we compute the HI power spectrum in redshift space $P_{\rm HI}^{RS}(k)$, using the plane parallel approximation to displace the galaxy positions: 
\begin{equation}\label{eq:disp}
    \vec{s}=\vec{r}+\frac{1+z}{H(z)}v(\vec{r})\cdot\hat{z}.
\end{equation}
In the above equation, $\vec{r}$ is the galaxy position in real-space, $v(\vec{r})$ is its peculiar velocity,  and we assume that the $z$-axis is the line of sight. The results are shown in the top sub-panels of figure~\ref{fig:Pk_RS}, for $z=0$ (top) and $z=4$ (bottom panels). As done already by other authors, it is interesting to compute the linear theory prediction for the Kaiser effect: 
\begin{equation}\label{eq:Kaiser}
    \frac{P_{\rm HI}^{RS}}{P_{\rm HI}}\sim1+\frac{2}{3}\beta +\frac{1}{5}\beta^2
\end{equation}
to test its validity against the simulation results.  In the above equation, $\beta=f/b_{\rm HI}$, with $f\simeq\Omega_m^{0.545}(z)$ the linear growth rate. For the bias $b_{\rm HI}$, we have used the approximate results of table~\ref{tab:bias}. Being the estimation of the bias slightly different for the MI and MII, we show two different limits in the lower sub-panels of figure~\ref{fig:Pk_RS}. These should be compared with the results from simulations: as expected, the minimum scale at which the Kaiser limit is valid increases with redshift. At $z=4$, linear theory works nicely down to $k\sim 0.5\;h\mathrm{Mpc}^{-1}$. 

\begin{figure}
\includegraphics[width=\columnwidth]{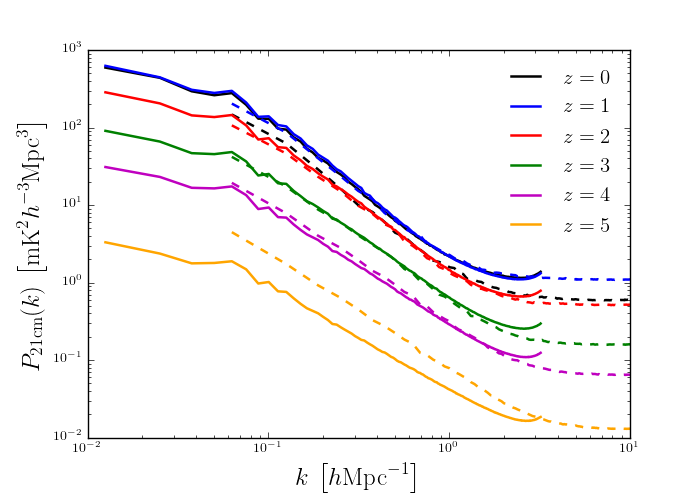}
\includegraphics[width=\columnwidth]{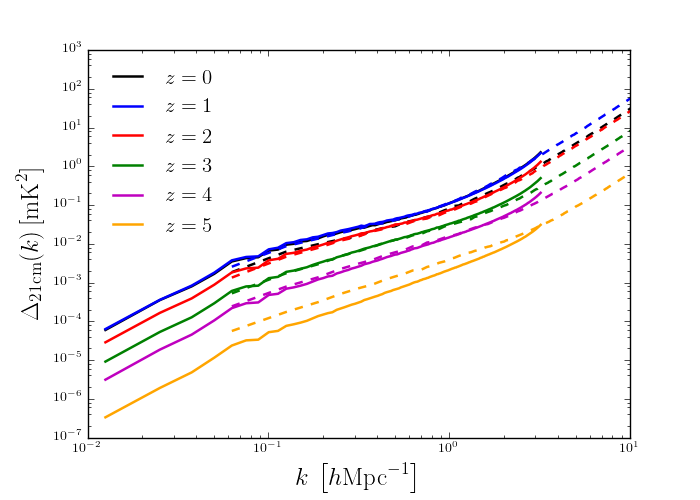}
\caption{The power spectrum of the $21$~cm signal $P_{21cm}(k)$ (top panel) and the $\Delta_{21cm}(k)\equiv P_{21cm}(k)k^3/2\pi^2$ (bottom panel) as predicted from the MI (solid lines) and the MII (dashed lines), at different redshifts (from $z=0$ to $5$, color coded as in legend).}\label{fig:Pk_21}
\end{figure}

From the power spectrum in redshift space measured in our simulations, we can extract  a prediction for the $21$~cm signal for future IM experiments. We rewrite the $21$~cm power spectrum of equation~\ref{eq:P_21tot} as \citep{Wyithe2010}:
\begin{equation}
    P_{21\,{\rm cm}}(k)=\bar{T}_b^2x_{\rm HI}^2P_{\rm HI}^{RS}(k),
\end{equation}
where we explicit the dependence of the brightness temperature contrast on the fraction of neutral atomic hydrogen $x_{\rm HI} \equiv \Omega_{\rm HI}/ \Omega_{\rm H}$. We estimate the hydrogen fraction as  $\Omega_{\rm H}=0.74 \Omega_b$, and use \citep{Furlanetto2006}
\begin{equation}
T_b=23.88 \left( \frac{\Omega_b h^2}{0.02}\right) \sqrt{\frac{0.15}{\Omega_m h^2}\frac{(1+z)}{10}}\; \mathrm{mK}
\end{equation}
The results are shown in the top panel of figure~\ref{fig:Pk_21}, for both the MI and the MII and for different redshifts. The $P_{21{\rm cm}}(k)$ decreases with redshift as does $\Omega_{\rm HI}$ (figure~\ref{fig:rho_HI_z}). For completeness, we show in the bottom panel of figure~\ref{fig:Pk_21} the quantity $\Delta_{21\,{\rm cm}}(k)\equiv P_{21\,{\rm cm}}(k)k^3/2\pi^2$. 
Our results, are consistent with those from other studies \citep[e.g.][]{VN2014}, with the caveat that the assumed value for $\Omega_{\rm HI}$ affects significantly the amplitude of $P_{21{\rm cm}}(k)$. This is especially the case at high redshift, where our semi-analytic model is offset low with respect to observational measurements (see section~\ref{sec:HI_rho}).
The behavior with redshift is influenced by the decrease of the cosmic HI density, but the trend is partially compensated by the increase of the bias $b_{\rm HI}$.


\section{Conclusions}
\label{sec:conlusion}

The upcoming new era of neutral hydrogen (HI) experiments bears the potential to significantly advance our knowledge of the Universe. $21$~cm Intensity Mapping (IM) surveys are being planned, to map HI within unprecedented volumes of the Universe, integrating the signal from hundreds of galaxies in large tri-dimensional pixels. A realistic modelling of the expected HI signal cannot refrain from understanding the role of HI in galaxy evolution. Moreover, to fully exploit forthcoming data for cosmological experiments, it is fundamental to elucidate how HI relates to the underlying dark matter distribution. 
To this aim, semi-analytic models (SAMs) represent a privileged and flexible tool to understand the main physical processes that regulate HI in galaxies, starting from a cosmological dark matter distribution. 

In this paper, we have used the GAlaxy Evolution and Assembly (GAEA) model for studying neutral hydrogen in the post reionization Universe. This state-of-the-art semi-analytic model comprises, among other physical prescriptions typically included, metal and energy recycling with a non-instantaneous approximation, an explicit treatment for the cold gas partition into atomic (HI) and molecular (H$_2$) hydrogen, and a star formation law based on the surface density of molecular hydrogen. GAEA has been run on merger trees extracted from the Millennium I  (MI) and the Millennium II (MII) N-body simulations. The latter samples a smaller volume but has a factor ten better resolution than the former. We have taken advantage of this duality to analyze the distribution of HI in galaxies, as a function of halo mass and cosmic epoch. 

\subsection{Modelling the HI distribution in dark matter haloes}

In the first part of our work, we have analyzed in detail predictions from the GAEA model with the goal to understand optimal methods to model the HI distribution. 
Given the large volumes involved, and the difficulties to model the formation and evolution of atomic hydrogen from first principles, most of the research in the field relies on statistical methods to populate dark matter haloes with HI (e.g. Halo Occupation Distribution approaches). In fact, these represent the only possible tools to construct large numbers of $21$~cm maps covering large cosmological volumes, without prohibitive computational costs. The approach also allows the exploration of different physical models for the HI evolution, as well as of different cosmological models. Our simulations are well suited for this task because they reproduce well the HI distribution measured in the local Universe. We have used our simulated catalogues to characterize in detail the HI distribution and explain the origin of the relation between HI mass and halo mass in our model. We summarize here our results.
\begin{itemize}
     \item The galaxy HI mass function (HIMF) is dominated by central galaxies at intermediate and large HI masses, and by satellites at low masses.  
    The total HIMF results in a complex convolution of galaxies residing in haloes with a broad range of masses. At low HI masses, the main contributions come from satellites in haloes of all masses, while at larger HI masses the main contribution comes from central galaxies of increasing halo mass. These general trends are shared by independent semi-analytic models \citep[e.g.][]{Lagos2011,Kim2017} and therefore do not depend on the particular physical prescriptions adopted. 
    \item The number density of galaxies with low HI masses does not evolve significantly with cosmic time, up to $z\sim5$. The mild dependence on redshift can be ascribed to the growing number of satellites towards later epochs. At large HI masses, the evolution as a function of redshift is much stronger and traces the hierarchical formation of progressively more massive haloes. Independent published theoretical models provide different predictions for the evolution of the HIMF with typically a stronger evolution of the number density at low HI masses \citep{Dave2017,Lagos2011,Baugh2019}. Observational measurements are still lacking beyond the local Universe, therefore future data will provide important constraints on our galaxy formation models for example confirming or refuting the importance of satellite galaxies for the evolution of gas at high redshift.
    \item Our model predicts a mild decline of the cosmic density of atomic hydrogen $\rho_{\rm HI}(z)$. This is in tension with observational data based on DLAs. A better and complete understanding of this disagreement is beyond the scopes of this work. We have demonstrated that this can be only in small part explained by the limited resolution of our simulations, and the expected increasing contribution from low-mass haloes at increasing cosmic epochs. 
In fact, very small haloes do not host significant amounts of HI because of cosmic reionization  \citep[see also ][]{VN2018}. The decreasing cosmic density of HI in our model should be probably ascribed to the presence of HI outside haloes that, based on hydro-dynamical simulations, can contribute to up to $\sim 20\%$ of the total HI at high redshift \citep{VN2014,Diemer2018,VN2018}. It is worth noting, however, that some of the simulations that exhibit a flatter, or even rising trend for the cosmic density of HI, are those that predict an increasing peak in the low mass end of the HIMF (see for example figure 7 and 12 of \citealt{Popping2014}, figure 4 in \citealt{Baugh2019}, or figure 7 and 9 in \citealt{Dave2017}).
\end{itemize}

The analysis of the HIMF lays the foundation to characterize the halo HI mass function $M_{\rm HI}(M_h)$, i.e. the total HI content of dark matter haloes as a function of halo mass. This is a key element in the construction of IM mock maps with HOD techniques, and several empirical relations can be found in the literature, ranging from simple power law behaviors to more complex parametrizations \citep[e.g.][]{Bagla2010,Barnes2010,Santos2015,VN2018,Baugh2019,Obuljen2019}. We have proposed a fitting formula based on our simulated data that is characterized by (i) a correction to the standard power law behavior at large halo mass, that is due to the effect of AGN feedback (as already proposed by \citealt{Baugh2019}) and (ii) a low mass cut off, expected from cosmic reionization. Our best fit parameters are listed in table~\ref{tab:M_HI}, and are given from redshift $z=0$ up to $z=5$. 
To better characterize the relation between HI and halo masses, we have analyzed how the scatter of the halo HI mass function depends on the formation history of haloes, using as a proxy the time when half of the mass was assembled. We find that the halo formation time is the main driver of the scatter of the halo HI mass function, with its normalization increasing with increasing formation time. HOD models typically do not account for this dependence on assembly bias, that is however important for cosmology precision. We have provided fitting functions for both central and satellite galaxies, and different assembly histories (table~\ref{tab:M_HI_cen}, \ref{tab:M_HI_sat} and \ref{tab:M_HI_assembly}). These relations can be applied to construct refined HOD models that can be used to generate $21$~cm mock maps. We intend to pursue this goal in future work. 

\subsection{Estimates for future Intensity Mapping experiments}

One of the main tools that will be used to analyze future IM experiments is the power spectrum of the neutral hydrogen distribution. In the second part of our work, we have analyzed the shape of the HI mass weighted power spectrum $P_{\rm HI}(k)$, as predicted by GAEA. Thanks to the flexibility of our semi-analytic model, the total HI distribution can be dissected as a function of different galaxy properties, quantifying the contribution from shot-noise and galaxy bias. Our main results are summarized in what follows.
\begin{itemize}
    \item Our model predicts a flattening of the power spectrum at small scales, in quantitative agreement with the expected value of the HI mass weighted galaxy shot-noise (see equation~\ref{eq:SN}). We have also computed the shot-noise level as proposed in  \citet{VN2018}, concentrating all HI inside dark matter haloes at their center, consistently with the shot noise definition of the halo model (see equation~\ref{eq:SN_1h}). 
    We predict a steadily decrease of the shot noise with redshift in this case.
    Our results confirms that shot noise will not limit the capability of measuring the Baryon Acoustic Oscillation (BAO) features with intensity mapping experiments. 

    \item We find that the HI bias, $b_{\rm HI}$, increases with redshift, and shows a characteristic dip at $k \sim 1 h{\rm Mpc}^{-1}$ in the local Universe, in qualitative agreement with observational measurements of \citet{Anderson2018}. This `feature', due to lack of HI clustering at those scales, has been investigated further by considering the different contribution from blue/active and red/passive galaxies.  The former dominate the total HI content of dark matter haloes, especially at intermediate halo mass, providing a more accurate description of the HI clustering; the latter, in contrast, dominate the central regions of the most massive haloes.
    \item  The amplitude of the power spectrum increases when selecting galaxies residing in progressively more massive haloes. At $z=0$ and for massive haloes, an increasing number of satellites reveals the 1-halo term contribution in the power spectrum. At $z=4$, these massive haloes are not yet assembled, and the resulting power spectrum reflects a highly biased shot-noise term. At low redshift, the HI power spectrum does not depend significantly on the HI threshold adopted, because HI galaxies of given mass are distributed across a wide range of halo masses. At all scales, an important contribution to the HI power spectrum is provided by satellite galaxies. HI poor galaxies dominate the HI power spectrum of satellites.
    At small scales, we have noted the role of `orphan galaxies' (i.e. galaxies whose parent dark matter substructure has been stripped below the resolution of the simulation): these peculiar satellites populate the most central regions of haloes and contribute to the total HI clustering. The importance of this contribution increases when lower resolution simulations are considered.
    
    \item We have computed the HI power spectrum in redshift space, as the measurement of the $21$~cm signal will map directly frequencies into redshifts. We have shown that the effect of peculiar velocities of galaxies produces the expected Kaiser effect at large scales, and the Finger-Of-God reduction of power on the smaller scales. Moreover, as done in \citet{VN2018}, we have tested the agreement with linear theory, comparing the ratio of the power spectrum in redshift space and real space with the Kaiser limit (see equation~\ref{eq:Kaiser}). We have  shown that, at redshift $z\gtrsim4$, linear theory works up to $k \sim 0.4\: h{\rm Mpc}^{-1}$;

\end{itemize}

Our model predictions are consistent with results from the Illustris simulation, described in \citet{VN2018}. 
The shot noise levels are consistent within a factor of two, although we predict a steadily decreasing shot noise with redshift, while \citet{VN2018} find that the shot noise decreases only for $z \geq 1$. Interestingly, \citet{Baugh2019} find a completely different behavior: their shot noise level increases with redshift. 
\citet{VN2018} show that, at low redshift, HI-poor haloes are more clustered than HI-rich haloes:  this agrees with our discussion on the red and blue galaxy population. Moreover,
our HI bias values are consistent with their results. The HI bias stays roughly constant also on the largest scales that we can probe with the MI simulation, as expected for theoretical predictions. From $z=1$ to $z=2$, we find a constant $b_{\rm HI}$ up to $k \sim 2 h{\rm Mpc}^{-1}$ while, at higher redshift, a scale dependence is already noticeable at $k \sim 0.3 h{\rm Mpc}^{-1}$, again in agreement with \citet{VN2018}. The  dip at $k \sim 1 h{\rm Mpc}^{-1}$ at $z=0$ is present also in the Illustris simulation. This feature is not present in the results of 
 \citet{Baugh2019} since it is linked to the 1-halo term, while they are computing the HI power spectrum placing the entire HI content of each halo at its centre of mass.
\citet{VN2018} demonstrated that the 1-halo term is important to have a proper description of the signal, especially in redshift space. Indeed, HOD techniques will need to incorporate prescriptions for the HI distribution inside haloes. Our results on of the role of satellites on small scales constitutes a step in this direction.
Finally, using the power spectrum in redshift space, we have computed the resulting $21$~cm signal in $\mathrm{mK}^2$, as a function of redshift. Our results depend on the evolution of the value of the HI neutral fraction $x_{\rm HI}$ and of the HI distribution across time. Future Intensity Mapping data from the SKA radio telescope and its pathfinders will allow us to test these predictions. On the other hand, the flexibility of SAMs will be crucial to interpret these data. The feedback process between data and galaxy evolution models will greatly enrich our understanding of the Universe and our capability to forecast and constrain cosmological scenarios.

\section*{acknowledgements}
 The authors acknowledge funding from the INAF PRIN-SKA 2017 project 1.05.01.88.04 (FORECaST). MS and MV are supported by the INFN INDARK PD51 grant. The authors acknowledge Volker Springel and Michael Boylan-Kolchin for providing the power spectra of the Millennium I and Millennium II.



\bibliographystyle{mnras}
\bibliography{biblio}


%
\appendix

\section{Conditional HI mass function at high redshift}\label{app:z4}
\begin{figure*}
\includegraphics[width=\columnwidth]{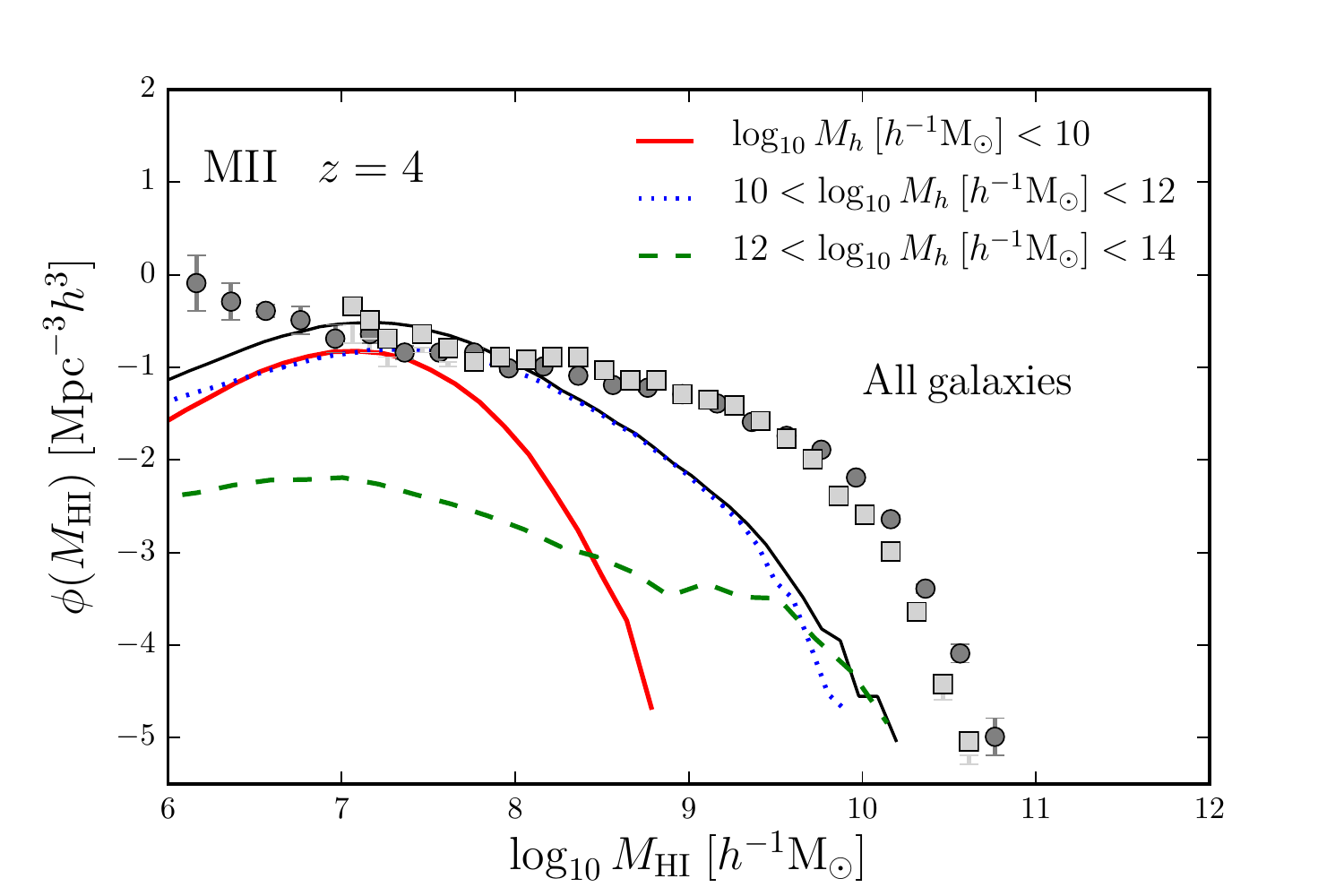}
\includegraphics[width=\columnwidth]{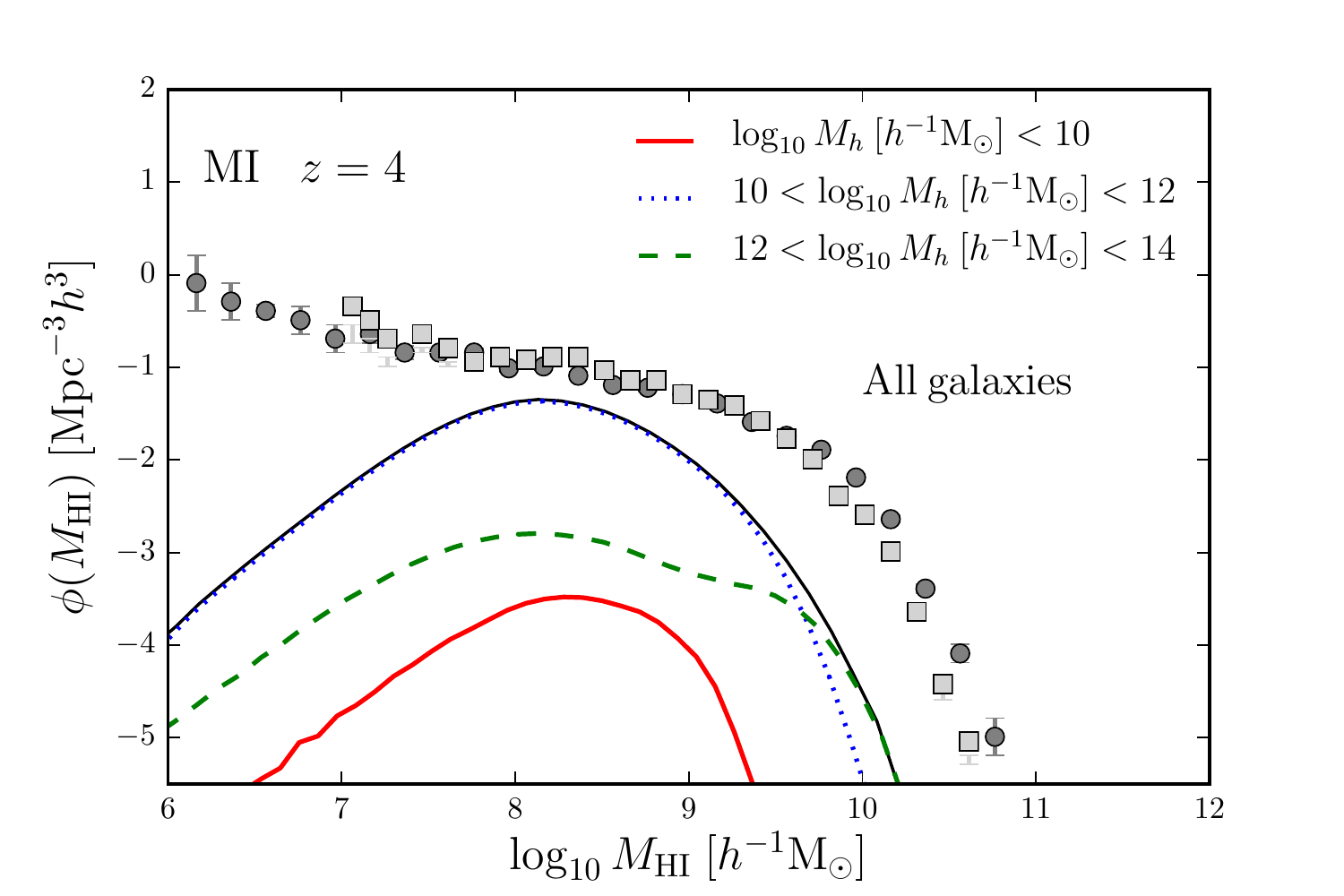}
\includegraphics[width=\columnwidth]{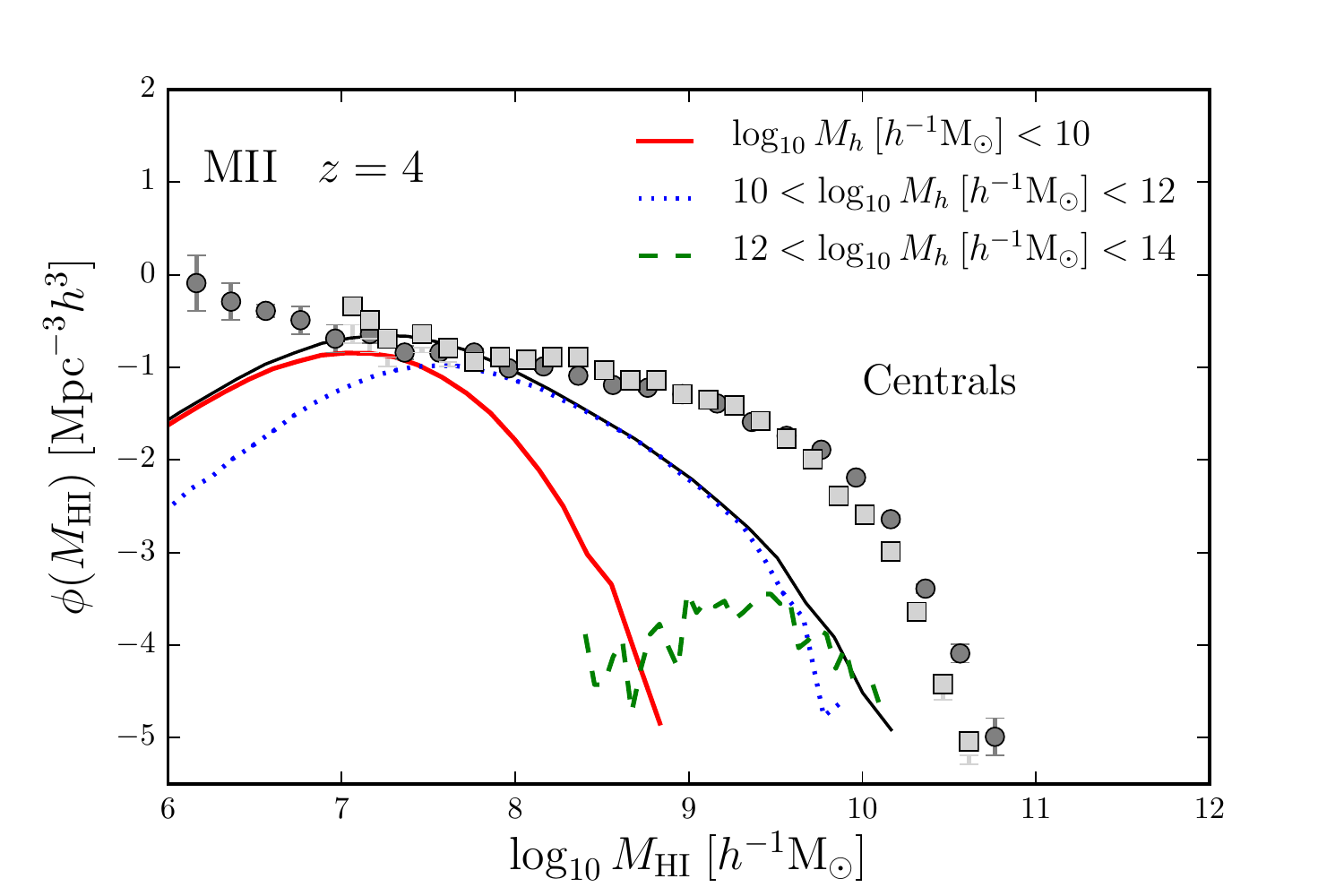}
\includegraphics[width=\columnwidth]{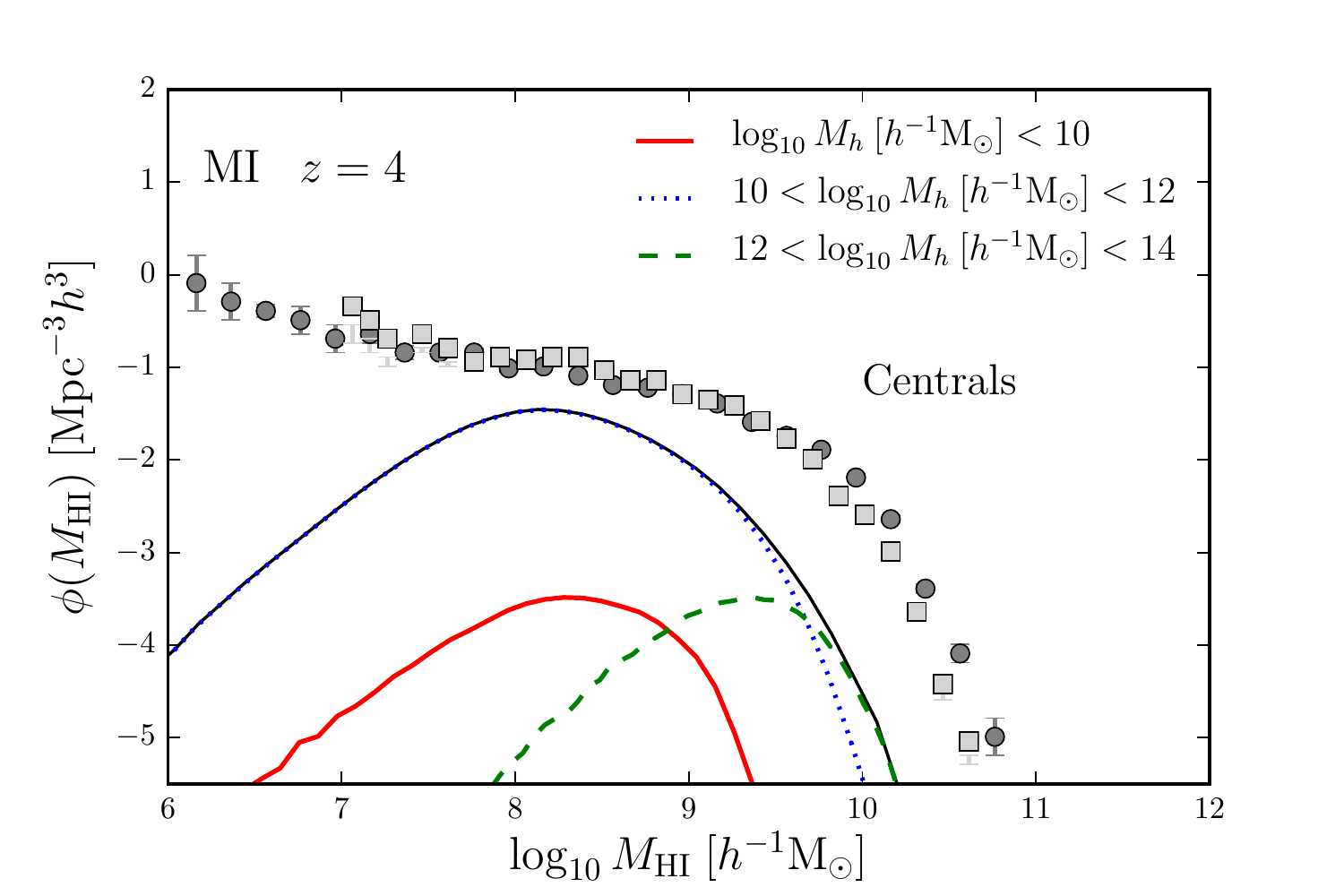}
\includegraphics[width=\columnwidth]{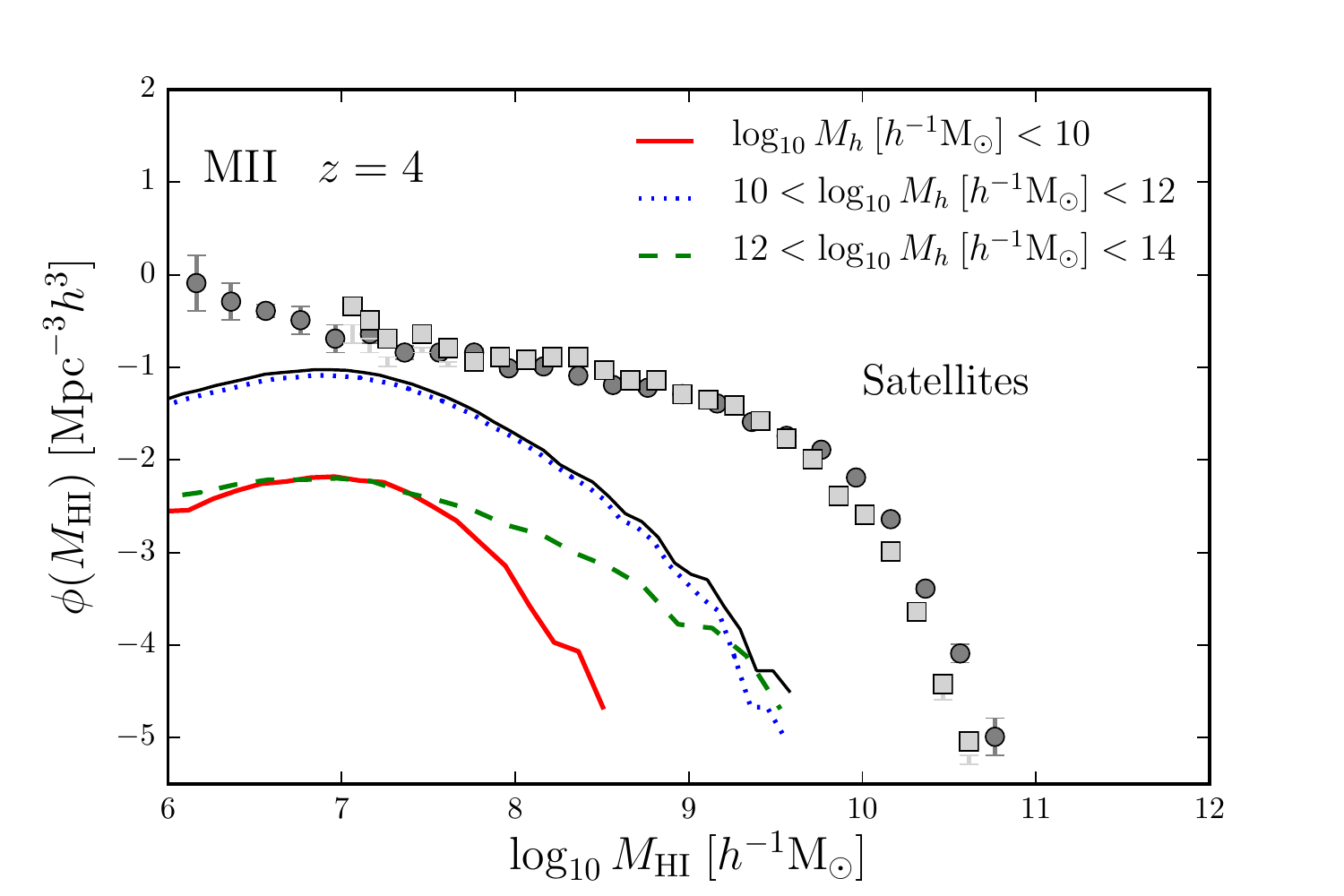}
\includegraphics[width=\columnwidth]{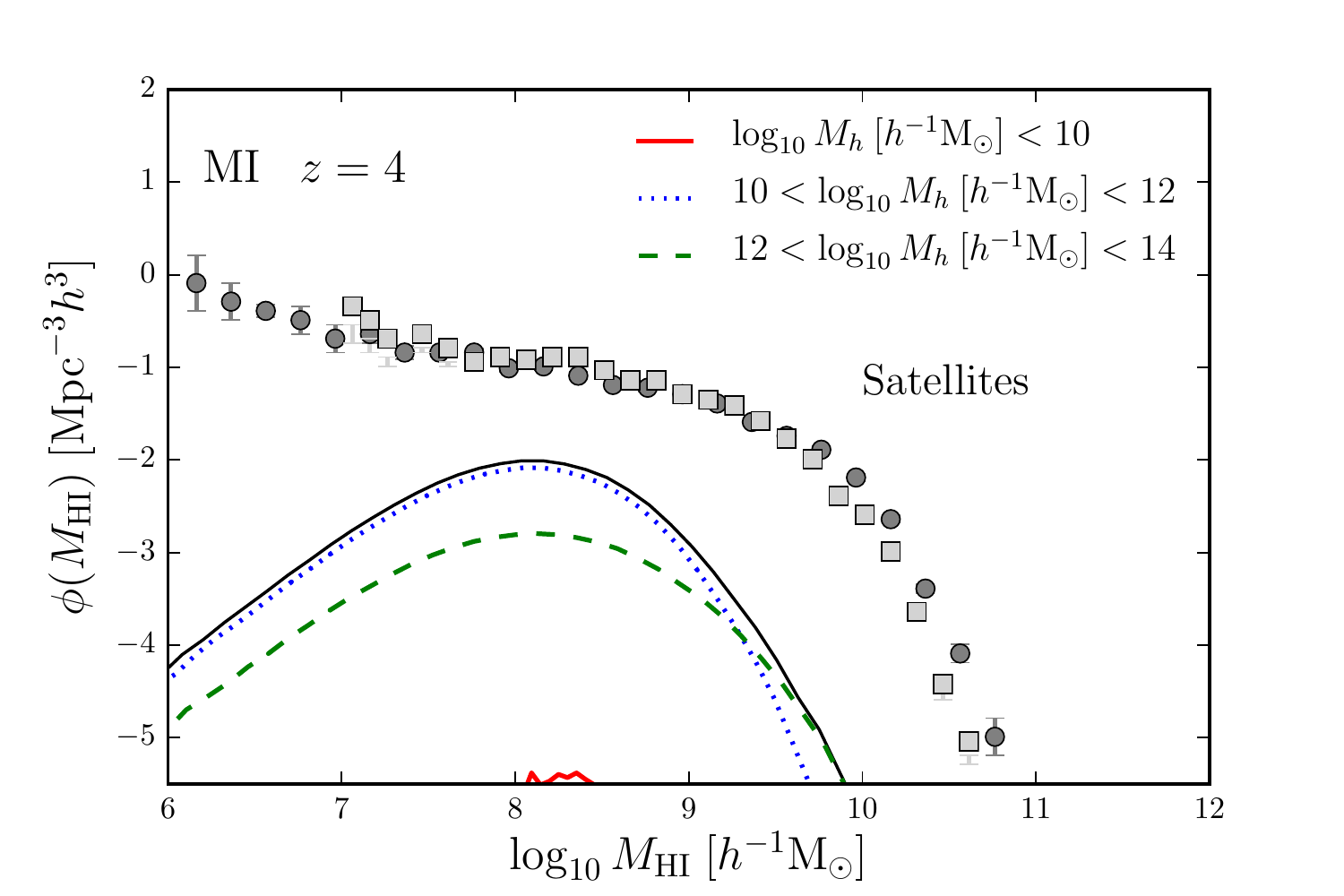}
\caption{The predicted HI conditional mass function for the MII (left column) and the MI (right column) at $z=4$. In the middle and bottom panels we separate the contribution from central and satellite galaxies, respectively. The solid black line in each panel is the sum of the contributions from different host dark matter haloes. To better appreciate the difference with $z=0$, we add with squares and circles the data measured by \citet{Zwaan2005} and \citet{Martin2010}, the blind HI surveys HIPASS \citep{Meyer2004} and ALFALFA \citep{Giovannelli2005} in the local Universe.}\label{fig:HI_mass_cond_z4}
\end{figure*}

In this appendix, we extend the analysis of section~\ref{sec:HI_cond} to higher redshift.
In figure~\ref{fig:HI_mass_cond_z4}, we show the HI mass function at $z=4$ for MI and MII (right and left panels, respectively). 
We show the HI mass function for all the galaxies in the top panels, for the central galaxies in the middle panels, and the satellite galaxies in the bottom panels. 
We further divide model galaxies according to their halo mass (see legend). 
At $z=4$, we find very few galaxies with $M_{\rm HI}\gtrsim 10^{10}\,{\rm M_{\sun}}$, and also very massive haloes ($M_{h} > 10^{14} h^{-1}{\rm M}_\odot$) are not yet formed. 
Despite these differences, the global relative contributions of haloes of different masses to the HI mass functions is similar to what found at $z=0$ (see figure~\ref{fig:HI_mass_cond}), especially for central galaxies. 
In the case of satellite galaxies, we notice a swap in the relative importance of haloes of mass $10^{12} < M_{h} [h^{-1}{\rm M}_\odot]< 10^{14}$ and $10^{10} < M_{h} [h^{-1}{\rm M}_\odot]< 10^{12}$, most likely due to the hierarchical growth of structures.

\section{HI halo mass function: fitting formulae}\label{app:M_HI_halo}
With the aim of extracting from semi-analytic models the properties that can be useful for HOD techniques to simulate $21$~cm maps for intensity mapping, we extend, in this appendix, the analysis of section~\ref{sec:MHI}. In particular, we model the HI halo mass function
$M_{\rm HI}(M_h)$ dividing central and satellite galaxies (see figure~\ref{fig:median_halo_HI_cen_sat}). 
For central galaxies we use the same formula as for the total, i.e. equation~\ref{eq:M_HI} and report the best fit values of the parameter in table~\ref{tab:M_HI_cen}. As redshift increases, we can see that $M_{\mathrm{break}}$ increases, as can also be seen in figure~\ref{fig:median_halo_HI_cen_sat}. 
The low values for $M_{\rm min}$ also at redshift zero, indicates that the low mass cut-off is mainly driven by satellites.
For the case of satellites, we use the standard formula \\ \citep[e.g.][]{Padmanabhan2017,Castorina2017,VN2018,Obuljen2019}
\begin{equation}
    M_{\mathrm{HI}}(M_h)=M_0 \left(\frac{M}{M_{\mathrm{min}}}\right)^{\alpha} e^{-\left(\frac{M_{\mathrm{min}}}{M_h}\right)^{\gamma}},\label{eq:M_HI_sat}
\end{equation}
where $M_0$,$ M_{\rm min}$, $\alpha$ and $\gamma$ are the fitted parameters. The best fit values are reported in 
table~\ref{tab:M_HI_sat}. 
\begin{table}
\caption{Best fit values for the parameters of the HI halo mass function $M_{\mathrm{HI}}(M_h)$ of equation~\ref{eq:M_HI} as a function of redshift when only centrals are considered (see figure~\ref{fig:median_halo_HI_cen_sat}). The cut-off parameter $\gamma$ is fixed here to $0.5$ as for the total HI halo mass function.}\label{tab:M_HI_cen}
\begin{tabular}{c|cccccc}
z & $a_1$ & $a_2$ & $\alpha$ & $\beta$ & $\log_{10}(M_{\mathrm{break}})$  & $\log_{10}( M_{\mathrm{min}})$  \\
& & & & & $[h^{-1}{\rm M}_\odot]$  &  $[h^{-1}{\rm M}_\odot]$ \\ \hline
0 & 2.9e-3 & 6.8e-5 & 0.41 & 0.85  & 10.66 & -1.98\\
1 & 1.6e-3 & 1.1e-4 & 0.56   &  0.43 & 11.86 & -2.99\\
2 & 1.3e-3 & 4.5e-4 &  0.74  &  0.25 &  12.26 & -3.01\\
3 & 2.2e-3 & -2.3e-4 & 0.46   &  0.15 & 12.27 & -3.75\\
4 & 3.3e-3 &  -1.1e-4 & 0.35  &  0.12 & 12.28 & -3.90\\
5 & 3.9e-3 & -1.5e-3 &  0.30  &  0.10  & 12.10 & -3.01 \\ \hline
\end{tabular}\end{table}

\begin{table}
\caption{Best fit values for the parameters of the HI halo mass function $M_{\mathrm{HI}}(M_h)$ of equation~\ref{eq:M_HI_sat} as a function of redshift when only satellites are considered (see figure~\ref{fig:median_halo_HI_cen_sat}). }\label{tab:M_HI_sat}
\begin{center}
\begin{tabular}{c|cccc}
z & $\log_{10}(M_0)$ & $\log_{10}(M_{\mathrm{min}})$  & $\alpha$ & $\gamma$  \\
& $[h^{-1}{\rm M}_\odot]$  &  $[h^{-1}{\rm M}_\odot]$ & & \\ \hline
0 & 9.30 &  12.0 & 0.81 & 0.70 \\
1 & 8.31 &  11.4 & 1.10 & 0.84  \\
2 & 7.66  & 11.00 & 1.16 & 1.05 \\
3 & 7.23  &  10.75 & 1.22 & 1.44 \\
4 & 7.74 &  10.95 & 1.05 & 0.90 \\
5 & 7.11 &  10.63 & 1.20 & 1.83 \\ \hline
\end{tabular}
\end{center}
\end{table}

\begin{figure*}
\includegraphics[width=\columnwidth]{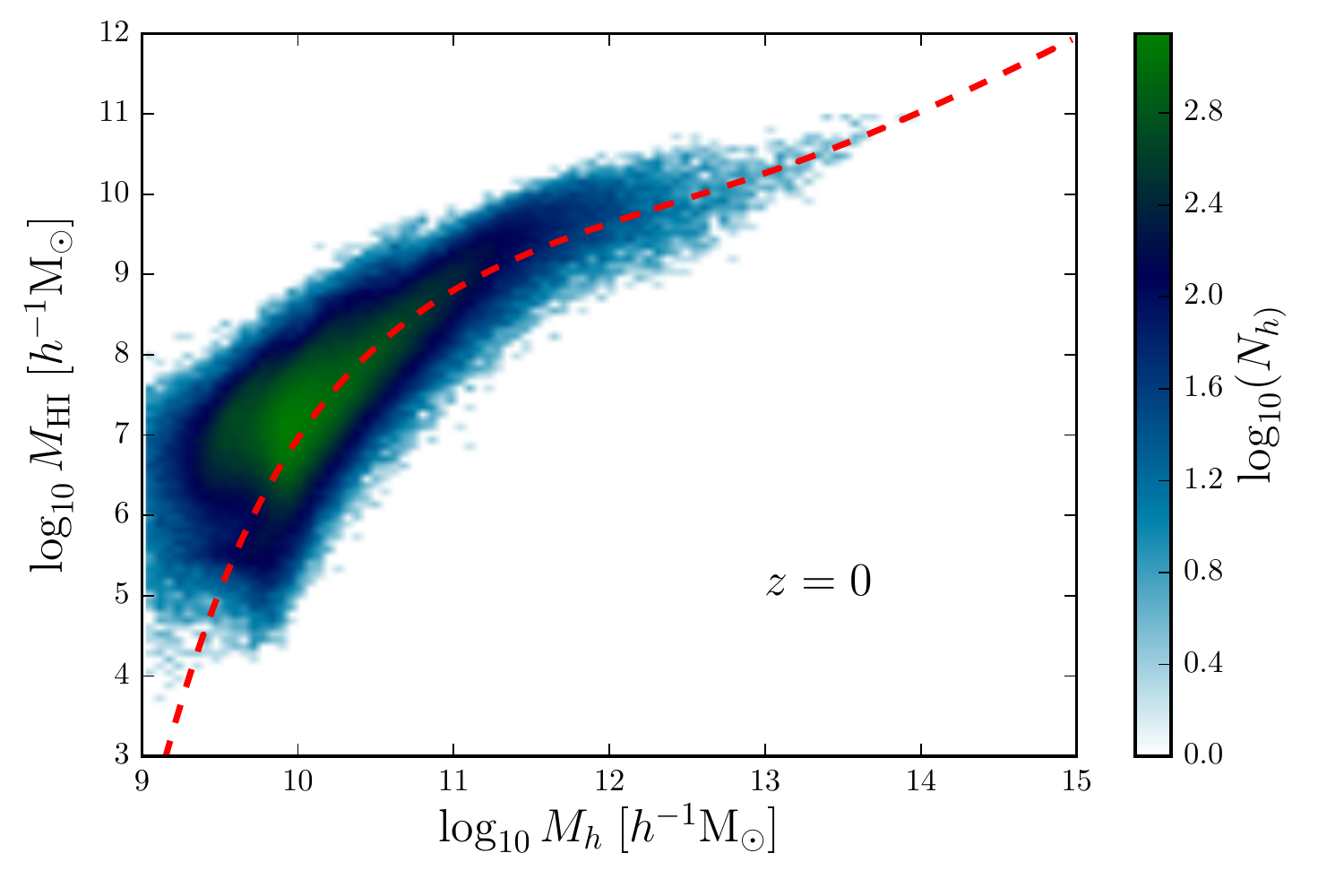}
\includegraphics[width=\columnwidth]{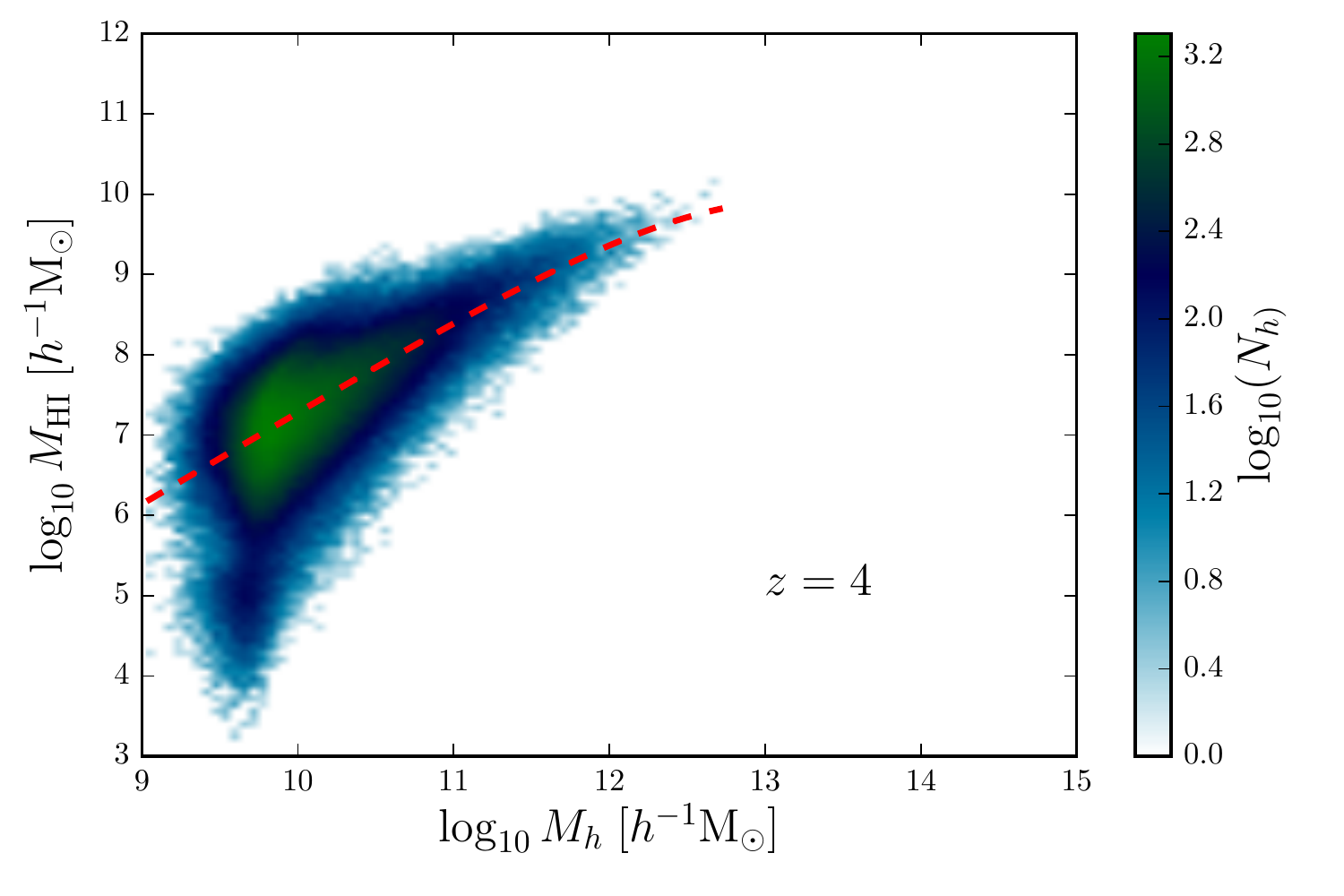}
\includegraphics[width=\columnwidth]{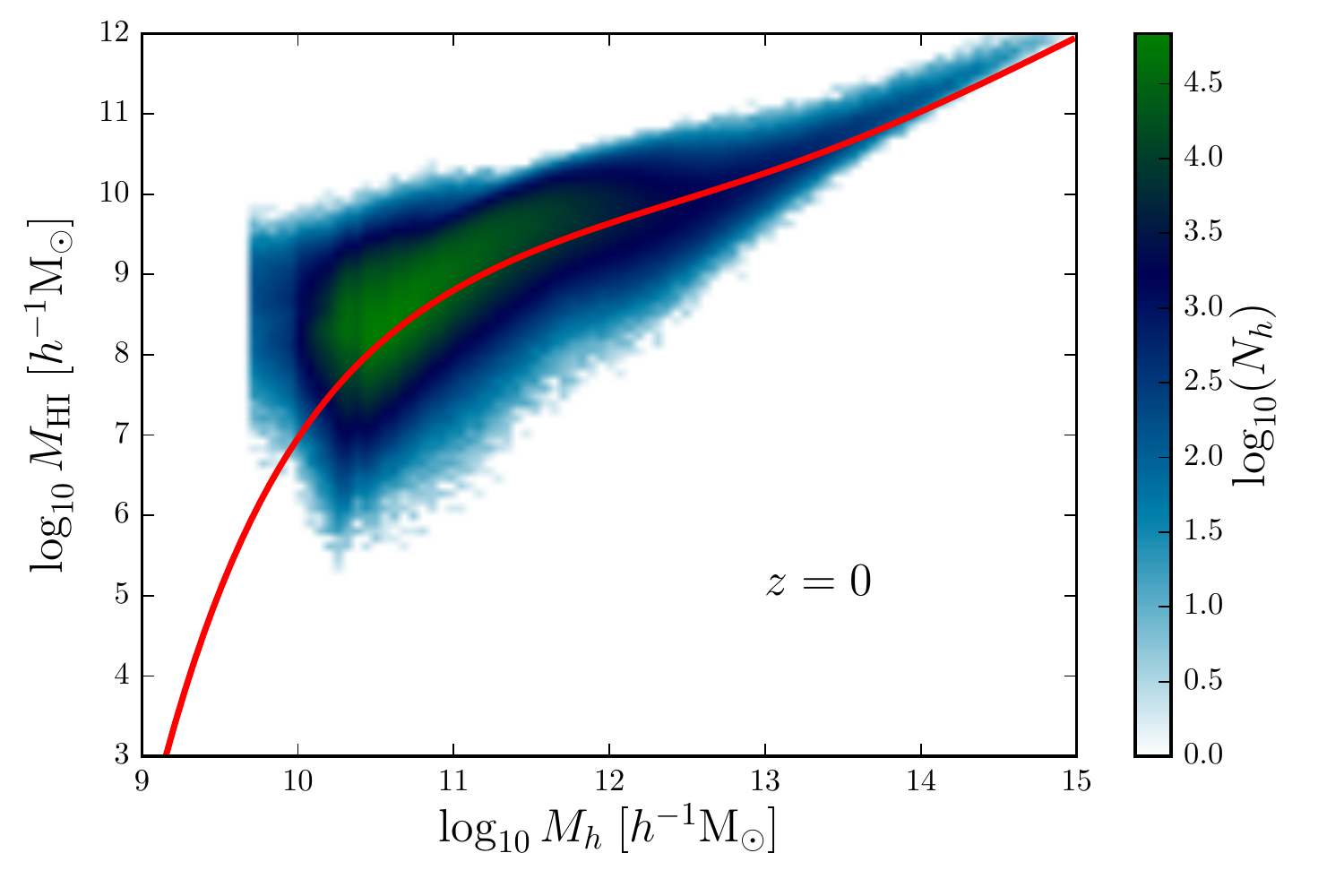}
\includegraphics[width=\columnwidth]{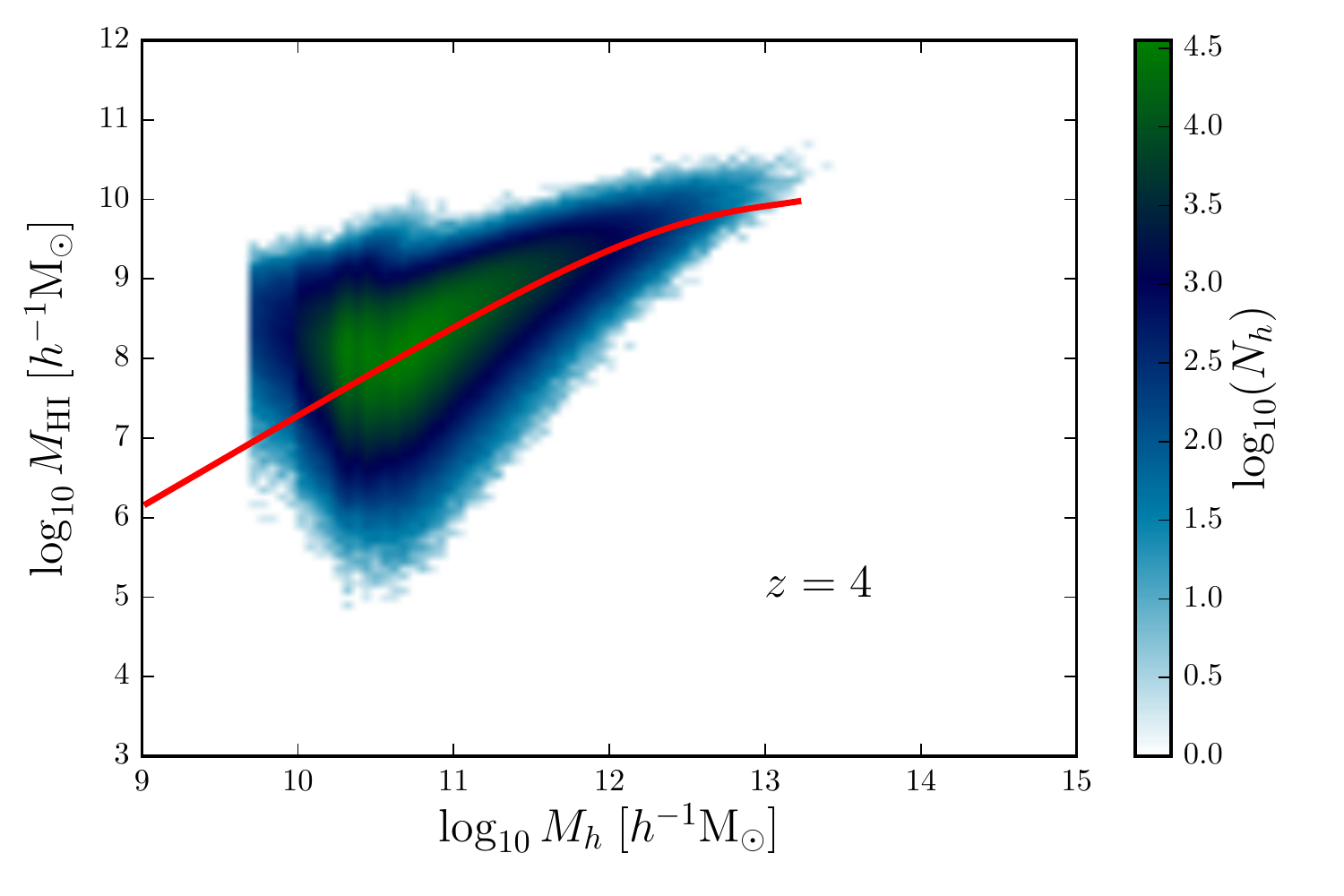}
\caption{The predicted HI halo mass function $M_{\rm HI}(M_h)$ at redshift $z=0$ and redshift $z=4$, color coded by the number of haloes in each bin. The red lines show the best fit obtained using the parametrization given in equation~\ref{eq:M_HI}. {\it Top panels}: MII. {\it Bottom panels:} MI. Note that the best fit value reported here is the one from the MII case, to show the agreement between the two simulations.}\label{fig:M_HI_scatter}
\end{figure*}

We analyze also the dependence on assembly bias (see figure~\ref{fig:assembly}). As described in section~\ref{sec:MHI} we use as a proxy for assembly history the redshift at which a halo has acquired half of its final mass, i.e. $z_{50}$. We define ``old/early'' assembled haloes as those whose $z_{50}$ falls within the $33$th percentile of the distribution, ``average age'' haloes as those whose $z_{50}$ is within the $33$th and the $66$th percentiles and ``young/late" assembly haloes as those whose $z_{50}$ ranges above the $66$th percentile.
In table~\ref{tab:M_HI_assembly} we report the fitting formula for the three cases. We use again equation~\ref{eq:M_HI}, fixing $\gamma=0.3$.
As discussed in section~\ref{sec:MHI}, the dependence on assembly history is related to the scatter of the $M_{\rm HI}(M_h)$ relation. To illustrate the scatter, in figure~\ref{fig:M_HI_scatter}, we show the HI halo mass function at redshift $z=0$ and $z=4$, color coded with the number of haloes in each bin, for both the MII and the MI. 
\begin{table}
\caption{Best fit values for the parameters of the HI halo mass function $M_{\mathrm{HI}}(M_h)$ of equation~\ref{eq:M_HI} as a function of assembly time (see figure~\ref{fig:assembly}). These value are obtained with a fixed value of the parameter $\gamma=0.3$.}\label{tab:M_HI_assembly}
\begin{tabular}{c|cccccc}
$z_{50}$ & $a_1$ & $a_2$ & $\alpha$ & $\beta$ & $\log_{10}(M_{\mathrm{break}})$  & $\log_{10}(M_{\mathrm{min}})$  \\
& & & & & $[h^{-1}{\rm M}_\odot]$  &  $[h^{-1}{\rm M}_\odot]$ \\ \hline
$\mathrm{early}$ & 1.6e-3 & -5.4e-5  & 0.80 & 0.29 & 12.06 & -7.82\\
$\mathrm{average}$  & 2.0e-3 & 7.5e-4 & 0.39  &  1.35 & 10.02 & -5.46\\
$\mathrm{late}$  & 2.5e-2 & 1.2e-03 & 0.25 & 1.60 & 8.24 & -3.89\\ \hline
\end{tabular}\end{table}

\label{lastpage}

\end{document}